\documentclass[11pt,a4paper]{article}                           

\usepackage{a4}          

\usepackage[UKenglish]{babel}
\usepackage{amsmath}
\usepackage{amssymb}
\usepackage{mathtools}
\usepackage[usenames]{color}
\usepackage{multirow}
\usepackage{graphicx}
\usepackage{epstopdf}
\usepackage{array}
\usepackage{hhline}
\usepackage{microtype}
\usepackage{booktabs}
\usepackage{multirow}

\usepackage[bf]{caption}
\usepackage{color}
\usepackage[usenames,dvipsnames]{xcolor}
\usepackage{tikz}
\usepackage{tikz-cd}
\usepackage{tikz-3dplot} 
\usepackage{pgfplots}
\usetikzlibrary{matrix,arrows,calc,chains,intersections,decorations.markings,matrix,shapes.geometric,positioning,plotmarks,scopes,patterns}
 
\usepackage{subfig}

\definecolor{darkgreen}{rgb}{0.05,0.6,0.1}

\usepackage{ifpdf}
\ifpdf
  \usepackage[pdftex,
    pdftitle={Massless matter in F-theory},
    pdfauthor={},
    pdfsubject={},
    bookmarksopen, bookmarksnumbered, bookmarksopenlevel=2]{hyperref}
\fi
\def\hybrid{\topmargin -20pt    \oddsidemargin 0pt
        \headheight 0pt \headsep 0pt
        \textwidth 6.25in       
        \textheight 9 in       
        \marginparwidth .875in
        \parskip 5pt plus 1pt 
          \jot = 1.5ex
   }
\hybrid
\numberwithin{equation}{section}
\numberwithin{table}{section}

\DeclareMathOperator{\coker}{Coker}

\DeclareMathOperator{\divisor}{div}
\DeclareMathOperator{\Hom}{Hom}
\DeclareMathOperator{\id}{id}

\DeclareMathOperator{\ord}{ord}
\DeclareMathOperator{\Rat}{Rat}

\newcommand{\beq}{\begin{equation}}  \newcommand{\eeq}{\end{equation}}
\newcommand{\bal}{\begin{aligned}}   \newcommand{\eal}{\end{aligned}}
\newcommand{\bea}{\begin{eqnarray}}  \newcommand{\eea}{\end{eqnarray}}

\newcommand{\CH}{{\rm{CH}}}

\newcommand{\Pic}{{\rm Pic}}
\newcommand{\Ker}{\rm Ker}

\newcommand{\tw}{\text{w}}

\usepackage{cancel}

\newcommand{\be}{\begin{equation}}
\newcommand{\ee}{\end{equation}}

\newcommand{\executeiffilenewer}[3]{%
 \ifnum\pdfstrcmp{\pdffilemoddate{#1}}%
 {\pdffilemoddate{#2}}>0%
 {\immediate\write18{#3}}\fi%
}

\begin{document}

\baselineskip=14pt
\parskip 5pt plus 1pt

\vspace*{-1.5cm}
\begin{flushright}    
  {\small

  }
\end{flushright}

\vspace{2cm}
\begin{center}        
  {\LARGE Chow groups, Deligne cohomology and massless matter in F-theory }
\end{center}

\vspace{0.75cm}
\begin{center}        
Martin Bies, Christoph Mayrhofer, Christian Pehle and Timo Weigand

\end{center}

\vspace{0.15cm}
\begin{center}        
  \emph{ Institut f\"ur Theoretische Physik, Ruprecht-Karls-Universit\"at, \\
             Heidelberg, Germany}
             \\[0.15cm]
 \end{center}

\vspace{2cm}


\begin{abstract}
We propose a method to compute the exact number of charged localized massless matter states
in an F-theory compactification on a Calabi-Yau 4-fold with non-trivial 3-form data. 
Our starting point is the description of the 3-form data via Deligne cohomology. A refined cycle map allows us to specify concrete elements therein in terms of  the second Chow group of the 4-fold, i.e. rational equivalence classes of algebraic 2-cycles.
We use intersection theory within the Chow ring to extract from this data a line bundle class on the curves in the base of the fibration on which charged matter is localized. The associated cohomology groups are conjectured to count the exact massless spectrum, in agreement with general patterns in Type IIB compactifications with 7-branes. 
We exemplify our approach by calculating the massless spectrum in an $SU(5) \times U(1)$ toy model based on an elliptic 4-fold with an extra section.
The explicit evaluation of the cohomology classes is performed with the help of the \emph{cohomCalg}-algorithm by Blumenhagen et al.

\end{abstract}

\thispagestyle{empty}
\clearpage
\setcounter{page}{1}


\newpage

\tableofcontents

\section{Introduction}

The central question of this work concerns the way how the exact charged massless spectrum is encoded in the 'gauge bundle data' of an  F-theory \cite{Vafa:1996xn,oai:arXiv.org:hep-th/9602114,oai:arXiv.org:hep-th/9603161} compactification to four dimensions. Apart from posing an exciting mathematical challenge by itself, the computation of the exact charged spectrum is of great relevance to phenomenological applications of F-theory as pioneered in \cite{oai:arXiv.org:0802.2969,oai:arXiv.org:0802.3391,Beasley:2008kw,Donagi:2008kj} and studied intensively in the recent literature. 
The massless gauge and matter degrees of freedom of an F-theory compactification are understood most generally by duality with M-theory. 
Non-abelian gauge bosons on a 7-brane along a complex codimension-one cycle on the base $B_3$ of the elliptic fibration $Y_4$ arise from the excitations of M2-branes wrapped along  combinations of vanishing $\mathbb P^1$s in the fiber; these intersect like the affine Dynkin diagram of the associated Lie algebra and give rise, in the zero-volume limit, to an ADE-type singularity in the fibre. Throughout this paper we will work with a Calabi-Yau resolution $\hat Y_4$ of $Y_4$, assuming it exists, corresponding to a non-vanishing volume of the fibre $\mathbb P^1$s.

Similarly, massless matter arises from suitable zero mode excitations of M2-branes wrapped around certain vanishing $\mathbb P^1$s in the elliptic fiber.
 Apart from so-called bulk matter, which propagates along the entire 7-branes of the F-theory compactification, extra charged M2-brane zero modes localise on complex curves $C_R$ in the base $B_3$, where one or two 7-brane loci (self-)intersect \cite{Katz:1996xe}. Over these curves the singularity type of the fibre enhances, i.e. one or more of the fiber $\mathbb P^1$s split.
 The representation $R$ of the gauge group $G$ in which the asscociated chiral ${\cal N}=1$ supermultiplets transform is endoced in the geometry of the new $\mathbb P^1$s over the curve $C_R$.
The precise number of chiral ${\cal N}=1$ supermultiplets in representation $R$ and $\bar R$  depends on additional 'gauge bundle data'. This data is encoded, by duality with M-theory, in the 3-form potential $C_3$ and its field strength $G_4 \in H^{2,2}(\hat Y_4)$.  

At the level of the chiral index of massless states, the specification of gauge flux $G_4$ is sufficient. This is because the chiral index of charged massless states localised on the matter curve $C_R$ is expressed as $\int_{A_R} G_4$, where $A_R$ is the matter surface consisting of the wrapped $\mathbb P^1$s fibered over $C_R$. While in absence of a microscopic understanding of the quantum theory of M2-branes in M-theory this formula has not been  derived  from first principle, it has been checked by duality with heterotic string theory \cite{oai:arXiv.org:0802.2969,oai:arXiv.org:0904.1218,oai:arXiv.org:1108.1794}, Type IIB with 7-branes \cite{Braun:2011zm,Krause:2011xj,oai:arXiv.org:1202.3138,Clingher:2012rg} and M-theory \cite{oai:arXiv.org:1111.1232}.

The computation of the exact massless spectrum requires a much finer specification of the 3-form data than in terms of $G_4 \in H^{2,2}(\hat Y_4)$. 
A definition of the M-theory 3-form data in topologically non-trivial situations has been given in \cite{Diaconescu:2003bm,Freed:2004yc,oai:arXiv.org:hep-th/0409158} in the language of Cheeger-Simons twisted differential characters \cite{Cheeger-Simons}. 
In this paper we will use an equivalent formulation \cite{Curio:1998bva,Donagi:1998vw} in terms of Deligne cohomology \cite{EsnaultDeligne}. 
The relevance of either of these two formulations for F/M-theory compactifications has been stressed  in various places in the recent F-theory literature including \cite{oai:arXiv.org:1203.6662,Clingher:2012rg,Anderson:2013rka}.

Deligne cohomology can be thought of as the generalisation of the Picard group of gauge equivalence classes of line bundles to a higher form gauge theory. In particular, we can think of the Deligne cohomology group $ H^{4}_D(X,\mathbb{Z}(2))$ as the object that fits into the short exact sequence
\bea
\begin{tikzcd}
0 \arrow{r} & J^2(Y_4) \arrow{r} & H^{4}_D(Y_4,\mathbb{Z}(2))
  \arrow{r}{\hat{c}_2}  & H^{2,2}_{\mathbb{Z}}(Y_4) \arrow{r} & 0.
\end{tikzcd}
\eea
The map $\hat c_2$ associates to an element in $ H^{4}_D(Y_4,\mathbb{Z}(2))$ a 4-form cohomology class which we interpret as a   $G_4$-flux. The kernel of this map consists of the flat $C_3$-field configurations with values in the so-called intermediate Jacobian $J^2(Y_4)$.
This generalises the more familiar relation between the gauge equivalence classes of line bundles $\Pic(X)$, the first Chern class $c_1$ assigning to each such equivalence class a curvature 2-form and the space of flat connections (or Wilson lines).

The starting point for our analysis is the observation that a convenient way to specify such 3-form data in $ H^{4}_D(\hat Y_4,\mathbb{Z}(2))$ is in terms of the Chow group $\CH^2(\hat Y_4)$, the group of rational equivalence classes of algebraic 2-cycles on $Y_4$.\footnote{Here and in the sequel by a $p$-cycle we mean a cycle of \emph{complex} dimension $p$.} 
This obervation is based on the existence of a well-known map from $\CH^2(\hat Y_4)$ to $ H^{4}_D(\hat Y_4,\mathbb{Z}(2))$  - the so-called refined cycle map $\hat \gamma$  \cite{EsnaultDeligne,GreenMurreVoisin} - which allows us to associate to each rational equivalence class of algebraic 2-cycles a piece of 3-form data. 

The relation between algebraic cycles and gauge data in F/M-theory as such is of course not new.
Indeed in \cite{Braun:2011zm} the existence of a map from the group of algebraic 2-cycles into $H^{2,2}(\hat Y_4)$ - the cycle map $\gamma$ - has been used to construct interesting $G_4$ fluxes. The intersection number in homology with the matter surfaces computes the chiral index of charged massless modes, as described above. 

Our approach is to exploit the refined cycle map to specify not only a gauge flux $G_4$, but a full set of 3-form data in $ H^{4}_D(\hat Y_4,\mathbb{Z}(2))$ via algebraic cycles up to rational equivalence. As will be reviewed in section \ref{subsec_DelChow},  rational equivalence is a refinement of the notion of homological equivalence (see e.g. \cite{EisenbudInt} for details). While two homologically equivalent algebraic cycles describe the same $G_4$-flux, two rationally equivalent algebraic cycles describe even the same 3-form data up to gauge equivalence and not only the same gauge flux. Therefore the 
 intersection product in the Chow ring \cite{FultonInt} - i.e. the intersection product  up to rational (as opposed to homological) equivalence  - is a legitimate operation if we are interested in preserving the full information about the 3-form data up to gauge equivalence and not merely topological information e.g. about the chiral index.


Indeed  we describe a well-defined procedure based on intersection theory on the Chow ring $\CH(\hat Y_4)$ as developed in \cite{FultonInt} which allows us to extract from an element in $\CH^2(\hat Y_4)$ a line bundle class on the matter curves $C_R$ with representative bundle $L_R$. We further conjecture that the localised charged massless matter multiplets of an F-theory compactification are counted by $H^i(C_R, L_R \otimes \sqrt{K_{C_R}})$ with  $\sqrt{K_{C_R}}$ the spin structure on $C_R$ determined by the holomorphic embedding of the matter curve into the Calabi-Yau 4-fold in an ${\cal N}=1$ supersymmetric configuration.
The very fact that the massless  localised charged matter multiplets are counted by cohomology groups of the form $H^i(C_R, {\cal L}  \otimes \sqrt{K_{C_R}})$ for some line bundle ${\cal L}$ is of course well-known from Type IIB models with intersecting 7-branes \cite{Katz:2002gh}. It was also derived in \cite{oai:arXiv.org:0802.2969,oai:arXiv.org:0802.3391} in more general F-theory models from the perspective of the twisted ${\cal N}=1$
gauge theory on a stack of 7-branes deformed locally into an intersecting configuration. 
However, a priori the relation between the line bundle ${\cal L}$ and the $G_4$-flux of a globally defined 4-fold is not obvious. Our analysis explains how to extract the bundle data on the matter curve for general 3-form configurations, without making use of any local approximations.

The way how to extract the line bundle $L_R$ on $C_R$ is essentially integration along the fiber, which, as noted above, is well-defined within rational equivalence groups. It bears some resemblance with the cylinder map exploited in the context of spectral covers 
\cite{Curio:1998bva,oai:arXiv.org:0904.1218,Donagi:2011jy,Donagi:2011dv,Clingher:2012rg}. Note in particular that in \cite{Marsano:2012yc} cohomology groups counting charged matter states in spectral cover models have been explicitly computed. 
Our approach is generally valid also in generic F-theory compactifications which cannot be described by spectral covers. In particular this includes e.g. the spectrum of charged singlet fields on matter curves away from a 7-brane with non-abelian gauge group as appearing in 4-dimensional F-theory compactications with extra abelian gauge groups \cite{Grimm:2010ez,Krause:2011xj,Grimm:2011fx,oai:arXiv.org:1202.3138,oai:arXiv.org:1208.2695,oai:arXiv.org:1210.6034,Mayrhofer:2012zy,Braun:2013yti,Borchmann:2013jwa,Cvetic:2013nia,Braun:2013nqa,Cvetic:2013uta,Borchmann:2013hta,Cvetic:2013qsa}.
We also point out that ref. \cite{oai:arXiv.org:1203.6662} approaches the counting of zero modes arising at curves of conifold singularities in M/F-theory by analysing the stable points of the superpotential and by describing the 3-form data in the language of Cheeger-Simons differential characters.

The assertion that the above line bundle $L_R$ on $C_R$ indeed counts the massless zero modes is not only natural, but also fully agrees with expectations both from a Type IIB and heterotic perspective. We exemplify this for  the special case of 3-form data underlying the $U(1)$ gauge flux of an F-theory model with extra abelian gauge symmetry. 
To demonstrate the applicability of our approach we compute the exact massless spectrum of an $SU(5) \times U(1)_X$ F-theory compactification of the type worked out in detail in \cite{Krause:2011xj}. The concrete 3-form data of this toy model is chosen such that its associated flux leads to three chiral generations of $SU(5)$ matter. The problem is reduced completely to the computation of a certain line bundle cohomology on a curve on the base $B_3$. 
If $B_3$ is holomorphically embedded into a smooth toric variety and for suitable 3-form data, theses cohomologies can be conveniently computed in concrete examples with the help of the \emph{cohomCalg} algorithm developed by Blumenhagen et al. \cite{Blumenhagen:2010pv, cohomCalg:Implementation, 2011JMP....52c3506J, Rahn:2010fm, Blumenhagen:2010ed} and standard techniques from homological algebra.

The remainder of this article is organised as follows:
To set the stage we recall in section \ref{rev-lineIIB} the counting of massless matter in Type IIB compactifications with 7-branes, which we are aiming to generalise to F/M-theory. In section \ref{ssec:GaugeDataDeligne} we review the connection between Deligne cohomology and the 3-form data up to gauge equivalence. Our approach to specify elements in the Deligne cohomology via rational equivalence classes of algebraic cycles is described in section \ref{subsec_DelChow}. For convenience of the reader we also include, in section \ref{subsec_PropChow}, some of the standard properties of Chow groups with focus on the well-known intersection product which we will need for our purposes. In section \ref{CohomForm} we use this machinery to extract from an element of $\CH^2(\hat Y_4)$ a line bundle equivalence class on a matter curve $C_R$, whose cohomology classes are naturally conjectured to count the massless charged ${\cal N}=1$ chiral supermutliplets. In section \ref{subsec_ApplU(1)} we specify this apporach to 3-form data underlying a $U(1)$ gauge flux and find full agreement with the corresponding well-established formulae in Type IIB.
Section \ref{sec3gen} applies our findings to an $SU(5) \times U(1)$ F-theory toy model. 
After defining the model in section \ref{subsecGenSetup},  we describe in general how in suitable configurations the cohomology classes defined in section \ref{subsec_ApplU(1)}  can be computed with the help of the \emph{cohomCalg} algorithm \cite{Blumenhagen:2010pv, cohomCalg:Implementation, 2011JMP....52c3506J, Rahn:2010fm, Blumenhagen:2010ed} and the use of Koszul spectral sequences.  This is described in section \ref{sub_cohom}.
Finally, in section \ref{sec:AConcreteExample} all of this is exemplified for a 3-form configuration which gives rise to three chiral generations of $SU(5) \times U(1)$ matter, plus vectorlike pairs, whose number we compute. 
An outlook is given in section \ref{sec_Concl.}. Details of the cohomology computations of section  \ref{sec3gen} and more background on the definition of Deligne cohomology are presented in the appendices.

\section{Gauge data from Deligne cohomology}

\subsection{Review of line bundles and their cohomologies in Type IIB} \label{rev-lineIIB}

We start by reviewing in this section the 
 description of gauge data in  Type IIB string theory. This well-known material will serve as a preparation for the definition of gauge in F/M-theory, which is probably less familiar to most readers.

Consider a Type IIB orientifold compactification including a set of stacks of $N_i$ 7-branes wrapping holomorphic cycles $D_i$ of complex dimension two  of the Calabi-Yau 3-fold $X_3$.\footnote{Note once more that throughout this paper a $p$-cycle will denote a cycle of complex dimension p.}
To define the internal gauge data one has to specify in general an element of the derived bounded category $D_b(X_3)$ of coherent sheaves on $X_3$.
A typical element of $D_b(X_3)$ corresponds to a bound state of coherent sheaves localized along the complex 2-cycles $D_i$, where physically the existence of a non-trivial bound state is associated with  a non-zero vacuum expectation value of some of the charged open string zero modes at the intersection of the branes. These correspond to the morphisms between the sheaves in the category of coherent sheaves. For a review of this picture we refer e.g. to \cite{Aspinwall:2004jr}.

For simplicity we focus in the sequel on special situations in which the morphisms are turned off and the individual sheaves can be described as line bundles on the respective 4-cycles $D_i$. In particular we assume vanishing gluing morphisms \cite{Donagi:2011jy,Donagi:2011dv}. In addition we focus on abelian gauge data.
With these simplifying assumptions specifying  the gauge data amounts to specifying an isomorphism class of line bundles  $L_i$ on $D_i$  with two representatives of the equivalence class being equal if the two associated line bundles differ only by a gauge transformation. 
As preparation for the definition of gauge data in F-theory we now briefly review how to specify such an equivalence class of line bundles.

Given a line bundle $L$ on a smooth complex projective
variety $X$, the first Chern class $c_1(L) = \frac{1}{2 \pi i } {\rm tr} F$ is an integral
$(1,1)$-form $c_1(L) \in H^{1,1}_{\mathbb{Z}}(X)$ with $H^{1,1}_{\mathbb Z}(X) = H^{1,1}(X) \cap H^2(X, \mathbb Z)$. Isomorphism classes
of holomorphic line bundles form the Picard group $\Pic(X)$. Its group
structure is given by the tensor product of line bundles, that is
for isomorphism classes $[L],[L']$ with representatives $L,L'$ in $\Pic(X)$, $[L]\cdot [L'] = [L
\otimes L']$. The inverse of a line bundle $L$ is the \emph{dual line bundle}
$L^\ast = \Hom(L,\mathcal{O}_X)$.
Since taking the first Chern class behaves additively with
respect to the tensor product, $c_1(L \otimes L') = c_1(L) + c_1(L')$, the map $L \mapsto c_1(L)$
defines a group homomorphism from the Picard group $\Pic(X)$ to
$H^{1,1}_\mathbb{Z}(X)$, which is in fact \emph{surjective} and
whose kernel is called $\Pic^0(X)$. In other words, there is an exact sequence
\begin{equation}
  \label{eq:ShortExactSequencePicard}
  0 \to \Pic^0(X) \to \Pic(X) \to H^{1,1}_\mathbb{Z}(X) \to 0.
\end{equation}
The Picard group $\Pic(X)$ can be identified with the first sheaf
cohomology group of $\mathcal{O}^\ast_X$, the sheaf of invertible holomorphic functions on
$X$,\footnote{See e.g.\ the appendix of \cite{Distler:1987ee} for a quick review.}
\begin{equation}
  \label{eq:6a}
  H^{1}(X,\mathcal{O}^\ast_X) \cong \Pic(X).
\end{equation}
Now, consider the exponential exact
sequence
\begin{equation}
  \label{eq:7}
  0 \to \mathbb{Z}_X \xrightarrow{2\pi i} \mathcal{O}_X \xrightarrow{
    exp } \mathcal{O}^\ast_X \to 0,
\end{equation}
where ${\mathbb Z}_X$ is the sheaf of integers over $X$ and ${\cal O}_X$ denotes the sheaf of holomorphic functions on $X$. The first map denotes multiplication with $2 \pi i$ and the second map is exponentiation.
Its associated long exact sequence in cohomology
\begin{equation}
  \label{eq:8}
  \cdots \to H^1(X,\mathbb{Z}) \to H^1(X,\mathcal{O}_X) \to
  H^1(X,\mathcal{O}^\ast_X) \xrightarrow{c_1} H^2(X,\mathbb{Z}) \to
  \cdots
\end{equation}
implies that  
\bea
\Pic^0(X) = \Ker(c_1 \colon H^1(X,\mathcal{O}^\ast_X) \to
H^2(X,\mathbb{Z}))
\eea
 can be identified with
$H^1(X,\mathcal{O}_X)/H^1(X,\mathbb{Z})$. Now by Hodge theory 
\begin{equation}
  \label{eq:9}
  H^1(X,\mathcal{O}_X) = H^{0,1}(X)
\end{equation}
and since complex conjugation simply exchanges the grading
$H^{1,0} = \overline{H^{0,1}}$, there is an isomorphism of real vector spaces
\begin{equation}
  \label{eq:10}
  H^1(X,\mathbb{R}) \subset H^1(X,\mathbb{C}) \to H^{0,1}(X).
\end{equation}
Therefore the lattice $H^1(X,\mathbb{Z}) \subset H^1(X,\mathbb{R})$
can be seen as a lattice in $H^{0,1}(X)$. In conclusion ${\Pic}^0(X) =
H^{0,1}(X)/H^{1}(X,\mathbb{Z})$ is a torus, also known as
the Jacobian $J^1(X)$ of $X$. It parametrizes, in physics language, the inequivalent Wilson lines, i.e. the flat connections.

If $X$ is simply connected, i.e. $\pi_1(X)=0$, the Jacobian $J^1(X)=0$ and a line bundle is, up to gauge equivalences, uniquely specified by its first Chern class. 
More generally, the details of the gauge data such as the spectrum of
massless states depend on finer data than merely an element in
$H^{1,1}_{\mathbb Z}$.

On a smooth irreducible projective variety 
 the group $\Pic(X)$ is isomorphic to the group ${\rm Cl}_1(X)$ of Cartier divisors modulo linear equivalence.
These in turn are isomorphic, on a smooth irreducible projective variety, to the group of Weil divisors modulo linear equivalence. The latter correspond to formal linear combinations of algebraic subvarieties of codimension 1, where two divisors are equivalent if they differ by the zeros and poles of a globally defined meromorphic function on $X$. This group is known as the first Chow group $\CH^1(X)$ of $X$.

A complication arises because due to the Free-Witten anomaly, the field strength of the gauge potential on a 7-brane along $D_i$ must be quantized such that \cite{Freed:1999vc}
\bea \label{FW1}
\frac{1}{2 \pi i} F_i + \frac{1}{2} c_1(K_{D_i}) \in H^{1,1}_{\mathbb Z}(D_i).
\eea
Here and throughout this paper we assume vanishing torsion $B$-field.
If $D_i$ is not spin, $F$ is therefore not the field strength of an integer quantised line bundle due to a $1/2$ shift. This will not pose any problems for us and we will, by abuse of language, still speak of a line bundle on $D_i$ (even though it is more correctly a ${\rm Spin}_{\mathbb C}$ bundle \cite{Freed:1999vc}).

To summarize, under the simplifying assumptions stated, we can think of the gauge data of a Type IIB compactification in terms of a set of divisor classes on each of the holomorphic 2-cycles $D_i$ wrapped by the 7-branes of the compactification. 

There are two types of massless open string modes - the so-called bulk modes whose wave-function is defined on the entire 2-cycle $D_i$ and the localized zero modes defined at the intersection curves $C_{ab} = D_a \cap D_b$ of two 2-cycles. The most general framework to describe these is in the language of ${\rm Ext}$ groups \cite{Katz:2002gh}. If all sheaves are given by line bundles $L_i$ it suffices to instead switch to the more familiar language of cohomology groups.
In particular, the localized  massless excitations of open strings stretched between $D_a$ and $D_b$ are counted by
\bea \label{cohomgroupIIB}
H^i(C_{ab}, L_{ab} \otimes \sqrt{K_{C_{ab}}}), \qquad\quad i=0,1.
\eea
Here $L_{ab} = L_a^*|_{C_{ab}} \otimes L_b |_{C_{ab}}$ and $\sqrt{K_{C_{ab}}}$ denotes the spin bundle on $C_{ab}$. 
Let us fix our conventions such that 
 $H^0(C_{ab}, L_{ab} \otimes \sqrt{K_{C_{ab}}})$ counts the number of ${\cal N}=1$ chiral multiplets in representation $({ \bf \overline N}_a, {\bf N}_b)$  with respect to the gauge group $U(N_a) \times U(N_b)$ on the two brane stacks;  
 $H^1(C_{ab}, L_{ab} \otimes \sqrt{K_{C_{ab}}})$ counts ${\cal N}=1$ chiral multiplets in representation $({\bf N}_a, {\bf \overline N}_b)$. If supersymmetry is broken spontaneously, the mass degeneracy between the bosonic and fermonic states within this chiral multiplet is lifted and in general only a chiral Weyl fermion remains massless.
The associated index is
 \bea
 \chi_{ab} = h^0(C_{ab}, L_{ab} \otimes \sqrt{K_{C_{ab}}}) - h^1(C_{ab}, L_{ab} \otimes \sqrt{K_{C_{ab}}}) = \int_{C_{ab}} c_1(L_{ab}).
 \eea
Note that the Freed-Witten condition (\ref{FW1}) ensures that  $L_{ab}$ is integer quantized on $C_{ab}$.
While the index depends only on the cohomology class of $c_1(L_a)$ and
$c_1(L_b)$, the actual cohomology groups are sensitive to the equivalence classes of $L_a$ and $L_b$ as elements  of ${\rm Pic}(D_a)$ and $ {\rm Pic}(D_b)$,
whose respective pullbacks to $C_{ab}$ define the line bundle $L_{ab}
\in {\rm Pic}({C_{ab}})$.

Finally, the spin structure appearing in (\ref{cohomgroupIIB}) is induced by the holomorphic embedding of $C_{ab}$ into $D_a$ (or, equivalently, into $D_b$), in the following sense: By adjunction
\bea
K_{C_{ab}} = K_{D_a}|_{C_{ab}} \otimes N_{C_{ab}/D_a}.
\eea
Each of the bundles $K_{D_a}|_{C_{ab}}$ and $N_{C_{ab}/D_a}$ in turn arise as the pullback of line bundles from the ambient Calabi-Yau 3-fold $X_3$. In fact,
\bea
 K_{D_a}|_{C_{ab}} = (K_{X_3} \otimes   {\cal O}_{X_3}(D_a))|_{C_{ab}}, \qquad \quad \quad
 N_{C_{ab}/D_a} = {\cal O}_{X_3}(D_b)|_{C_{ab}}
\eea
and thus\footnote{For a Calabi-Yau $X_3$ clearly $K_{X_3}$ is trivial, but for later reference we keep in the subsequent formalae.}
  \bea
  K_{C_{ab}} =  {\cal M}|_{C_{ab}}, \qquad \quad       {\cal M} =  K_{X_3} \otimes   {\cal O}_{X_3}(D_a)  \otimes  { \cal O}_{X_3}(D_b) .
 \eea
The appearing  line bundles on $X_3$ are uniquely determined by their first Chern class since $h^{1}(X_3)=0$.
If $c_1({\cal M})$ is even, the bundle ${\cal M}^{1/2}$ is well-defined, and the claim is that the spin structure appearing in (\ref{cohomgroupIIB})
is
\bea \label{spinstructure}
\sqrt{K_{C_{ab}} } = {\cal M}^{1/2}|_{C_{ab}}.
\eea
Even if $ {\cal M}^{1/2}$ per se is not well-defined as a line bundle on $X_3$, 
the Freed-Witten quantization condition will always produce suitable combinations of bundles which are integer quantized.
Of special interest to us is e.g. the  
 situation in which the line bundles $L_a$ and $L_b$ themselves arise by pullback of line bundles on $X_3$. By slight abuse of notation let us denote these by ${\cal O}_{X_3}(L_a)$ and ${\cal O}_{X_3}(L_b)$. In this case the Freed-Witten quantization condition  ensures that $ c_1( {\cal O}_{X_3}(L_a)  ) +  c_1( {\cal O}_{X_3}(L_b)  ) + \frac{1}{2} c_1({\cal M})$ is integer on $X_3$. The pullback of the appearing combination of bundles to $C_{ab}$ then gives the argument appearing in the cohomology groups (\ref{cohomgroupIIB}). 
Similar reasoning can be applied to more general situations.

\subsection{Gauge data in F/M-theory via Deligne cohomology}
\label{ssec:GaugeDataDeligne}

In F-theory compactifications, the gauge data is encoded in a rather different way.
Let $Y_4$ denote an elliptically fibered Calabi-Yau 4-fold with projection
\bea
\pi: Y_4 \rightarrow B_3.
\eea
While the generic fiber of $Y_4$ is a smooth elliptic curve, its topology  is more complicated in higher codimension due to fiber degenerations  over the discriminant locus $\Delta \subset B_3$.
We will assume that a smooth resolution $\hat Y_4$ of the singular 4-fold $Y_4$ exists in which the singular fibers are replaced by chains of $\mathbb P^1$s whose intersection structure reflects the gauge and matter degrees of freedom. Recent work on the explicit construction of such a 4-fold resolution in the context of 4-dimensional F-theory compactifications includes \cite{Blumenhagen:2009yv,Grimm:2009yu,Chen:2010ts,oai:arXiv.org:1011.6388,Knapp:2011wk,Esole:2011sm,oai:arXiv.org:1108.1794,Krause:2011xj,oai:arXiv.org:1111.1232,oai:arXiv.org:1202.3138,Lawrie:2012gg,Hayashi:2013lra,Hayashi:2014kca} and references therein, to which we refer for more details on the fiber geometry reviewed in the next few paragraphs.

In general, the discriminant $\Delta$ splits into a number of irreducible components $\Delta_i$ as well as a remaining component $\Delta'$. Over each $\Delta_i$ the topology of the resolved fiber reproduces the affine Dynkin diagram of a Lie algebra $\mathfrak{g}_i$. This gives rise to vector multiplets in the adjoint representation of $\mathfrak{g}_i$ propagating on the component $\Delta_i$. In Type IIB language, such a discriminant component $\Delta_i$ is to be identified with the location of a corresponding stack of 7-branes even though the types of gauge algebras $\mathfrak{g}_i$ are more general than in perturbative Type IIB compactifications. The non-abelian gauge theory on $\Delta_i$ can be described in terms very similar to the conventional treatment of the worldvolume theory on a stack of Type IIB 7-branes.
Indeed the localization of the gauge degrees of freedom on $\Delta_i$ allowed the authors of \cite{oai:arXiv.org:0802.2969,oai:arXiv.org:0802.3391} to invoke a description in terms of a topologically twisted eight-dimensional ${\cal N}=1$ supersymmetric gauge theory. In this approach, gauge bundle data are specified essentially in the IIB language of line (or vector) bundles defined on $\Delta_i$, reproducing the  results of \cite{Katz:2002gh} on the cohomology groups counting massless matter at  the intersection of $\Delta_i$ with other 7-branes on $B_3$.

 In addition, however, the discriminant $\Delta$ contains a remaining component $\Delta'$ over which the fiber acquires an $I_1$-singularity corresponding to a self-intersecting $\mathbb P^1$. This so-called $I_1$-locus would correspond, in Type IIB language, to a complicated (in general non-perturbative) bound state  of all {\it single-wrapped} 7-branes as well as the O7-plane. 
Crucially, in a general F-theory compactification it is not possible to identify the individual cycles on $B_3$ on which such single 7-branes are wrapped and a direct definition of the gauge bundle data e.g. via the approach of \cite{oai:arXiv.org:0802.2969,oai:arXiv.org:0802.3391}  is not possible.
In particular, it is a priori not clear how to encode gauge bundle data via line bundles defined globally on a 7-brane as the very location of such single 7-branes is obscured.

In higher  codimension some of the fibre components can factorize further. 
In codimension two this happens in the fibre over curves in the base along which two components of the discriminant intersect or the same component self-intersects.
In  typical examples the fibre structure in codimension-two mimics the (affine) Dynkin diagram of a  higher-rank algebra into which the gauge algebras associated with the two intersecting fiber components are embedded. In the F-theory limit of vanishing fibre volume, M2-branes wrapping suitable linear combinations of $\mathbb P^1s$ in the fiber give rise to massless matter in some representation $R$ of the participating gauge algebras. Therefore, to each representation $R$ of massless matter one can associate a complex 2-cycle $A_R$ - the \emph{matter surface} -  given by the fibration of a certain linear combination of fiber $\mathbb P^1$s over a \emph{matter curve} $C_R$ in the base. 
More details will be given in section \ref{CohomForm}.

In F-theory additional data is
provided by specifying the 3-form configuration $C_3$ and its associated $4$-form flux $G_4 \in H^{2,2}(\hat Y_4)$, whose
existence is inferred from duality with M-theory \cite{oai:arXiv.org:hep-th/9605053,oai:arXiv.org:hep-th/9606122,oai:arXiv.org:hep-th/9908088}.
The 3-form $C_3$ gives rise, among others, to the degrees of freedom associated with abelian gauge bosons upon dimensional reduction. The full 3-form configuration therefore encodes the gauge data of the compactification on $\hat Y_4$.

The choice of $G_4$ is subject to two transversality constraints which ensure that it has 'one leg along the fiber' \cite{oai:arXiv.org:hep-th/9908088}.
More precisely, $G_4$ flux must satisfy the cohomological relations
 \bea
\int_{\hat Y_4} G_4 \cup \pi^*\omega_4 = 0 = \int_{\hat Y_4} G_4  \cup   [Z] \cup \pi^*\omega_2
 \eea
for every element $\omega_4 \in H^4(B_3)$ and $\omega_2 \in H^2(B_3)$ and with $[Z] \in H^{1,1}(\hat Y_4)$ denoting the class of the zero section of the fibration $\hat Y_4$. The product is the intersection product in cohomology on $\hat Y_4$.

Similarly to the half-integer quantisation shift (\ref{FW1}) for the gauge flux in 7-brane language, $G_4$ is in general an element of $H^{2,2}_{\frac{1}{2}{\mathbb Z}}(\hat Y_4)$ because it is subject to the quantisation condition \cite{oai:arXiv.org:hep-th/9609122}
\bea \label{FW2}
G_4 + \frac{1}{2} c_2(\hat Y_4) \in H^{2,2}_{{\mathbb Z}}(\hat Y_4).
\eea

Specifying such a $G_4$ flux allows one to compute the chiral index of massless matter in representation $R$
by integrating $G_4$ along the complex
two-cycle $A_R \in Z_2(Y)$ associated to the matter representation on  $C_{R}$ \cite{oai:arXiv.org:0802.2969, oai:arXiv.org:0904.1218, Braun:2011zm,oai:arXiv.org:1108.1794,Krause:2011xj, oai:arXiv.org:1111.1232,oai:arXiv.org:1202.3138,oai:arXiv.org:1203.6662},
\begin{equation}
  \label{eq:4}
  \chi_{R} = \int_{A_R} G_4.
\end{equation}
The fact that $\frac{1}{2} \int_{A_R} c_2(\hat Y_4) \in {\mathbb Z}$ guarantees that this quantity is integer \cite{oai:arXiv.org:1011.6388,oai:arXiv.org:1202.3138,oai:arXiv.org:1203.4542}. 

However, $G_4$ does not contain sufficient data to compute  
the number of chiral  ${\cal N}=1$ multiplets in representation $R$ and $\bar R$ separately.
This is to be expected from the situation in Type IIB, where, as reviewed above, we need only the first Chern class
$c_1(L_{ab})$ obtained from the map $c_1 \colon \Pic(C_{ab}) \to
H^{1,1}_\mathbb{Z}(C_{ab})$ to compute the chiral index, but the actual line
bundle $L_{ab}$ to compute  $h^i(C_{ab},L_{ab} \otimes K_{C_{ab}}^{1/2})$. 
Therefore the information about the M-theory 3-form encoded in $G_4 \in H^{2,2}(\hat Y_4)$
must be refined in the same way as giving a class $[L_{ab}] \in
\Pic(C_{ab})$ refines the first Chern class $c_1(L_{ab}) \in
H^{1,1}_\mathbb{Z}(C_{ab})$. It turns out that from a mathematical
standpoint natural candidates for such a refinement live in the fourth
\emph{Deligne cohomology} class of $\hat Y_4$.
Indeed the relevance of Deligne cohomology  - or equivalently of the theory of Cheeger-Simons differential characters \cite{Cheeger-Simons} -  to describe gauge data in M/F-theory has been explored in several places in the literature - see  \cite{Diaconescu:2003bm,Freed:2004yc,oai:arXiv.org:hep-th/0409158,oai:arXiv.org:1203.6662} and, respectively, 
\cite{Donagi:2011jy, Clingher:2012rg,Anderson:2013rka}.

For ease of presentation we ignore the half-integer shift of $G_4$ for the being and focus on situations in which $G_4 \in H^{2,2}_{\mathbb Z}(\hat Y_4)$. As long as we are making this simplification we can restrict ourselves to studying \emph{integral} Deligne cohomology. We will come back to the half-integer shift in quantisation in more general setups 
at the end of section \ref{sub_connDelChow}.

Before we come to a definition of Deligne cohomology, let us state
some of its properties. Deligne cohomology on a smooth complex projective variety $X$  is graded by two integers $p,\,q$ specifying the Deligne cohomology group $H^{p}_D(X,\mathbb{Z}(q))$, but we will only be interested in
the case $p = 2q$. The Deligne cohomolgy group
$H^{2}_D(X,\mathbb{Z}(1))$ is isomorphic to the Picard group
\begin{equation}
H^{2}_D(X,\mathbb{Z}(1)) \cong H^1(X,\mathcal{O}^\ast_X) \cong \Pic(X).
\end{equation}
As reviewed around equ.(\ref{eq:ShortExactSequencePicard}) it therefore fits into the short exact sequence
\begin{equation} \label{ses1}
  0 \to J^1(X) \to H^2_D(X,\mathbb{Z}(1)) \xrightarrow{c_1} H^{1,1}_\mathbb{Z}(X) \to 0,
\end{equation}
with 
\begin{equation} \label{J1def}
  J^1(X) =  \frac{H^{1}(X,\mathbb{C})}{H^{1,0}(X,\mathbb{C})
    + H^{1}(X,\mathbb{Z})} = H^{0,1}(X,\mathbb{C})/H^{1}(X,\mathbb{Z})
\end{equation}
the first intermediate Jacobian.

We would now like to describe the 'gauge data' associated not with a $1$-form potential, but with a $(2p-1)$-form potential, where the case  $p=2$ of interest to us corresponds to the M-theory 3-form potential $C_3$ with field strength $G_4$, the 4-form flux.
The general idea is to generalize (\ref{ses1}) to a short exact sequence
\begin{equation}
\label{eq:deligne-short-exacta}
  0 \to J^p(X) \to H^{2p}_D(X,\mathbb{Z}(p)) \xrightarrow{\hat c_p}
  H^{p,p}_\mathbb{Z}(X) \to 0.
\end{equation}
Here $H^{p,p}_\mathbb{Z}(X) = H^{2p}(X, \mathbb{Z}) \cap H^{p,p}(X, \mathbb{C})$ is the group of \emph{Hodge cycles}.
Just like the first Chern class maps the Picard group to  $H^{1,1}_\mathbb{Z}(X)$ by assigning to a line bundle its curvature, the above sequence associates to an element of $H^{2p}_D(X,\mathbb{Z}(p))$ a corresponding field strength with values in $H^{p,p}_\mathbb{Z}(X)$.
This is provided by the \emph{surjective} map\footnote{When the context admits it we will simply abbreviate $\hat c_{X,p}$ by $\hat c_p$. } 
\begin{equation}
\label{eq:deligne-hodge}
\hat c_{X,p} \colon H^{2p}_D(X,\mathbb{Z}(p)) \xrightarrow{}
H^{p,p}_\mathbb{Z}(X). 
\end{equation}
The kernel of this map is the space of flat $(2p-1)$-form connections
and mathematically given by
the \emph{intermediate Jacobian} $J^p(X)$, defined as
\begin{equation}
\label{eq:6b}
  J^p(X) =  \frac{H^{2p-1}(X,\mathbb{C})}{F^pH^{2p-1}(X,\mathbb{C})
    + H^{2p - 1}(X,\mathbb{Z})}.
\end{equation}
Here the so-called Hodge filtration $F^p H^k(X)$ 
\begin{equation}
F^p H^k(X) = \bigoplus_{p' \geq p} H^{p',k-p'} =
H^{k ,0} \oplus H^{k-1,1} \oplus \ldots \oplus H^{p,k-p}
\end{equation}
generalizes the group $H^{1,0}(X, \mathbb C)$ appearing in (\ref{J1def}).

Specializing to the case relevant for M/F-theory on a Calabi-Yau 4-fold $\hat Y_4$, integral Deligne cohomology $H^{4}_D(\hat Y_4,\mathbb{Z}(2))$ is an
extension of the group of \emph{Hodge cycles}  $H^{2,2}_\mathbb{Z}(\hat Y_4)$
by the \emph{intermediate Jacobian} $J^2(\hat Y_4)$.
By means of the surjective map $\hat c_2$
it is possible to pick for any 4-form flux $G_4 \in H^{2,2}_\mathbb{Z}(\hat Y_4)$ a refinement ${\cal A}_G \in
H^{4}_D(\hat Y_4,\mathbb{Z}(2))$ such that $$\hat c_2({\cal A}_G) = G_4 \in H^{2,2}_{\mathbb{Z}}(\hat Y_4).$$ 
By analogy with the special case $p=1$ we  interpret an element ${\cal A}_G \in
H^{4}_D(\hat Y_4,\mathbb{Z}(2))$ as fixing an equivalence class of gauge bundle data modulo gauge equivalence.

The crucial insight of Deligne was that an object that fits into the
short exact sequence (\ref{eq:deligne-short-exacta}) indeed exists and can be defined formally as the hypercohomology of 
the \emph{Deligne-Beilinson complex}
\begin{equation}
  \label{eq:15}
  \mathbb{Z}(p)_D = 0 \to \mathbb{Z} \xrightarrow{(2\pi i)^p} \mathcal{O}_X \to
  \Omega^1_X \to \cdots \Omega^{p-1}_X.
\end{equation}
Since all we need from this well-known construction for our purposes is the existence of $H^{4}_D(\hat Y_4,\mathbb{Z}(2))$ we relegate a brief account of the Deligne-Beilinson complex and its associated hypercohomology to appendix \ref{app-DeligneCohom}.

Let us stress that, as encoded in (\ref{eq:deligne-short-exacta}),  the surjective map $\hat c_p$ is not an isomorphism. Specifying a 4-form flux $G_4 \in H^{2,2}_{\mathbb Z}(\hat Y_4)$ is in general by no means sufficient to uniquely determine the gauge bundle data parametrised by $H_D^{2,2}(\hat Y_4,\mathbb{Z}(2))$ (in just the same manner as specifying the first Chern class of a line bundle does in general not specify the equivalence class of the line bundle).
For practical purposes we therefore need a method to actually define the gauge bundle data ${\cal A}_G \in H^{4}_D(\hat Y_4,\mathbb{Z}(2))$ more finely than merely by specifying its associated 4-form flux $G_4 = \hat c_2({\cal A}_G)$.
This is what we turn to now.

\subsection{Specifying Deligne cohomology via Chow groups}
\label{subsec_DelChow}

Our approach to defining an element ${\cal A}_G \in H^{4}_D(\hat Y_4,\mathbb{Z}(2))$  is to specify a suitable equivalence class of  algebraic cycles $A_G$ and to extract from this an object ${\cal A}_G$  via the so-called refined cycle map. 
Before describing this, let us first review the correspondence between algebraic cycles and $G_4$-fluxes.
Indeed, in the recent $F$-theory literature it was suggested to use algebraic
cycles of complex dimension $2$ to specify $G_4$-fluxes (see \cite{Braun:2011zm,oai:arXiv.org:1108.1794,Krause:2011xj,
oai:arXiv.org:1111.1232,oai:arXiv.org:1202.3138,Intriligator:2012ue} for early examples of $G_4$ gauge fluxes in fully-fledged 4-folds). Recall that the group of
\emph{algebraic cycles} $Z^p(X)$  of complex codimension $p$ on a variety $X$ of  complex dimension $d$
is defined to be the free group generated by irreducible subvarieties
of codimension $p$ of $X$. An element $C \in Z^p(X)$ is a finite sum
\begin{equation}
C = \sum_i n_i C_i
\end{equation}
of \emph{not necessarily smooth} irreducible subvarieties $C_i$ of
$X$, for some integers $n_i$.

To understand the connection between algebraic cycles and $G_4$-fluxes, recall furthermore that
 to any algebraic cycle in $Z^p(X)$ one can naturally
associate a cocycle in $H^{2p}(X,\mathbb{Z})$ in such a way that the
resulting \emph{cycle map}
\begin{equation}
\gamma_{X,p} \colon Z^p(X) \to H^{p,p}_{\mathbb{Z}}(X) \label{defcyclemap}
\end{equation}
is a group homomorphism, i.e. $\gamma_{X,p}(C + C') =
\gamma_{X,p}(C) + \gamma_{X,p}(C')$.\footnote{When the context is clear, we will drop the subscripts on the cycle map and denote it by $\gamma_p$ or simply $\gamma$.} 
It suffices to define the cycle map for
a possibly singular subvariety $V$ of codimension $p$ and extend it
linearly. If $V$ is smooth,  integration over $V$ gives, by Poincar\'e duality, an element
in $H^{2p}(X,\mathbb{Z})$. It is harder, but equally possible to define the
cycle map for singular subvarieties and to ensure that the image
indeed lies in $H^{p,p}(X)$, see chapter $11$ in \cite{VoisinHodgeTheoryI}.
Thus the cycle map applied to an algebraic 2-cycle $A_G \in Z^2(\hat Y_4)$ on a Calabi-Yau 4-fold $\hat Y_4$ yields
a candidate for a $G_4$-flux, since $\gamma(A_G)$ is an element of $H^{2,2}_\mathbb{Z}(\hat Y_4)$. 

As we have discussed in the preceding section, to get a handle on the gauge theoretic data in F-theory, we need to have access to more than the $G_4$-flux alone. Instead we need to specify an element in Deligne cohomology $\mathcal{A}_G \in H^{4}_D(\hat Y_4,\mathbb{Z}(2))$ of the Calabi-Yau 4-fold $\hat Y_4$. Luckily, as we will review in more detail below, for an algebraic variety $X$ it is also possible to define a \emph{refined cycle map} (see e.g. p.123 in \cite{GreenMurreVoisin})
\begin{equation}
\hat {\gamma}_{X,p} \colon Z^p(X) \to H^{2p}_D(X,\mathbb{Z}(p))
\label{eq:refined-cycle-map}
\end{equation}
to Deligne cohomology with the property that the cycle map $\gamma_{X,p} \colon Z^p(X) \to H^{p,p}_{\mathbb{Z}}(X)$ factors as $\gamma_{X,p} = \hat{c}_{X,p} \circ \hat{\gamma}_{X,p}$, where
$\hat{c}_{X,p} \colon H^{2p}_D(X,\mathbb{Z}(p)) \to H^{p,p}_{\mathbb{Z}}(X)$ is the map
defined in equation (\ref{eq:deligne-hodge}).

The refined cycle 
map $\hat{\gamma}_{X,p}$ allows us to access the refined gauge theoretic information
contained in Deligne cohomology from algebraic geometry. 
That is, whenever we are able to specify an algebraic cycle $A_G \in Z^2(\hat Y_4)$ within the 4-fold $\hat Y_4$, we can map 
it to an element ${\cal A}_G \in H^4_D(\hat Y_4, \mathbb{Z}(2))$. This map need in general not be surjective and therefore it may not give  access to all possible configurations of gauge data; nonetheless we can at least specify examples of gauge configurations in this manner.

However, specifying explicit algebraic cycles introduces a huge amount of redundancy as two different algebraic cycles may map to the same element in  ${\cal A}_G \in H^4_D(\hat Y_4, \mathbb{Z}(2))$ and therefore describe the same gauge configuration up to gauge equivalence. 
We will at least partially reduce this redundancy by specifying not an algebraic cycle, but rather a suitable equivalence class of algebraic cycles and define a map from this equivalence class to $H^{2p}_D(X,\mathbb{Z}(p))$. The equivalence relation is supposed to at least partially correspond to the gauge equivalence between gauge data. For the special case $p=1$, the 
correct equivalence relation between algebraic codimension 1-cycles which captures gauge equivalence is known to be given by rational equivalence. Even more strongly, the first Chow group  $\CH^1(X)$ of algebraic codimension 1-cycles modulo rational equivalence is isomorphic to 
$H^{2}_D(X, \mathbb{Z}(1)) = \Pic(X)$.
For $p=2$ an isomorphism between the analogous second Chow group $\CH^2(X)$ of rationally equivalent codimension 2-cycles and $H^{4}_D(X, \mathbb{Z}(2))$ is unfortunately not known to be given. Nevertheless our approach will be to define elements of $H^{4}_D(X, \mathbb{Z}(2))$ by specifying an element of the group $\CH^2(X)$. 
The refined cycle map (\ref{eq:refined-cycle-map}) indeed maps two rationally equivalent algebraic cycles to the same element in  $H^{4}_D(X, \mathbb{Z}(2))$ and therefore descends to a refined cycle map 
\begin{equation}
\hat{\gamma}_{X,p} \colon \CH^p(X) \to H^{2p}_D(X,\mathbb{Z}(p)).
\label{eq:refined-cycle-mapb}
\end{equation}
Let us repeat this important statement: We can 'move' within the rational equivalence class of algebraic cycles without changing the associated gauge data (up to gauge equivalence).  In particular, all operations such as forming intersections
which are well-defined as operations on $\CH(X)$ are suitable manipulations which do not  change the gauge data. Indeed, the intersection product on $\CH(X)$ will allow us to extract in a natural way the gauge data relevant for the counting of massless charged zero modes.

For the reader's convenience we  will now give a brief overview of some of the relevant mathematical notions. For more details on this in principle well-known material see \cite{EisenbudInt} and \cite{FultonInt}.

\subsubsection{Chow groups}
The Chow group $\CH(X)$ of a $d$-dimensional smooth projective variety $X$ has properties
similar to both the cohomology and homology groups of a topological
space. Its elements are algebraic cycles subject to a certain
equivalence relation. Working with plain algebraic cycles turns out to be inconvenient. Just
as in homology it is desirable to impose an equivalence relation on
algebraic cycles, so that they can be ``moved'' within the containing
algebraic variety.

The kernel of the cycle map $\gamma_{X,p}$ defined in (\ref{defcyclemap}) consists of algebraic cycles
that are homologous to zero. We are interested in an equivalence
relation imposed on algebraic cycles $C,C' \in Z^p(X)$ that is finer
than declaring them to be equivalent $C \sim C'$ if their image under the cycle map coincides $\gamma_{X,p}(C) =
\gamma_{X,p}(C')$, i.e. if they are homologous. Intuitively speaking complex geometry is rigid and thus algebraic cycles should not merely be homologous to be considered the same.

A good notion of equivalence for algebraic cycles is \emph{rational
  equivalence}. It is a straightforward generalization of the notion
of \emph{linear equivalence} for Weil divisors. While there is a more geometric definition of rational equivalence
available, let us first give one that makes the connection to linear
equivalence obvious. 

Given an invertible function $r \in K(W)^\ast$ on a $(k+1)$-dimensional
subvariety $W$ of $X$, it is possible to associate to it a $k$-cycle
\begin{equation}
\label{eq:divisor}
 [ \divisor(r)] = \sum \ord_V(r) [V],
\end{equation}
 where the sum runs over all codimension-one subvarieties $V$ of $W$
and $\ord_V(r)$ is the \emph{order} of $r$ in $V$. For a
function $f \in \mathcal{O}_X$ on an \emph{affine} variety $X$ the
divisor of $f$ is easy to grasp. Recall that by Krull's Hauptidealsatz
the isolated components of $f$ are all of codimension one and so the sum 
$$[\divisor(f)] = \sum_V \ord_V(f) [V]$$
runs over a finite number of irreducible codimension one
subvarieties $V$ of $X$ and $\ord_V(f)$ is simply the order of
vanishing at $V$.
This construction can be generalized beyond affine varieties.

A $k$-cycle $A \in Z^k(X)$ is called \emph{rationally equivalent} to
zero, $A \sim 0$, if it can be written as a finite sum  $A = \sum
[\divisor(r_i)]$, where $r_i \in K(W_i)^\ast$ are invertible rational
functions in some $(k+1)$-dimensional subvarieties $W_i$ of $X$.
The $k$-cyles rationally equivalent to zero form a subgroup
$\Rat_k(X).$\footnote{If $A \sim 0$ and $B \sim 0$, then also $A + B \sim
  0$. Since $[\divisor(r^{-1})] = - [\divisor(r)]$, $A 
\sim 0$ implies
  $-A \sim 0$.}
The \emph{Chow group} $\CH_k(X)$ is the group of rational equivalence classes
\begin{equation}
  \label{eq:chowgrp-def}
  \CH_k(X) = Z_k(X) / \Rat_k(X).
\end{equation}
It can be graded by either dimension or codimension. The grading by
codimension will be denoted by a superscript $\CH^p(X)$, the grading by
dimension by a subscript $\CH_m(X)$.

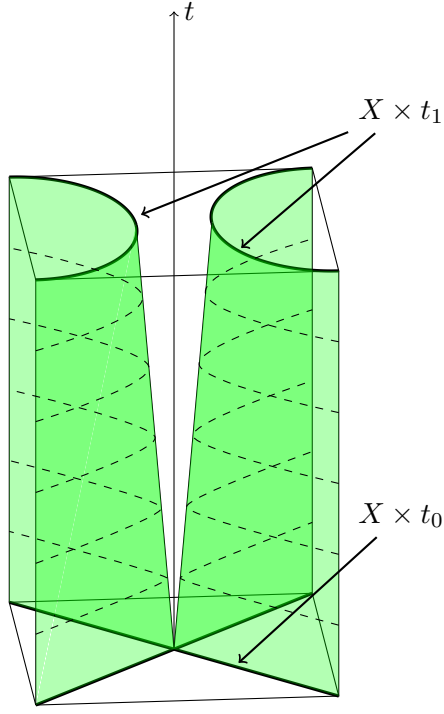
\begin{figure}[tb]
\begin{center}

\tdplotsetmaincoords{70}{85}
\tdplotsetrotatedcoords{0}{0}{0}%

\begin{tikzpicture}[scale=0.5,tdplot_main_coords];

\pgfmathsetmacro{\w}{8} 
\pgfmathsetmacro{\d}{8} 
\pgfmathsetmacro{\h}{12} 

\draw (\w/2,-\d/2,0)--(\w/2,\d/2,0)--(- \w/2, \d/2,0)--(- \w/2,- \d/2,0)--(\w/2,-\d/2,0);
\draw (\w/2,- \d/2,\h)--(\w/2,\d/2,\h)--(-\w/2,\d/2,\h)--(-\w/2,-\d/2,\h)--(\w/2,-\d/2,\h);
\draw (\w/2,-\d/2,\h)--(\w/2,-\d/2,0);
\draw (-\w/2,-\d/2,\h)--(-\w/2,-\d/2,0);
\draw (-\w/2,\d/2,\h)--(-\w/2,\d/2,0);
\draw (\w/2,\d/2,\h)--(\w/2,\d/2,0);


\draw[->] (0,0,0)--(0,0,1.5*\h) node [right] {$t$};

\draw[domain=180:270,smooth,variable=\t,black,very thick]
      plot ( {\d/2*cos(\t)}, {\w/2+(\w/2 -1)*sin(\t)}, {\h} );
\draw[domain=270:360,smooth,variable=\t,black,very thick]
      plot ( {\d/2*cos(\t)}, {\w/2+(\w/2 -1)*sin(\t)}, {\h} );
\draw[domain=0:90,smooth,variable=\t,black,very thick]
      plot ( {\d/2*cos(\t)}, {-\w/2+(\w/2 -1)*sin(\t)}, {\h} );
\draw[domain=90:180,smooth,variable=\t,black,very thick]
      plot ( {\d/2*cos(\t)}, {-\w/2+(\w/2 -1)*sin(\t)}, {\h} );

\draw (0,0,0)--(0,-1,\h);
\draw (0,0,0)--(0,1,\h);

\draw[very thick, black] (\w/2,-\d/2,0)--(-\w/2,\d/2,0);
\draw[very thick, black] (\w/2,\d/2,0)--(-\w/2,-\d/2,0);


\fill[green,opacity=.30]
(\w/2,-\d/2,0)
--(0,0,0)
--(0,-1,\h)
plot[smooth, samples=100, domain=0:90] ( {\d/2*cos(\x)}, {-\w/2+(\w/2 -1)*sin(\x)}, {\h})
--(\w/2,-\d/2,0);

\fill[green,opacity=.30]
(-\w/2,-\d/2,0)
--(0,0,0)
--(0,-1,\h)
plot[smooth, samples=100, domain=90:180] ( {\d/2*cos(\x)}, {-\w/2 + (\w/2 -1)*sin(\x)}, {\h})
--(-\w/2,-\d/2,0);


\fill[green,opacity=.30]
(\w/2,\d/2,0)
--(0,0,0)
--(0,1,\h)
plot[smooth, samples=100, domain=-90:0] ( {\d/2*cos(\x)}, {\w/2+(\w/2-1)*sin(\x)}, {\h})
--(\w/2,\d/2,0);

\fill[green,opacity=.30]
(-\w/2,\d/2,0)
--(0,0,0)
--(0,1,\h)
plot[smooth, samples=100, domain=-90:-180] ( {\d/2*cos(\x)}, {\w/2+(\w/2-1)*sin(\x)}, {\h})
--(-\w/2,\d/2,0);

\pgfmathsetmacro{\newh}{\h-1};
\foreach \number in {2, 4,..., \newh}{
			\pgfmathsetmacro{\tension}{\number / \h}
			\draw[black,dashed] plot [smooth, tension=\tension] coordinates {(\w/2,-\d/2,\number) (0,-\number/\h,\number) (-\w/2,-\d/2,\number)};
			\draw[black,dashed] plot [smooth, tension=\tension] coordinates {(\w/2,\d/2,\number) (0,\number/\h,\number) (-\w/2,\d/2,\number)};
}

\node (A) at (0,3*\d/4,1.25*\h) {$X \times t_1$};
\node (B) at (0,3*\d/4,0.3*\h) {$X \times t_0$};

\draw[->,thick] (A)--(0.25*\w, 0.2*\d,\h);
\draw[->,thick] (A)--(-0.1*\w, -0.1*\d,\h);
\draw[->,thick] (B)--(0.15*\w, 0.2*\d,0);

\end{tikzpicture}
\end{center}
\caption{Rational equivalence between the union of two lines in $\mathbb{CP}^2$ and a hyperbola. This picture is inspired by \cite{EisenbudInt}.}
\label{figure2001}
\end{figure}

An equivalent (\cite{FultonInt}, §1.6, \cite{EisenbudInt}, §1.1.2) and more geometric definition of rational equivalence is as follows: Two algebraic cycles $C, C' \in Z(X)$ are rationally equivalent if there is a rationally parameterized family of cycles interpolating between them, in other words if there is a cycle on $\mathbb{P}^1 \times X$ whose restriction to two fibers $\{t\} \times X$ and $\{t'\} \times X$ are $C$ and $C'$. We picture the rational equivalence between a hyperbola and the union of two lines in $\mathbb{CP}^2$ in \autoref{figure2001}.
This intuition can be made precise by defining in analogy to
cohomology a ``boundary map'' $\partial_X \colon Z(X \times
\mathbb{P}^1) \to Z(X)$: Let $V$ be an irreducible subvariety of $X
\times \mathbb{P}^1$, set $\partial_X(V) = 0$, if the projection $\pi
\colon V \to \mathbb{P}^1$ is not dominant\footnote{Recall that a
  rational map is dominant if there exists an equivalent map whose
  image is dense in its codomain.} and set $\partial_X(V) = V_0 - V_\infty$ otherwise,
where $V_p = \pi^{-1}(p) \subset X \times \{p\} \cong X$ are the
fibers over points $p$ in $\mathbb{P}^1$. Denote the subgroup of
$Z(X)$ generated by cycles of the form $V_0 - V_\infty$ by $\Rat(X)$
and the subgroup generated by codimension $p$ cycles by $\Rat^p(X)$.

The \emph{Chow group} $\CH(X)$ can then equivalently defined to be the group of rational equivalence classes
\begin{equation}
  \CH(X) = Z(X) / \Rat(X) = \coker(\partial_X \colon Z(X \times \mathbb{P}^1) \to Z(X)).
\end{equation}
With these definitions in place it can be demonstrated that the cycle map $\gamma_{X,p} \colon Z^p(X) \to H^{2p}(X,\mathbb{C})$ descends to a map $\CH^p(X) \to H^{2p}(X,\mathbb{C})$ and rational equivalence classes of
algebraic cycles that are homologous to zero, i.e. in the kernel of
$\gamma_{X,p}$, form a subgroup $\CH^p_{\text{hom}}(X) \subset \CH^p(X)$.

\subsubsection{Connection between Deligne Cohomology and Chow Groups} \label{sub_connDelChow}
As we have seen above, the first Chow group of $\CH^1(X)$ of a smooth projective variety $X$ and the Picard group $\Pic(X)$ of $X$ are isomorphic. Moreover the cycle map $\gamma_{1} \colon \CH^1(X) \to H^{1,1}_\mathbb{Z}(X)$ factors through
this isomorphism via the map $c_1 \colon \Pic(X) \to
H^{1,1}_\mathbb{Z}(X)$ that associates to each isomorphism class of
line bundles the first Chern class. The kernel $\CH^1_{\text{hom}}(X)$ of
the cycle map, i.e. the homologically trivial
cycles, are mapped isomorphically to the Jacobian $J^1(X)$ of $X$ via
the Abel-Jacobi map. To summarize one has the following commuting
diagram:
\[
\begin{tikzcd}
  0 \arrow{r} & \CH^1_{\text{hom}}(X) \arrow{r} \arrow{d}{AJ} & \CH^1(X)
  \arrow{r}{\gamma_{1}}  \arrow{d}{} & H^{1,1}_{alg}(X)
  \arrow{r} \arrow{d} & 0 \\
  0 \arrow{r} & J^1(X) \arrow{r} & \Pic(X)
  \arrow{r}{c_1}  & H^{1,1}_{\mathbb{Z}}(X) \arrow{r} & 0
\end{tikzcd}
\]

Now in section \ref{ssec:GaugeDataDeligne} we asserted that the lower horizontal sequence
$0 \to  J^1(X) \to \Pic(X)
 \to H^{1,1}_{\mathbb{Z}}(X)\to 0$
is a special case of the short exact sequence
\begin{equation}
  0 \to J^p(X) \to H^{2p}_D(X,\mathbb{Z}(p)) \xrightarrow{\hat c_p}
  H^{p,p}_\mathbb{Z}(X) \to 0.
\end{equation}
It is therefore natural to ask whether there is an analogue of the whole
diagram. We already mentioned that there is a cycle map $\gamma_p \colon
\CH^p(X) \to H^{p,p}_\mathbb{Z}(X)$ and it is also possible to define a
generalized Abel-Jacobi map $AJ \colon \CH^p_{\text{hom}}(X) \to J^p(X)$. So what we are looking
for is a refinement $\hat{\gamma}_p \colon \CH^p(X) \to
H^{2p}_D(X,\mathbb{Z}(p))$ of the cycle map to Deligne cohomology
such that the following diagram commutes:

\[
\begin{tikzcd}
  0 \arrow{r} & \CH^p_{\text{hom}}(X) \arrow{r} \arrow{d}{AJ} & \CH^p(X)
  \arrow{r}{\gamma_p}  \arrow{d}{\hat{\gamma}_p} & H^{p,p}_{alg}(X)
  \arrow{r} \arrow{d} & 0 \\
  0 \arrow{r} & J^p(X) \arrow{r} & H^{2p}_D(X,\mathbb{Z}(p))
  \arrow{r}{\hat{c}_p}  & H^{p,p}_{\mathbb{Z}}(X) \arrow{r} & 0
\end{tikzcd}
\]

Such a refined cycle map indeed exists (see \cite{EsnaultDeligne} \S $7$).
For the subsequent analysis, however, only its existence is of relevance and in particular the fact that rationally equivalent algebraic cycles map to the same Deligne cohomology class under $\hat \gamma_p$.

The existence of the refined cycle map $\hat \gamma_p$ for $p=2$ therefore allows us to define an element of 
the Deligne cohomology group $H^4_D(\hat Y_4, \mathbb{Z}(2))$ on our Calabi-Yau 4-fold $\hat Y_4$ by specifying a Chow class $\alpha_G \in \CH^2(\hat Y_4)$. Thus, given a complex 2-cycle $A_G \in Z^2(\hat Y_4)$ whose rational equivalence class is $\alpha_G$, the object $\hat \gamma_{2}(\alpha_G) \in  H^4_D(\hat Y_4, \mathbb{Z}(2))$ defines a certain type of F/M-theory gauge data in the sense described in section \ref{ssec:GaugeDataDeligne}. The $G_4$ flux associated with this gauge data is the element 
\bea
G_4 = \gamma_2(A_G) = \hat{c}_2 \circ \hat{\gamma}_2 (\alpha_G) \in H^{2,2}_{\mathbb Z}(\hat Y_4).
\eea
We stress, however, that the specification of $\alpha_G$ via $A_G$ contains much more information than merely the $G_4$-flux. This construction is the direct generalization of the specification of an equivalence class of line bundles as an element of $\Pic(X)$ by specifying a divisor class $\CH^1(X)$. While $\Pic(X)$ and $\CH^1(X)$ are isomorphic, $\CH^2(X)$ and $H^4_D(X, \mathbb{Z}(2))$ need in general not be isomorphic. 

In practice this is not a serious drawback. Practically all $G_4$ gauge fluxes defined in the recent $F$-theory literature are in fact in the image of the cycle map $\gamma \colon \CH^2(\hat{Y}_4) \to H^{2,2}_\mathbb{Z}(X)$. Whenever this is the case the refined cycle map $\gamma \colon \CH^2(\hat{Y}_4) \to H^{4}_D(X,\mathbb{Z}(2))$ defines a compatible Deligne cohomology class. Moreover the Hodge conjecture states that if we allow algebraic cycles with rational coefficients, the cycle map will in fact be surjective. On the other hand counter examples for the case of integral coefficients are known. Since $\hat{c}_2$ is surjective, showing surjectivity of the refined cycle map for rational coefficients is indeed a one-million-dollar problem. The refined cycle map will also not be injective in general, that is distinct equivalence classes of algebraic cycles might map to the same Deligne cohomology class. This just means that we might not detect all equivalent gauge data in the Chow group. What is important for us is that since the refined cycle map is well-defined we will get the same Deligne cohomology class irrespective of the representative algebraic cycle we pick.

To reiterate, by the refined cycle map to every rational equivalence class of algebraic 2-cycles there exists an equivalence class of gauge data as an element in $H^4_D(\hat Y_4, \mathbb{Z}(2))$, and  manipulations of representative cycles $A_G$ which do not change their rational equivalence class $\alpha_G$ do not change the associated element in $H^4_D(\hat Y_4, \mathbb{Z}(2))$.

So far we have ignored the fact that $G_4$ is really an element of $H^{2,2}_{\frac{1}{2}{\mathbb Z}}(\hat Y_4)$ due to the quantization condition (\ref{FW2}). This shift is taken into account by considering the  Deligne cohomology group 
$H^4_D(\hat Y_4, \frac{1}{2}\mathbb{Z}(2))$ with half-integer coefficients.
Similarly, via the refined cycle map one can specify  an element therein via an element in the second Chow group with suitably half-integer as opposed to integer coefficients.
For the associated flux $G_4 = \gamma(\alpha_G)$ to be quantised according to  (\ref{FW2}), the Chow group element must satisfy the condition $\gamma(\alpha_G) + \frac{1}{2} c_2(\hat Y_4) \in H^{2,2}_{\mathbb Z}(\hat Y_4)$.
Note that the second Chern class of the tangent bundle to $\hat Y_4$ defines an element of ${\CH}^2(\hat Y_4)$ \cite{EisenbudInt}, which by abuse of notation we also denote by $c_2(\hat Y_4)$. Then the qunatisation condition is fulfilled if we pick $\alpha_G + \frac{1}{2} c_2(\hat Y_4)$  to be an element of the integer second Chow group.
With the understanding that the coefficients of the Chow and Deligne cohomology groups must be chosen appropriately we will not make the half-integer shift explicit in what follows and continue to talk about $H^4_D(\hat Y_4, \mathbb{Z}(2))$.



\subsection{Properties of Chow groups} \label{subsec_PropChow}

In this section we  describe
operations on Chow groups with the understanding
that these operations could also be directly expressed in Deligne cohomology as a result of the existence of the refined cycle map. The
advantage of specifying the gauge data via Chow groups is that Chow groups are, at least in
principle, suitable for explicit computation and accessible to methods of
algebraic geometry.
This will turn out useful when it comes to
explicitly extracting the data needed to compute the charged massless matter spectrum of an F-theory compactification. As we will see, we will have to make heavy use of intersection theory within $\CH(X)$. In this section we review some of the key elements of this well-developed subject \cite{FultonInt,EisenbudInt} which will be of direct relevance for section \ref{CohomFormSec}.
The reader familiar with this material may wish to directly jump to section \ref{CohomFormSec}.

\subsubsection{Functoriality}
In order to work with Chow groups it is necessary to understand their
behaviour under morphisms between varieties $f \colon X \to Y$. Again it
is helpful to compare the situation to cohomology and homology.
Homology is a covariant functor from topological spaces to
abelian groups, in other words there is a pushforward map
\begin{equation}
f_\ast \colon H_m(X) \to H_m(Y)
\end{equation}
associated to any morphism $f \colon X \to Y$.
We would expect that by analogy there is a pushforward map $f_\ast
\colon \CH_m(X) \to \CH_m(Y)$ preserving dimension. If $f$ has
\emph{relative dimension} $e$, then $f_\ast$
maps $\CH^p(X) \to \CH^{p-e}(Y)$.

Similarly cohomology is a contravariant functor, i.e. there exists a
pullback map
\begin{equation}
f^\ast \colon H^m(Y) \to H^m(X)
\end{equation}
associated to any morphism $f \colon X \to Y$. If we think of Chow
groups as analogous to cohomology groups, we would think that there is
a pullback map
\begin{equation}
f^\ast \colon \CH^p(Y) \to \CH^p(X)
\end{equation}
preserving codimension.
It turns out that both expectations are justified, with some
restrictions, as we will see
shortly.

The pushforward between Chow groups $f_\ast \colon \CH_p(X) \to \CH_p(Y)$ is
defined if $f \colon X \to Y$ is a proper\footnote{Since we will be
  only concerned with complex algebraic varieties, \emph{proper} can
  be understood here in the sense of the complex topology on $X$ and
  $Y$, i.e. the map $\pi \colon X \to Y$ is proper if the preimage of a
  compact set is compact. In algebraic geometry a morphism $\pi \colon
X \to Y$ is called proper if it is universally closed and separated.} morphism between algebraic
varieties: Let $V$ be a subvariety of $X$, then $f(V)$ is a subvariety
of $Y$ of dimension $\dim(f(V)) \leq \dim(V)$. A guess would be that
the cycle class of $V
$ in $\CH(X)$ should
be sent to the cycle class of $f(V)$  in $\CH(Y)$. However this would not
respect rational equivalence as it does not account for multiplicities
in multi-sheeted covers for example.

\begin{figure}[tb]
\begin{center}
\begin{tikzpicture}[dot/.style={circle,inner sep=1pt,fill,label={#1},name=#1},
  extended line/.style={shorten >=-#1,shorten <=-#1},
  extended line/.default=1cm]

\draw[thick] (0,0)--(12,0);

\node (Start) at (0,3.7) {};

\node (a) at (2.5,1.95) {};
\node (b) at (3,2.5) {};
\node (c) at (4,3.6) {};
\node (d) at (7,3.5) {};
\node (e) at (7,3) {};
\node (f) at (7,1.8) {};
\node (g) at (9,3.25) {};
\node (h) at (9.5,1.75) {};
\node (End) at (12,1.7) {};

\draw [black,thick,blue] plot [smooth, tension=0.6] coordinates {(Start) (c) (d) (g) (e) (b) (a) (f) (h) (End)};

\fill[black] (a) circle (2pt) node [above] {$a$};
\fill[black] (b) circle (2pt) node [above] {$b$};
\fill[black] (c) circle (2pt) node [above] {$c$};
\fill[black] (d) circle (2pt) node [above left] {$d$};
\fill[black] (e) circle (2pt) node [above left] {$e$};
\fill[black] (f) circle (2pt) node [above left] {$f$};
\fill[black] (g) circle (2pt) node [above right] {$g$};
\fill[black] (h) circle (2pt) node [above right] {$h$};

\draw[thick, extended line =0.5cm] (c)--(a);
\draw[thick, extended line = 0.5cm] (d)--(f);
\draw[thick, extended line = 0.5cm] (g)--(h);

\node (label1) [below] at (0,3.7) {$\mathcal{M}$};
\node (label2) [above] at (0,0) {$\mathcal{M}^\prime$};

\draw[dashed] (2.5,1.95)--(2.5,0);
\draw[dashed] (3,2.5)--(3,0);
\draw[dashed] (4,3.6)--(4,0);
\draw[dashed] (7,3.5)--(7,0);
\draw[dashed] (9,3.25)--(9,0);
\draw[dashed] (9.5,1.75)--(9.5,0);

\fill[black] (2.5,0) circle (2pt) node [below] {$a^\prime$};
\fill[black] (3,0) circle (2pt) node [below] {$b^\prime$};
\fill[black] (4,0) circle (2pt) node [below] {$c^\prime$};
\fill[black] (7,0) circle (2pt) node [below] {$d^\prime$};
\fill[black] (9,0) circle (2pt) node [below] {$g^\prime$};
\fill[black] (9.5,0) circle (2pt) node [below] {$h^\prime$};

\draw[->,very thick] (5.5,1.5)--(5.5,0.25);

\end{tikzpicture}
\end{center}
\caption{If $a + b + c \sim d +e+f \sim 2g+h$, then the pushforwards are equivalent cycles, i.e. $a^\prime + b^\prime + c^\prime \sim 3 d^\prime \sim 2g^\prime + h^\prime$. This picture is based on \cite{EisenbudInt}.}
\label{figure2000}
\end{figure}
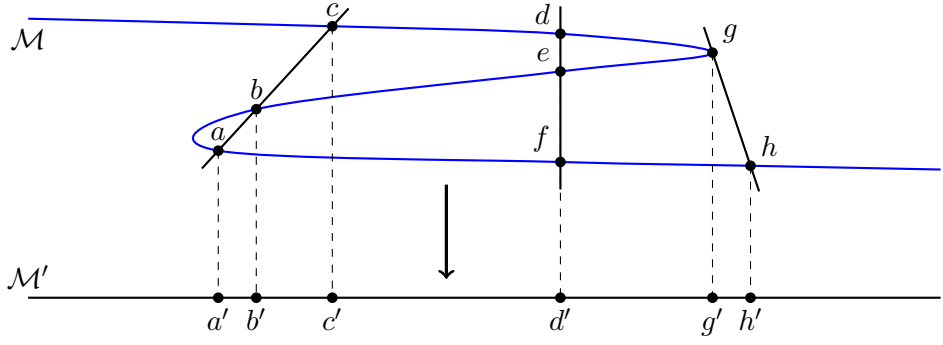

Observe that as long as $\dim(V) = \dim(f(V))$ the map $f\lvert_V \colon V \to f(V)$ is
\emph{generically finite} in the sense that the extension of function
fields $K(V)/K(f(V))$ is of finite degree $n$. Geometrically this
means that generically there are $n$ points in the fibre over a point
in the base $f(V)$. The geometric interpretation suggests taking
the degree of the covering into account. Define $f_\ast(V) = 0$
if the dimension of $f(V)$ is strictly smaller than the dimension of
$V$ and  $f_\ast V = n f(V)$ if $f \lvert_V$ has degree $n$. This
definition is extended linearly to all cycles on $X$ by defining
\begin{equation}
f_\ast( \sum m_i V_i ) = \sum m_i f_\ast  V_i .
\end{equation}
It turns out that this definition is compatible with rational
equivalence (\cite{FultonInt}, Chapter 1) so we get a pushforward map
\begin{equation}
f_\ast \colon \CH_m(X) \to \CH_m(Y)
\end{equation}
which is additive,
\begin{equation}
  f_\ast(\alpha + \beta) = f_\ast(\alpha) + f_\ast(\beta).
\end{equation}
We give a picture of this situation in \autoref{figure2000}.

A good pullback map $f^\ast\colon \CH(Y) \to \CH(X)$ defined on cycles has to preserve algebraic
equivalence and should be geometric in the sense that if, for a
subvariety $V$ of $Y$ of codimension $p$, $f^{-1}(V)$ is \emph{generically reduced} of
codimension $p$, then the pullback of the algebraic cycle $[V]$ should
be the class of the inverse image $f^{-1}(V)$: $f^\ast[V] = [f^{-1}(V)]$. This 
requirement turns out to determine $f^\ast$ uniquely, at least as long
as $f \colon X \to Y$ is a \emph{generically separable} map between
smooth algebraic varieties. Indeed, if $f \colon X \to Y$ is such a
map, any cycle $\alpha \in
\CH^p(Y)$ can be represented by an algebraic cycle $A = \sum_i n_i A_i
\in Z^p(Y)$, such that $f^{-1}(A_i)$ is generically reduced of
codimension $p$ for all $i$ and the class $\sum n_i [f^{-1}(A_i)]$ in
$\CH^p(X)$ is independent of the choice of $A$.

Define the pullback $f^\ast(\alpha) = \sum_i n_i [f^{-1}(A_i)]$. It is
then possible to show that the pullback is additive, i.e.
\begin{equation}
  f^\ast(\alpha + \beta) = f^\ast(\alpha) + f^\ast(\beta).
\end{equation}

\subsubsection{Intersection product} \label{subsec_Intersectio}
One of the main motivations for working with Chow groups is to study
the intersection of irreducible subvarieties $A,B$ in $X$. It should
be defined in such a way that it is a refinement of the \emph{cup
  product}  $\cup \colon H^i(X) \otimes
H^j(X) \to H^{i+j}(X)$ in cohomology. In other words the intersection
product should equip $\CH^p(X)$ with a ring structure and the cycle map
$\gamma_X \colon \colon \CH^p(X) \to H^{2p}(X)$ should be a ring
homomorphism, i.e.
\begin{equation}
\gamma_X(\alpha \cdot \beta) = \gamma_X(\alpha) \cup \gamma_X(\beta)
\end{equation}
for cycles $\alpha, \beta \in \CH(X)$.

There are different approaches to defining such an intersection
product. What follows is a brief overview of the approach presented in
Fulton's book on the subject \cite{FultonInt}. The general idea is to
mimic the definition of the cup-product in cohomology. Recall
(see for example \cite{BredonAT}, Chapter IV) that it can be defined to
be induced by the composition
\begin{equation}
C^\bullet(X) \times C^\bullet(X) \to C^\bullet(X \times X) \xrightarrow{\Delta^\ast} C^\bullet(X)\label{eq:12}
\end{equation}
on cochain complexes $C^\bullet(X)$, where the first morphism is the K\"unneth map and the
second is the pullback induced by the diagonal map $\Delta \colon X \to X \times
X$. The role of the K\"unneth map is taken on by the exterior product
$\times \colon Z_k(X) \otimes Z_l(Y) \to Z_{k+l}(X \times Y)$ of
algebraic cycles that maps $A = \sum n_i A_i \in Z_k(X)$ and $B = \sum
m_i B_i \in Z_l(Y)$ to $A
\times B = \sum n_i m_j (A_i \times B_j) \in Z_{k + l}(X \times
X)$, where $A_i
\times B_j$ is the Cartesian product of the irreducible subvarieties
$A_i$ and $B_j$. This product descends to
a well defined exterior product $\times \colon \CH_k(X) \otimes \CH_l(Y)
\to \CH_{k + l}(X \times Y)$, see \cite{FultonInt} Chapter 1. 
The definition of an analogue of pullback to the diagonal $C^\bullet(X \times X)
\to C^\bullet(X)$ will occupy us for the rest of this section. Notice
that if $X$ is a smooth algebraic variety the diagonal embedding
$\Delta \colon X \to X \times X$ is a regular embedding. So what we are
looking for is to define a morphism $f^\ast \colon \CH(Y) \to \CH(X)$ for any \emph{regular embedding} $f \colon X \to Y$.

We will construct such a \emph{Gysin homomorphism} $f^\ast \colon \CH(Y) \to \CH(X)$ in several steps, beginning with the case of a vector bundle $\pi \colon E \to X$ of rank $r$ on $X$. The flat pullback map $\pi^\ast \colon \CH_{k - r}(X) \to \CH_{k}(E)$ is an isomorphism for all $k$ (\cite{FultonInt}, Chapter 3). So if $s \colon X \to E$ is the zero section of $E$, in particular $\pi \circ s = \id_X$, then it is possible to define the \emph{Gysin homomorphism}
\begin{equation*}
  s^\ast \colon \CH_k(E) \to \CH_{k-r}(X),
\end{equation*}
where $s^\ast$ is the inverse of the pullback map $\pi^\ast$
\begin{equation}
  s^\ast(\alpha) = (\pi^\ast)^{-1}(\alpha).
\end{equation}

A closed embedding $i \colon X \to Y$ of varieties is a \emph{regular
  embedding of codimension} $d$ if the ideal sheaf $\mathcal{I}$
defining $X$ as a subvariety of $Y$ is locally generated by a regular
sequence of length $d$.\footnote{A sequence $a_1, \ldots,
  a_d \in A$  in a ring $A$ is called regular if $a_i$ is a non-zero
  divisor in $A / (a_1, \ldots, a_{i-1})$} Geometrically this
means that $X$ is a \emph{local complete intersection} in
$Y$. Relevant examples include the diagonal embedding $\delta
\colon X \to
X \times X$ if $X$ is smooth and for any morphism $f \colon X
\to Y$ to a non-singular variety $Y$ the induced regular graph
morphism $\gamma_f \colon X \to X \times Y$ given by $x \mapsto (x,
f(x))$.
If $i \colon X \to Y$ is a regular embedding of codimension $d$, the conormal sheaf $\mathcal{I}/\mathcal{I}^2$ is a locally free sheaf of rank $d$ and the \emph{normal bundle} $N_XY$ to $X$ in $Y$ is the vector bundle whose sheaf of sections is dual to $\mathcal{I}/\mathcal{I}^2$.

The normal cone $C_XY$ of a closed subscheme $X$ in $Y$ is the
spectrum of the graded ideal\footnote{Take for example the ideal $I = (xy)$ in $R = \mathbb{C}[x,y]$, $V = Spec(R/I)$ is the union of the coordinate axes. The normal cone is $V \times \mathbb{A}^1$.} $\sum_{n \geq 0} \mathcal{I}^{n}/
\mathcal{I}^{n+1}$. If $i \colon X \to Y$ is a regular embedding,
then the normal cone $C_XY$ is naturally isomorphic to the normal
bundle $N_XY$. Intuitively the normal cone $C_X Y$ is a substitute for
the notion of ``tubular neighbourhood'' from differential geometry.

In particular we can associate to any algebraic cycle on $Y$ an
algebraic cycle on $C_X Y$
\begin{equation}
  \sigma \colon Z_k(Y) \to Z_k(C_X Y)
\end{equation}
by the formula
\begin{equation}
  \sigma [V] = [C_{V \cap X} V] 
\end{equation}
for any $k$-dimensional subvariety $V$ of $Y$ and to all $k$-cycles by
linearity (\cite{FultonInt}, Chapter 5.2). This
is well defined since $C_{V \cap X} V$ is a subvariety of pure
dimension $k$ (\cite{FultonInt} loc. cit.). 

The \emph{specialization homomorphism} $\sigma$ descends to the level of Chow groups (\cite{FultonInt}, loc. cit.):
\begin{equation}
  \sigma \colon \CH_k(Y) \to \CH_k(C_{X}Y).
\end{equation}

Let $i \colon X \to Y$ be a regular embedding of codimension $d$, with
normal bundle $N = N_XY$. Then the \emph{Gysin homomorphism} $i^\ast
\colon \CH_k(Y) \to \CH_{k-d}(X)$ is defined to be the composition $i^\ast =
s^\ast_{N_X Y} \circ \sigma$,
\begin{equation*}
  \CH_k(Y) \xrightarrow{\sigma} \CH_{k}(N_X Y) \xrightarrow{s^\ast_{N}} \CH_{k-d}(X),
\end{equation*}
where $s_N \colon X \to N_{X}Y$ denotes the zero section of the normal
bundle, $s^\ast_N$ is the induced isomorphism on Chow groups and
$\sigma$ is the specialization homomorphism defined above. 

The \emph{intersection product} is then the composition of the exterior
product with the Gysin homomorphism $\Delta^\ast$ induced by the
diagonal embedding $\Delta \colon X \to X \times X$,
\begin{equation*}
\CH_k(X) \otimes \CH_l(X) \xrightarrow{\times} \CH_{k + l}(X \times X) \xrightarrow{\Delta^\ast} \CH_{k + l - n}(X).
\end{equation*}
If $\alpha$ and $\beta$ are cycles in $\CH_k(X)$ and $\CH_l(X)$
respectively, their intersection product in $\CH_{k + l}(X)$ will be denoted by $\alpha
\cdot \beta$.

This notion of intersection product generalizes as follows. 
Suppose that $f \colon X \to Y$ is a morphism to a
smooth variety $Y$ of dimension $n$. Then the graph morphism $\gamma_f \colon X
\to X \times Y$ given by $x \mapsto (x, f(x))$ is a regular
embedding of codimension $n$. Notice that if $f$ is the identity morphism, the graph morphism is just the diagonal embedding above. Define the intersection product by the
composition of the exterior product with the Gysin morphism
$\gamma_f^{\ast}$, 
\begin{equation} \label{intdefseq}
  \CH_k(X) \otimes \CH_l(Y) \xrightarrow{\times} \CH_{k + l}(X \times Y)
  \xrightarrow{\gamma^\ast_f} \CH_{k + l - n}(X),
\end{equation}
and denote the product of two cycles $\alpha$ and $\beta$ by 
\bea \label{cdotf}
\alpha \cdot_f \beta.
\eea
Since we will make have use of these intersection products in the
sequel, we state some of their properties (for proofs see
Chapter $8$ of \cite{FultonInt}). For $\alpha = [X]$, denote $[X]
\cdot_{f} \beta = f^\ast \beta$. This definition generalizes the Gysin
morphism defined above.

The intersection product is associative, i.e. for $X \xrightarrow{f} Y \xrightarrow{g} Z$, with $Y$ and $Z$
non-singular varieties, we have
\begin{equation}
  x \cdot_f (y \cdot_g z) =  (x \cdot_f y) \cdot_{gf} z,
\end{equation}
with $x \in \CH(X),\, y\in \CH(Y)$ and $z \in \CH(Z)$. In particular,
for $y = [Y]$, we have
\begin{equation} \label{composite-int}
  x \cdot_f g^\ast z =  x \cdot_{gf} z,
\end{equation}
since $x \cdot_f [Y] = x.$
Finally, we will need the \emph{projection formula}
\begin{equation}
  \label{eq:projection-formula}
  f_{\ast}(x \cdot_{gf} z) = f_\ast(x) \cdot_g z.
\end{equation}
In particular, if we are given a fibre square
\begin{equation}
  \label{eq:projection-formula-fibre-square}
\begin{tikzcd}
X \times_{Z} Y \arrow{r}{p} \arrow{d}{q} & X \arrow{d}{f} \\
Y \arrow{r}{g}                           & Z
\end{tikzcd}  
\end{equation}
and algebraic cycles $\alpha \in \CH(X \times_{Z} Y)$ and $\beta \in \CH(Z)$, then
\begin{equation}
  \label{eq:fibre-square-projection-formula}
  q_\ast(\alpha \cdot_{fp} \beta) = q_\ast(\alpha \cdot_{gq} \beta) = q_\ast(\alpha) \cdot_g \beta .
\end{equation}

\section{Cohomology formulae for F-theory compactifications} \label{CohomFormSec}

\subsection{A general cohomology formula} \label{CohomForm}

Let us recap the main results of the previous sections:
To define the gauge bundle data of an F-theory compactification on a resolved Calabi-Yau 4-fold $\hat Y_4$ we propose to specify an element 
\bea
\alpha_G \in \CH^2(\hat Y_4)
\eea
of the rational equivalence class of complex-dimension 2-cycles on $\hat Y_4$.
With the help of the refined cycle map from ${\rm CH}^2({\hat Y_4})$  to $H^4_{D}(\hat Y_4, \mathbb Z(2))$, this yields a specific type of 3-form data up to gauge transformations. 
A representative of the equivalence class $\alpha_G$ in turn is specified by an explicit 2-cycle
\bea
A_G \in Z^2(\hat Y_4)
\eea
whose associated Dolbeault cohomology class 
\bea
G_4 \equiv \gamma(A_G) \in H^{(2,2)}_{\mathbb{Z}}(\hat Y_4)
\eea
 is typically referred to as $G_4$-flux in the F-theory literature.


Our main interest is in the fibre structure over the curves $C_R$ in the base $B_3$ where massless matter in representation $R$ of some gauge group is localised. As briefly reviewed at the beginning of section \ref{ssec:GaugeDataDeligne}, such matter curves correspond to loci where
two components of the discriminant of the elliptic fibration intersect, or where a discriminant component self-intersects.
The latter includes in particular configurations 
 without any non-abelian gauge groups, where an essentially local approach to the gauge bundle data is not applicable. Indeed an important advantage of our approach is that it provides a way how to extract the massless matter from the gauge bundle data also in such situations.

To each matter representation $R$ one can associate a so-called matter surface given by a complex 2-cycle
\bea
A_R = \sum_i n_{R,i} A_R^i \in Z^2(\hat Y_4),
\eea
 where each $A_R^i$ is given by a $\mathbb P^1$-fibration over the curve $C_R$.\footnote{More precisely, to each of the dim$(R)$ weights  $\beta^k$, $k=1, \ldots, {\rm dim}R$, of the representation $R$ one associates a different matter surface $A_{R,k}$. Since nothing depends on which $A_{R,k}$ is used to compute the matter multiplicities, we  simply make a convenient choice and suppress the index $k$ in the sequel.}  
The support of $A_R$ is the union of the support of the individual $A_R^i$,
\bea
|A_R| = \bigcup_i |A_R^i|.
\eea
It is fibred over the matter curve in the base via
\bea
\pi_R: |A_R| \rightarrow C_R.
\eea 
We will denote by
$\alpha_R \in \CH^2(\hat Y_4)$
the rational equivalence class with representative $A_R$. 

In the sequel it will be useful to view $A_R$ as a cycle on the restriction of the elliptic fibration 
to the curve $C_R$ with projection
\bea
\pi|_{C_R}: \hat Y_4|_{C_{R}} \rightarrow C_R. 
\eea
 The preimage of a generic point on $C_R$ with respect to $\pi|_{C_R}$ is given by the union of a number $\mathbb P^1$s with multiplicities.
Note that strictly speaking $\hat Y_4|_{C_{R}}$ is singular.
Nonetheless we can view the Chow class of $A_R$ as an element of ${\rm CH}_2(\hat Y_4|_{C_{R}})$.

This structure defines in a natural way a gauge equivalence class of line bundles on the matter curve $C_R$.
Let
\bea \label{defiotaR}
\iota_R:  \hat Y_4|_{C_{R}}  \rightarrow \hat Y_4
\eea
denote the embedding of  $\hat Y_4|_{C_{R}}$ into $\hat Y_4$. 
If we view $\alpha_R$ as an element of ${\rm CH}_2(\hat Y_4|_{C_{R}})$, we are precisely in the situation described around (\ref{intdefseq}), with the role of the map $f: X \rightarrow Y$ played by 
$\iota_R:  \hat Y_4|_{C_{R}}  \rightarrow \hat Y_4$. In particular  we can form the intersection product 
\bea \label{concreteIntProd1}
\alpha_R \cdot_{\iota_R} \alpha_G \in  \CH_0(\hat Y_4|_{C_{R}}).
\eea
Its push-forward to the matter curve $C_R$ defines an element
\bea \label{PiRARAG}
{\pi|_{C_R}}_* ( \alpha_R \cdot_{\iota_R} \alpha_G) \in  \CH_0(C_R).
\eea

At this stage we must recall that $\alpha_G$ must be chosen as an element in the Chow ring with half-integer coefficients such that $\alpha_G + \frac{1}{2} c_2(\hat Y_4)$ is an integer Chow cycle class. This raises the question how (\ref{PiRARAG}) is quantised. For all concrete examples studied so far, $\frac{1}{2} \int_{A_R} c_2(\hat Y_4) \in \mathbb Z$ \cite{oai:arXiv.org:1011.6388,oai:arXiv.org:1202.3138,oai:arXiv.org:1203.4542}. 
One therefore would hope that $c_2(\hat Y_4)$ viewed as a Chow class is such that $\frac{1}{2} \alpha_R \cdot_{\iota_R} c_2(\hat Y_4)$ is integer for every matter surface. Whenever this is the case also (\ref{concreteIntProd1}) and (\ref{PiRARAG}) are integral.
For $c_2(\hat Y_4)$ of the specific form described in \cite{oai:arXiv.org:1011.6388,oai:arXiv.org:1203.4542}  considerations similar to the ones in the next section indicate  that for suitably quantized $\alpha_R$ (\ref{concreteIntProd1}) and (\ref{PiRARAG}) are integral. Obviously none of those complications arise if $G_4$ is an integral cycle defined by an integral Chow cycle $\alpha_G \in \CH^2(\hat{Y}_4)$.

The question of the integrality of (\ref{concreteIntProd1}) and (\ref{PiRARAG}) is important because the \emph{integer} Chow group $ \CH_0(C_R) \equiv \CH^1(C_R)$ is isomorphic to $\Pic(C_R)$. Therefore  (\ref{PiRARAG}) specifies an equivalence class of line bundles on $C_R$ for suitably quantized $\alpha_R$. 
Let us denote by $A_{R,G} \in Z_0(C_R)$ a representative of ${\pi|_{C_R}}_* ( \alpha_R \cdot_{\iota_R} \alpha_G)$, which we are always free to choose. Then 
\bea
L_{G,R} = {\cal O}_{C_R} (A_{R,G})
\eea
 is a line bundle on $C_R$ of degree
 \bea
 {\rm deg}(L_{G,R}) = \int_{\hat Y_4} \gamma(A_R) \cup \gamma(A_G).
 \eea

 Let us stress that this procedure yields a line bundle on the matter curve $C_R$, but in general not on a higher-dimensional
 subvariety of the base $B_3$. This, however, is sufficient to compute the localized massless matter states.
 In view of the expression (\ref{cohomgroupIIB}) counting localized charge matter zero modes in Type IIB compactifications, it is natural to identify $L_{G,R}$ with the line bundle on the matter curve appearing in the generalization of (\ref{cohomgroupIIB}) to F-theory.
 Our conjecture is therefore the massless matter   in representation $R$ localized on $C_R$ is in 1-1 correspondence with the cohomology groups
\bea \label{cohomgroups1}
H^i(C_R, L_{G,R} \otimes \sqrt{K_{C_R}}), \qquad i=0,1.
\eea
More precisely, if the compactification preserves ${\cal N}=1$ supersymmetry, 
${\rm dim} \,  H^i(C_R, L_R^G \otimes K_{C_R}^{1/2})$ counts the number ${\cal N}=1$ chiral multiplets in representation $R$ for $i=0$ and those in representation $\bar R$ for $i=1$.
As in Type IIB the spin structure $\sqrt{K_{C_R}}$ must be picked in agreement with the holomorphic embedding of the matter curve, e.g. as a complete intersection, into $B_3$.

Note that the chiral index associated with the cohomology groups (\ref{cohomgroups1}),
\bea \label{G4index}
\chi_{R,G} =  {\rm deg}(L_{G,R})  = \int_{\hat Y_4} \gamma(A_R) \cup \gamma(A_G),   
\eea
is  - since by construction $\gamma(A_G) = G_4$  - simply the usual expression  $\int_{A_R} G_4$ that follows by duality with the heterotic string \cite{Donagi:2009ra}, with Type IIB \cite{Braun:2011zm,Krause:2011xj,oai:arXiv.org:1202.3138} and via M-theory anomaly cancellation \cite{Grimm:2011fx}. 

It is important to appreciate that the intersection product (\ref{concreteIntProd1}) underlying the definition of the line bundle $L_{G,R}$ via (\ref{PiRARAG}) is stable with respect to rational equivalence on $\hat Y_4$. This important property was the reason for using the somewhat technical formalism described in section \ref{subsec_PropChow}. This means that the result of (\ref{PiRARAG})  does not depend on the concrete representative $A_G \in \CH^2(\hat Y_4)$ as long as its rational equivalence class is given by $\alpha_G$. This is to be contrasted with the formula (\ref{G4index}) for the chiral index, which is invariant under deformations of $A_G$ which do not change its cohomology class $\gamma(A_G)$ in $H^{2,2}_{\mathbb Z}(\hat Y_4)$. Expression (\ref{PiRARAG}) contains therefore, in general, much more information.
In particular  it is in general important to know how to perform intersection products in concrete examples within the Chow ring of the elliptic fibration $\hat Y_4$.

To conclude this subsection, let us comment on a special class of gauge bundle data which, though non-generic, is of some practical relevance. Namely, suppose the Chow class $\alpha_G$ descends by pullback from a Chow class in a higher-dimensional ambient space of the compactification manifold $\hat Y_4$.
Indeed the elliptically fibred 4-fold $\hat Y_4$ can often be described as a hypersurface or complete intersection with a higher-dimensional variety.
Consider for example  the situation in which $\hat Y_4$ is embedded into a smooth 5-fold $X_5$ via an inclusion\footnote{More generally, $\hat Y_4$ could be embedded also as a complete intersection into a higher dimensional ambient space.}
\bea \label{j-incl}
j: \hat Y_4 \rightarrow X_5.
\eea
Suppose furthermore we are interested in a special class of bundle data of the form
\bea \label{aGpullback}
\alpha_G = j^* \tilde \alpha_G, \qquad\quad \quad \tilde \alpha_G \in \CH^2(X_5).
\eea
Then with the help of (\ref{composite-int}) the intersection product (\ref{concreteIntProd1}) becomes
\bea \label{prodonX}
\alpha_R \cdot_{\iota_R} \alpha_G  = 
\alpha_R \cdot_{\iota_R} j^* \tilde \alpha_G =
\alpha_R \cdot_{j  \, \iota_R}  \tilde \alpha_G \in \CH_0(\hat Y_4|_{C_R}).
\eea
In particular, (\ref{prodonX})  is stable  under deformations of $\tilde \alpha_G$ which respect  rational equivalence on $X_5$.
This observation is of practical relevance because in many situations the Chow ring of the space $X_5$ may be far better under control than that of its hypersurface $\hat Y_4$.
For example, if $X_5$ is a smooth toric variety, then (\cite{FultonTor}, §5.2)  
\bea
\CH^\ast(X_5) \simeq H^\ast(X_5,\mathbb{Z})
\eea
and in evaluating the intersection we can exploit \emph{cohomological} relations on $X_5$ for $ \tilde \alpha_G$. We stress, however, that the class of gauge bundle data of the type (\ref{aGpullback}) is a very special subset of the general possibilities, albeit one with many applications in the recent F-theory literature.

\subsection{Application to $U(1)$ gauge data}
\label{subsec_ApplU(1)}

To exemplify the definition of the line bundles $L_{G,R}$ and the computation of the matter multiplicities we now specify the 4-fold $\hat Y_4$ a little further.
We are interested in an F-theory compactification which exhibits, possibly in addition to some non-abelian gauge group, one or several abelian gauge group factors.
The presence of non-Cartan $U(1)$ gauge group factors is tied to the existence of rational sections of the fibration \cite{oai:arXiv.org:hep-th/9602114,oai:arXiv.org:hep-th/9603161} so that the elliptic fibration $\hat Y_4$ must exhibit a  Mordell-Weil group with a non-trivial finite part.\footnote{The construction of 4-dimensional F-theory compactifications with abelian gauge group factors via rational sections has recently received a lot of attention \cite{Grimm:2010ez,Krause:2011xj,Grimm:2011fx,oai:arXiv.org:1208.2695,oai:arXiv.org:1210.6034,oai:arXiv.org:1202.3138,Mayrhofer:2012zy,Braun:2013yti,Borchmann:2013jwa,Cvetic:2013nia,Braun:2013nqa,Cvetic:2013uta,Borchmann:2013hta,Cvetic:2013qsa}. For earlier work on multi-section fibrations and abelian gauge groups  in 6-dimensional compactifications of F-theory see \cite{Klemm:1996hh,Aldazabal:1996du,Candelas:1997pv,Berglund:1998va}.}
To each abelian gauge group factor $U(1)_A$ one can associate a rational equivalence class 
\bea
\tw_A \in \CH^1(\hat Y_4)
\eea
whose associated cohomology class $\gamma(\tw_A) \in H^{1,1}_{\mathbb{Z}}(\hat Y_4)$ is determined via the Shioda map \cite{Shioda}  from the rational sections of the elliptic fibration such that
\bea
\int_{\hat Y_4} \gamma(\tw_A) \cup [Z]  \cup \pi^*\omega_4  = 0 = \int_{\hat Y_4} \gamma(\tw_A)  \cup  \pi^*\omega_6 \quad \forall \omega_4 \in H^4(B_3), \quad \omega_6 \in H^6(B_3)
\eea
with $[Z] \in H^{1,1}(\hat Y_4)$ denoting the class of the zero section of the fibration $\hat Y_4$. 

The physical significance of this transversality condition is that it guarantees that  the Kaluza-Klein expansion $C_3 = A_A \wedge \tilde \tw_A + \ldots$ (with $\tilde \tw_A$ a representative of the class $\gamma(\tw_A)$) of the M-theory 3-form $C_3$ gives rise to a 1-form $A_A$  in the external dimensions which is identified with the $U(1)_A$ gauge potential. 
Let us denote by 
\bea
W_A = \sum_j m_{A,j} W_A^j  \, \in  \, Z_1(\hat Y_4)
\eea
an explicit cycle representative of the Chow class $\tw_A$ such that $W_A^j$ is an algebraic cycle in $\hat Y_4$. 
This suffices to define a certain type of $U(1)_A$ gauge data as follows:
Consider a complex 2-cycle $F \in Z_2(B_3)$ with associated Chow class $f \in \CH^1(B_3)$.
Then the object
\bea \label{alphaFX}
\alpha_{F,A} = \tw_A \cdot_\pi f \,   \in \,   \CH^2(\hat Y_4) 
\eea
  - again in the notation introduced (\ref{cdotf}) with the role of $f: X \rightarrow Y$ taken by $\pi: \hat Y_4 \rightarrow B_3$ -  
specifies a '$U(1)_A$ bundle' whose associated 4-form class
\bea
G_4^A = \pi^\ast\gamma(f) \cup \gamma(\tw_A) \in H^{2,2}(\hat Y_4)
\eea
is of the form of the $U(1)_A$-flux as introduced in \cite{Grimm:2010ez,Braun:2011zm,Krause:2011xj,Grimm:2011fx}. 
The definition of the line bundle on $C_R$ via (\ref{PiRARAG}) is now facilitated by the projection formula (\ref{eq:fibre-square-projection-formula}), which allows us to write
\bea \label{refproda}
{\pi|_{C_R}}_* ( \alpha_R \cdot_{\iota_R} \alpha_{F,A}) = {\pi|_{C_R}}_* ( \alpha_R \cdot_{\iota_R}  ( \tw_A \cdot_\pi f )) = 
 {\pi|_{C_R}}_* ( \alpha_R \cdot_{\iota_R} \tw_A ) \cdot_{\iota_R|_{B_3}} f.
\eea
Here $\iota_R|_{B_3}$ denotes the restriction of the embedding $\iota_R$ defined in (\ref{defiotaR}) to the base $B_3$ of the fibration.

This can be further evaluated in the physically interesting situation in which $\hat Y_4$ is embedded, as in (\ref{j-incl}), into a smooth toric ambient space $X_5$. Indeed in many relevant examples,  the Chow class $\tw_A$ associated with the Shioda map has the property of being a pullback from $X_5$,
\bea
\tw_A = j^*\tilde \tw_A.
\eea
With the help of (\ref{prodonX}) we have
\bea
{\pi|_{C_R}}_* ( \alpha_R \cdot_{\iota_R} \tw_A ) = {\pi|_{C_R}}_* ( \alpha_R \cdot_{j \iota_R} \tilde \tw_A ).
\eea 
The intersection in brackets is essentially 'integration along the fibre' and since $\tw_A$ arises by pullback from a toric ambient space $X_5$, we are allowed to perform the intersection up to \emph{homological} equivalence on this ambient space.
We are therefore precisely in the situation described in great detail in \cite{Braun:2011zm,Krause:2011xj}. The intersection in the fibre yields here $q_A(R)$ points over the curve $C_R$.
Therefore
\bea
{\pi|_{C_R}}_* ( \alpha_R \cdot_{\iota_R} \tw_A ) = {\pi|_{C_R}}_* ( \alpha_R \cdot_{j \iota_R} \tilde \tw_A ) = q_A(R) \, [C_R],  \, \in \, \CH_1(C_R),
\eea 
where by $[C_R]$ we denote the Chow class of the curve $C_R$ viewed as an element of $\CH_1(C_R)$. 
The numerical prefactor $q_A(R) \in \mathbb Z$, which, as we said, results from the multiplicities of the intersection in the fibre, is physically interpreted as the $U(1)_A$ charge of the matter representation $R$ \cite{Braun:2011zm,Krause:2011xj}.
The remaining intersection with $f$ must in general be performed within rational equivalence on $B_3$, and it is here where the use of homological intersection theory in general comes to an end.
Let us pick an explicit representative $A_{R,A} \in Z_0(C_R)$ of the Chow class $[C_R]  \cdot_{\iota_R|_B} f$. It defines a line bundle $ {\cal O}_{C_R} (A_{R,A})$. Then the massless matter states correspond to the cohomology classes
\bea  \label{finform2}
H^i(C_R, L_{R,A} \otimes \sqrt{K_{C_R}}), \qquad \quad L_{R,A} = [{\cal O}_{C_R} (A_{R,A})]^{\otimes q_A(R)}.
\eea 

In geometries with a well-defined Sen limit, this agrees with the expression for the massless matter states in Type IIB orientifolds.

\section{A 3-generation $\mathbf{SU(5) \times U(1)_X}$ example} \label{sec3gen}

\subsection{The general setup} \label{subsecGenSetup}

To illustrate the presented technology we compute the massless charged spectrum of an F-theory compactification with $SU \left( 5 \right) \times U \left( 1 \right)_X$ gauge group.
The simplest such compactification corresponds to what was called in \cite{Grimm:2010ez} a  $U(1)$ restricted Tate model, which describes
a compactification Calabi-Yau 4-fold $Y_4$  given by a specific type of elliptic fibration with Mordell-Weil group of rank one over a base $B_3$. The fibration is described as the hypersurface equation 
\begin{equation}
 P_T = \{y^2 s + a_1 x y z s + a_3 y z^3 = x^3 s^2 + a_2 x^2 z^2 s + a_4 x z^4\}.
\end{equation}
Here $[x : y : z: s]$ denote homogeneous coordinates on the toric fibre ambient space subject to the scaling relations $(x,y,z,s) \simeq (\lambda^2 \mu^{-1} x, \lambda^3  \mu^{-1} y, \lambda^1 z,  \mu s)$.
The Tate polynomials $a_i$ are sections of ${K}_{B_3}^{-i}$.
For 
\bea
a_2 =a_{2,1} w, \qquad a_3 = a_{3,2} w^2, \qquad a_4= a_{4,3} w^3
\eea
with generic $a_1, a_{2,1}, a_{3,2}, a_{4,3}$ the fibration exhibits an $SU(5)$ singularity in the fibre over the base surface
\bea
\{ p \in B_3: w=0 \}, \qquad \quad w \in H^0({\cal O}_{B_3}(D_{\rm GUT})).
\eea
We will use the resolution of this singularity as worked out in detail in  \cite{Krause:2011xj}.
Restricting to one of the existing six inequivalent triangulations, the resolved 4-fold is described by the hypersurface
\begin{equation}\label{tate_proper_transform}
 \begin{split}
  y^2\,s\,e_3\,e_4 &+ a_1\,x\,y\,z\,s + a_{3,2}\,y\,z^3\,e_0^2\,e_1\,e_4 \\
                   &= x^3\,s^2\,e_1\,e_2^2\,e_3 + a_{2,1}\,x^2\,z^2\,s\,e_0\,e_1\,e_2 + a_{4,3}\,x\,z^4\,e_0^3\,e_1^2\,e_2\,e_4.
 \end{split}
\end{equation}
The resolution has introduced four resolution divisors $E_i: \{e_i=0\}$, $i=1,\ldots,4$, and $e_0$ is the proper transform of $w$.
The scaling relations of the homogeneous coordinates can be found in table 1 of \cite{Krause:2011xj}, to which we refer for more details of the geometry.

In addition to the holomorphic section $Z:  \{z=0\}$ the fibration exhibits a rational section $S: \{s=0\}$. 
Let us denote their rational equivalence classes again by $Z$ and $S$, viewed as elements of $\CH^1(\hat Y_4)$, and similarly for $E_i$. 
By slight abuse of notation we also denote the rational equivalence class associated with the anti-canonical divisor by $\bar{K}_{B_3}$.
Then the object \cite{Krause:2011xj}
\bea
\tw_X = 5 (S - Z - \bar{K}_{B_3}) + \sum_i l_i E_i  \in \CH^1(\hat Y_4), \qquad\quad l_i = (2,4,6,3),
\eea
defines, via the cycle map, an element $\gamma(\tw_X) \in H^{1,1}(\hat Y_4)$ such that expansion of the M-theory 3-form $C_3$ in terms of a representative of $\gamma(\tw_X)$
gives rise to the gauge potential $A_X$ associated with a $U(1)_X$ gauge group.

The base $B_3$ contains four matter curves supporting charged matter in representations ${\bf 10_{-1}}$, ${\bf \bar 5_{3}}$, ${\bf 5_{2}}$ and ${\bf 1_5}$ plus the respective conjugates, where the subscripts denote the $U(1)_X$ charges. In agreement with their physical interpretation as Standard Model matter and, respectively, Higgs fields in an SU(5) GUT model, we will denote the field ${\bf \bar 5_{3}}$ as ${ \bf \bar 5_m}$ and  ${\bf 5_{2}}$ as ${\bf  5_H}$. The matter curves are  given as complete intersections on $B_3$,
\bea
C_R=   \{p \in B_3, \,  U_R \, (p)=  V_R \, (p) =0 \}
\eea
with 
\bea
U_R \in H^0(B_3, {\cal O}_{B_3}(D_R^a)), \qquad V_R \in H^0(B_3, {\cal O}_{B_3}(D_R^b)).
\eea
Specifically,
\bea \label{mattercurves1}
&C_{\bf 10}: \{w= 0 \} \cap \{ a_1 = 0 \}, \qquad 
C_{\bf \bar 5_m}: \{w= 0 \} \cap \{ a_{3,2} = 0 \}, \nonumber \\
&C_{\bf 5_H}: \{w= 0 \} \cap \{ a_1 a_{4,3} - a_{2,1} a_{3,2} = 0 \}, \qquad
C_{\bf 1}: \{ a_{3,2} = 0\} \cap \{ a_{4,3} = 0 \}.
\eea


We are interested in the massless matter spectrum induced by the gauge configuration encoded in the rational equivalence class
\bea \label{alphaXa}
\alpha_X =  \tw_X \cdot_\pi f \in \CH^2(\hat Y_4) \qquad\quad \quad {\rm for } \quad  f \in \CH^1(B_3),
\eea
 see the discussion around (\ref{alphaFX}). The associated $G_4 = \gamma(\alpha_X) \in H^{2,2}(\hat Y_4)$ is the gauge flux associated with $U(1)_X$. 
Recall that this flux is subject to the quantization condition $G_4 + \frac{1}{2} c_2(\hat Y_4) \in H^4_{\mathbb Z}(\hat Y_4)$ \cite{oai:arXiv.org:hep-th/9609122}.
In the present context, a sufficient condition  for this is \cite{Krause:2011xj,oai:arXiv.org:1202.3138}
$ \gamma(f) + \frac{1}{2} \gamma(D_{\rm GUT}) \in H^{1,1}_{\mathbb Z}(B_3)$. Since $h^1(B_3)=0$ this is equivalent, at the level of the Picard group, to
\bea \label{quantex1}
f + \frac{1}{2} D_{\rm GUT} \in {\rm Pic}(B_3).
\eea
A detailed analysis of the quantization condition for $G_4$ in more general setups has been performed in \cite{oai:arXiv.org:1011.6388,oai:arXiv.org:1203.4542}.

The object $f \in \CH^1(B_3)$ defines an equivalence class of line bundles on $B_3$. Let us denote by ${\cal L}$ a representative of this line bundle class with $c_1({\cal L}) = \gamma(f)$.
Then, according to equ. (\ref{finform2}), the massless matter on the respective matter curves is counted by the following cohomology groups:
\bea
{\bf 10} &\leftrightarrow& H^i(C_{\bf 10}, {\cal L}^{-1}|_{C_{\bf 10}}  \otimes \sqrt{K_{C_{\bf 10}} } ), \\
{\bf \bar 5_m} &\leftrightarrow& H^i(C_{\bf \bar 5_m}, {\cal L}^{3}|_{C_{\bf \bar 5_m}}  \otimes \sqrt{K_{C_{\bf \bar 5_m}} } ), \\
{\bf 5_H}& \leftrightarrow& H^i(C_{\bf  5_H}, {\cal L}^{2}|_{C_{\bf  5_H}}  \otimes \sqrt{K_{C_{\bf  5_H}} } ), \\
{\bf 1} &\leftrightarrow& H^i(C_{\bf 1}, {\cal L}^{5}|_{C_{\bf 1}}  \otimes \sqrt{K_{C_{\bf 1}} } ).
\eea
The choice of spin structure is, as described around eq. (\ref{spinstructure}), the one inherited from the global embedding of the matter curves (\ref{mattercurves1}) as a complete intersection of 2-cycles with divisor classes $D^a_R$ and $D^b_R$.
For example, if $c_1(K_{B_3})  + D^a_R + D^b_R$ is an even class, it is given, by
  slight abuse of notation, by $\sqrt{K_{C_R}} = {\cal O}_{B_3}(\frac{1}{2}(K_{B_3}  + D^a_R + D^b_R))|_{C_R}$ with $D^a_R$ and $D^b_R$ the two divisor classes that define the complete intersection matter curves (\ref{mattercurves1}). 

The computation of the massless spectrum therefore reduces to evaluating the cohomologies of the pullback of a line bundle on $B_3$ to a codimension-two complete intersection matter curve.

To simplify things further we describe the base $B_3$ as a hypersurface in a smooth and compact normal toric variety $X_\Sigma$, 
\begin{equation} \label{Qpoly}
 B_3 = \left\{ p \in X_\Sigma \; , \; Q \left( p \right) = 0 \right\}, \qquad \quad  Q \in H^0 \left( X_\Sigma , \mathcal{O}_{X_\Sigma} \left( D_{B_3} \right) \right), 
 \end{equation}
 such that the divisor class $f \in \CH^1(B_3)$ is itself the pullback of a class $\tilde f \in \CH^1(X_\Sigma)$. The same holds for the classes defining the matter curves as complete intersections, which we will denote by the same letter.

Let ${\cal O}_{X_{\Sigma}}(\tilde f)$ be the line bundle on $X_\Sigma$ with first Chern class $\gamma(\tilde f)$.
We are then interested in the cohomologies 
\bea \label{cohom-bundles}
H^i(C_R, L_R |_{C_R}),  \qquad\quad  L_R = {\cal O}_{X_\Sigma}\Big( q_R \tilde f +   \frac{1}{2}(K_{X_\Sigma} + D_{B_3} + D^a_R + D^b_R)\Big). 
\eea
Note that $L_R$ is guaranteed to be integer quantized.
In section \ref{sub_cohom} we will describe how to evaluate such cohomology classes, and then apply this general procedure in section \ref{sec:AConcreteExample} to a toy model of the type described above that leads to three chiral generations of $SU(5) \times U(1)_X$ matter.

\subsection{Pullback cohomologies via \emph{cohomCalg} and the Koszul spectral sequence}
\label{sub_cohom}

According to equ.(\ref{cohom-bundles}) our task is quite generally to 
compute cohomology groups $H^i(C,{\cal L}|_C)$ for a line bundle  ${\cal L} = {\cal O}_{X_\Sigma}(D)$ on a smooth and compact normal toric variety $X_{\Sigma}$ and $C$ a smooth subvariety of $X_\Sigma$ given as a complete intersection.
In the case of interest to us, ${X_\Sigma}$ is of complex dimension 4 and  $C$ is a smooth codimension 3 locus given as a complete intersection
\bea
C := \left\{ p \in X_\Sigma \; , \; \tilde{s}_1 \left( p \right) = \tilde{s}_2 \left( p \right) = \tilde{s}_3 \left( p \right) = 0 \right\}
\eea
with
\bea
\tilde{s}_i \in H^0 \left( X_\Sigma, \mathcal{O}_{X_\Sigma} \left( S_i \right) \right).
\eea
This problem consists of two parts: First compute the cohomology groups $H^i(X_\Sigma, {\cal L} )$ and then deduce the desired cohomologies of the pullback bundle ${\cal L}|_C$ via the standard technology of Koszul spectral sequences.
 
On smooth and compact normal toric varieties $X_\Sigma$ the cohomology classes of holomorphic line bundles can be computed very efficiently with the help of the \emph{cohomCalg} algorithm developed by Blumenhagen and collaborators \cite{Blumenhagen:2010pv, cohomCalg:Implementation, 2011JMP....52c3506J, Rahn:2010fm, Blumenhagen:2010ed}. Many applications of this algorithm have been described in detail in \cite{Blumenhagen:2010ed} and \cite{RahnPhd}. 
The algorithm \cite{Blumenhagen:2010pv}, implemented in \cite{cohomCalg:Implementation}, provides a basis of the cohomology groups $H^i(X_\Sigma, {\cal L} )$ as so-called rationoms. For example, 
on  $X_\Sigma = \mathbb{CP}^3$ the cohomology classes of $\mathcal{L} = \mathcal{O}_{\mathbb{CP}^3} \left( 1 \right)$ are easily computed to be
\bea
H^0 \left( \mathbb{CP}^3, \mathcal{L} \right) = \left\{ \alpha_1 x_1 + \alpha_2 x_2 + \alpha_3 x_3 \; , \; \alpha_i \in \mathbb{C} \right\}, \quad H^1 \left( \mathbb{CP}^3, \mathcal{L} \right) = H^2 \left( \mathbb{CP}^3, \mathcal{L} \right) = 0
\eea
Many more examples can be found in \cite{Blumenhagen:2010pv, 2011JMP....52c3506J, Rahn:2010fm, Blumenhagen:2010ed,RahnPhd}.

To compute from $H^i(X_\Sigma, {\cal L} )$  with ${\cal L} = {\cal O}_{X_\Sigma}(D)$ the cohomologies of the pullback bundle ${\cal L}|_C$, it is standard to invoke 
the sheaf exact Koszul sequence
\bea
0 \to \mathcal{L}^\prime \stackrel{\alpha}{\rightarrow} \mathcal{V}_2 \stackrel{\beta}{\rightarrow} \mathcal{V}_1 \stackrel{\gamma}{\rightarrow} \mathcal{L} \stackrel{r}{\rightarrow} \left. \mathcal{L} \right|_C \to 0 \label{equ:Koszul3CoDimension}
\eea
with
\bea
\mathcal{L}' &=& \mathcal{O}_{X_\Sigma} \left( D - S_1 - S_2 - S_3 \right), \\
\mathcal{V}_2 &= &\mathcal{O}_{X_\Sigma} \left( D - S_2 - S_3 \right) \oplus \mathcal{O}_{X_\Sigma} \left( D - S_1 - S_3 \right) \oplus \mathcal{O}_{X_\Sigma
			      } \left( D - S_1 - S_2 \right), \\
\mathcal{V}_1 &=& \mathcal{O}_{X_\Sigma} \left( D - S_1 \right) \oplus \mathcal{O}_{X_\Sigma} \left( D - S_2 \right) \oplus \mathcal{O}_{X_\Sigma} \left( D -            S_3 \right).
\eea
As is well-known, by application of the splitting principle on can replace (\ref{equ:Koszul3CoDimension}) by three short sheaf exact sequences
\begin{align}
    & 0 \to \mathcal{L}^\prime \stackrel{\alpha}{\rightarrow} \mathcal{V}_2 \to \mathcal{I}_1, \\
		& 0 \to \mathcal{I}_1 \to \mathcal{V}_1 \to \mathcal{I}_2 \to 0, \\
		& 0 \to \mathcal{I}_2 \to \mathcal{L} \stackrel{r}{\rightarrow} \left. \mathcal{L} \right|_C \to 0.
\end{align}
To each of those there exists an associated long exact sequence in cohomology. Oftentimes exactness of these sequences is sufficient to determine $h^i(C,{\cal L}|_C)$ uniquely, and in those cases the Koszul extension of \emph{cohomCalg} \cite{cohomCalg:Implementation} can be used to evaluate $h^i(C,{\cal L}|_C)$. Unfortunately in many situations these exactness properties are not enough, and to determine $H^i(C,{\cal L}|_C)$ from the Koszul spectral sequence
one has to explicitly construct the maps in this sequence. This approach goes beyond the Koszul extension of \emph{cohomCalg} and is very similar to the techniques presented in \cite{Anderson:2013qca}. 

The mappings in equ. (\ref{equ:Koszul3CoDimension}) are induced from the Koszul complex over $\tilde{s}_1$, $\tilde{s}_2$ and $\tilde{s}_3$ \cite{eisenbud1995commutative, cox2011toric}. 
The appearing sheaf homomorphisms are thus induced by means of the natural restriction of functions from the global-section valued
\bea
\alpha = \left( \begin{array}{c} \tilde{s}_1 \\ - \tilde{s}_2 \\ \tilde{s}_3 \end{array} \right), \qquad
\beta = \left( \begin{array}{ccc} 0 & - \tilde{s}_3 & - \tilde{s}_2 \\ - \tilde{s}_3 & 0 & \tilde{s}_1 \\ \tilde{s}_2 & \tilde{s}_1 & 0 \end{array} \right), \qquad
\gamma = \left( \tilde{s}_1, \tilde{s}_2, \tilde{s}_3 \right).
\eea

From equ. (\ref{equ:Koszul3CoDimension}) the cohomologies of $\left. \mathcal{L} \right|_C$ can now be computed via the associated Koszul spectral sequence. For the codimension 3 case, this spectral sequence converges on the $E_4$-sheet and we display the sheets $E_0$ to $E_4$ in \autoref{figure5}, \autoref{figure6}, \autoref{figure7}, \autoref{figure8} and \autoref{figure9}.
\begin{figure}[tb]
\begin{center}
\begin{tikzpicture}[every node/.style={anchor=center}]

\matrix (m) [matrix of nodes, row sep= 1.5em, column sep=1em, column 1/.style={nodes={minimum width = 3em},column sep = 0em}]
    { 
       $\mathcal{L}^\prime$ & $\check{C}^0 \left( \mathcal{U}, \mathcal{L}^\prime \right)$ & $\check{C}^1 \left( \mathcal{U}, \mathcal{L}^\prime \right)$ & $\check{C}^2 
			 \left( \mathcal{U}, \mathcal{L}^\prime \right)$ & $\check{C}^3 \left( \mathcal{U}, \mathcal{L}^\prime \right)$ & $\check{C}^4 \left( \mathcal{U}, \mathcal{L}^
			 \prime \right)$ \\
		   $\mathcal{V}_2$ & $\check{C}^0 \left( \mathcal{U}, \mathcal{V}_2 \right)$ & $\check{C}^1 \left( \mathcal{U}, \mathcal{V}_2 \right)$ & $\check{C}^2 \left( \mathcal
			 {U}, \mathcal{V}_2 \right)$ & $\check{C}^3 \left( \mathcal{U}, \mathcal{V}_2 \right)$ & $\check{C}^4 \left( \mathcal{U}, \mathcal{V}_2 \right)$ \\
			 $\mathcal{V}_1$ & $\check{C}^0 \left( \mathcal{U}, \mathcal{V}_1 \right)$ & $\check{C}^1 \left( \mathcal{U}, \mathcal{V}_1 \right)$ & $\check{C}^2 \left( \mathcal
			 {U}, \mathcal{V}_1 \right)$ & $\check{C}^3 \left( \mathcal{U}, \mathcal{V}_1 \right)$ & $\check{C}^4 \left( \mathcal{U}, \mathcal{V}_1 \right)$ \\
			 $\mathcal{L}$ \vphantom{${E}_{(4)}^{0,4}$} & $\check{C}^0 \left( \mathcal{U}, \mathcal{L} \right)$ & $\check{C}^1 \left( \mathcal{U}, \mathcal{L} \right)$ & $
			 \check{C}^2 \left( \mathcal{U}, \mathcal{L} \right)$ & $\check{C}^3 \left( \mathcal{U}, \mathcal{L} \right)$ & $\check{C}^4 \left( \mathcal{U}, \mathcal{L} \right
			 )$ \\
		   \hphantom{$\mathcal{V}_2$} & $H^0$ & $H^1$ & $H^2$ & $H^3$ & $H^4$ \\
		};

\draw[color=black, thick] ([xshift=-0.5em]m-1-1.north east)--([xshift=-0.5em]m-5-1.south east);
\draw[color=black, thick] ([yshift=-0.75em]m-4-1.south west)--([yshift=-0.75em]m-4-6.south east);
\draw[color=black, thick] ([yshift=-0.95em]m-4-1.south west)--([yshift=-0.95em]m-4-6.south east);

\draw[->,thick] (m-1-1)-- (m-2-1);
\draw[->,thick] (m-2-1)-- (m-3-1);
\draw[->,thick] (m-3-1)-- (m-4-1);

\path[->,thick] (m-1-2) edge (m-2-2) edge node[right] {$\alpha^0$} (m-2-2);
\path[->,thick] (m-1-3) edge (m-2-3) edge node[right] {$\alpha^1$} (m-2-3);
\path[->,thick] (m-1-4) edge (m-2-4) edge node[right] {$\alpha^2$} (m-2-4);
\path[->,thick] (m-1-5) edge (m-2-5) edge node[right] {$\alpha^3$} (m-2-5);
\path[->,thick] (m-1-6) edge (m-2-6) edge node[right] {$\alpha^4$} (m-2-6);

\path[->,thick] (m-2-2) edge (m-3-2) edge node[right] {$\beta^0$} (m-3-2);
\path[->,thick] (m-2-3) edge (m-3-3) edge node[right] {$\beta^1$} (m-3-3);
\path[->,thick] (m-2-4) edge (m-3-4) edge node[right] {$\beta^2$} (m-3-4);
\path[->,thick] (m-2-5) edge (m-3-5) edge node[right] {$\beta^3$} (m-3-5);
\path[->,thick] (m-2-6) edge (m-3-6) edge node[right] {$\beta^4$} (m-3-6);

\path[->,thick] (m-3-2) edge (m-4-2) edge node[right] {$\gamma^0$} (m-4-2);
\path[->,thick] (m-3-3) edge (m-4-3) edge node[right] {$\gamma^1$} (m-4-3);
\path[->,thick] (m-3-4) edge (m-4-4) edge node[right] {$\gamma^2$} (m-4-4);
\path[->,thick] (m-3-5) edge (m-4-5) edge node[right] {$\gamma^3$} (m-4-5);
\path[->,thick] (m-3-6) edge (m-4-6) edge node[right] {$\gamma^4$} (m-4-6);

\path[->,thick] (m-1-2) edge (m-1-3) edge node[above] {$\delta$} (m-1-3);
\path[->,thick] (m-1-3) edge (m-1-4) edge node[above] {$\delta$} (m-1-4);
\path[->,thick] (m-1-4) edge (m-1-5) edge node[above] {$\delta$} (m-1-5);
\path[->,thick] (m-1-5) edge (m-1-6) edge node[above] {$\delta$} (m-1-6);

\path[->,thick] (m-2-2) edge (m-2-3) edge node[above] {$\delta$} (m-2-3);
\path[->,thick] (m-2-3) edge (m-2-4) edge node[above] {$\delta$} (m-2-4);
\path[->,thick] (m-2-4) edge (m-2-5) edge node[above] {$\delta$} (m-2-5);
\path[->,thick] (m-2-5) edge (m-2-6) edge node[above] {$\delta$} (m-2-6);

\path[->,thick] (m-3-2) edge (m-3-3) edge node[above] {$\delta$} (m-3-3);
\path[->,thick] (m-3-3) edge (m-3-4) edge node[above] {$\delta$} (m-3-4);
\path[->,thick] (m-3-4) edge (m-3-5) edge node[above] {$\delta$} (m-3-5);
\path[->,thick] (m-3-5) edge (m-3-6) edge node[above] {$\delta$} (m-3-6);

\path[->,thick] (m-4-2) edge (m-4-3) edge node[above] {$\delta$} (m-4-3);
\path[->,thick] (m-4-3) edge (m-4-4) edge node[above] {$\delta$} (m-4-4);
\path[->,thick] (m-4-4) edge (m-4-5) edge node[above] {$\delta$} (m-4-5);
\path[->,thick] (m-4-5) edge (m-4-6) edge node[above] {$\delta$} (m-4-6);

\end{tikzpicture}
\end{center}
\caption{The $E_0$-sheet of the Koszul spectral sequence for codimension $3$.}
\label{figure5} 
\end{figure}
\begin{figure}[tb]
\begin{center}

\begin{tikzpicture}[every node/.style={anchor=center}]

\matrix (m) [matrix of nodes, row sep= 1.5em, column sep=1.5em, column 1/.style={nodes={minimum width = 3em},column sep = 0em}]
    { 
       $\mathcal{L}^\prime$ & $\check{H}^0 \left( \mathcal{U}, \mathcal{L}^\prime \right)$ & $\check{H}^1 \left( \mathcal{U}, \mathcal{L}^\prime \right)$ & $\check{H}^2 
			\left( \mathcal{U}, \mathcal{L}^\prime \right)$ & $\check{H}^3 \left( \mathcal{U}, \mathcal{L}^\prime \right)$ & $\check{H}^4 \left( \mathcal{U}, \mathcal{L}^
			\prime \right)$ \\
		  $\mathcal{V}_2$ & $\check{H}^0 \left( \mathcal{U}, \mathcal{V}_2 \right)$ & $\check{H}^1 \left( \mathcal{U}, \mathcal{V}_2 \right)$ & $\check{H}^2 \left( \mathcal{
			U}, \mathcal{V}_2 \right)$ & $\check{H}^3 \left( \mathcal{U}, \mathcal{V}_2 \right)$ & $\check{H}^4 \left( \mathcal{U}, \mathcal{V}_2 \right)$ \\
			$\mathcal{V}_1$ & $\check{H}^0 \left( \mathcal{U}, \mathcal{V}_1 \right)$ & $\check{H}^1 \left( \mathcal{U}, \mathcal{V}_1 \right)$ & $\check{H}^2 \left( \mathcal{
			U}, \mathcal{V}_1 \right)$ & $\check{H}^3 \left( \mathcal{U}, \mathcal{V}_1 \right)$ & $\check{H}^4 \left( \mathcal{U}, \mathcal{V}_1 \right)$ \\
			$\mathcal{L}$ \vphantom{${E}_{(4)}^{0,4}$} & $\check{H}^0 \left( \mathcal{U}, \mathcal{L} \right)$ & $\check{H}^1 \left( \mathcal{U}, \mathcal{L} \right)$ & $
			\check{H}^2 \left( \mathcal{U}, \mathcal{L} \right)$ & $\check{H}^3 \left( \mathcal{U}, \mathcal{L} \right)$ & $\check{H}^4 \left( \mathcal{U}, \mathcal{L} \right)
			$ \\
		  \hphantom{$\mathcal{V}_2$} & $H^0$ & $H^1$ & $H^2$ & $H^3$ & $H^4$ \\
		};

\draw[color=black, thick] ([xshift=-0.5em]m-1-1.north east)--([xshift=-0.5em]m-5-1.south east);
\draw[color=black, thick] ([yshift=-0.75em]m-4-1.south west)--([yshift=-0.75em]m-4-6.south east);
\draw[color=black, thick] ([yshift=-0.95em]m-4-1.south west)--([yshift=-0.95em]m-4-6.south east);

\draw[->,thick] (m-1-1)-- (m-2-1);
\draw[->,thick] (m-2-1)-- (m-3-1);
\draw[->,thick] (m-3-1)-- (m-4-1);

\path[->,thick] (m-1-2) edge (m-2-2) edge node[right] {$\tilde{\alpha}^0$} (m-2-2);
\path[->,thick] (m-1-3) edge (m-2-3) edge node[right] {$\tilde{\alpha}^1$} (m-2-3);
\path[->,thick] (m-1-4) edge (m-2-4) edge node[right] {$\tilde{\alpha}^2$} (m-2-4);
\path[->,thick] (m-1-5) edge (m-2-5) edge node[right] {$\tilde{\alpha}^3$} (m-2-5);
\path[->,thick] (m-1-6) edge (m-2-6) edge node[right] {$\tilde{\alpha}^4$} (m-2-6);

\path[->,thick] (m-2-2) edge (m-3-2) edge node[right] {$\tilde{\beta}^0$} (m-3-2);
\path[->,thick] (m-2-3) edge (m-3-3) edge node[right] {$\tilde{\beta}^1$} (m-3-3);
\path[->,thick] (m-2-4) edge (m-3-4) edge node[right] {$\tilde{\beta}^2$} (m-3-4);
\path[->,thick] (m-2-5) edge (m-3-5) edge node[right] {$\tilde{\beta}^3$} (m-3-5);
\path[->,thick] (m-2-6) edge (m-3-6) edge node[right] {$\tilde{\beta}^4$} (m-3-6);

\path[->,thick] (m-3-2) edge (m-4-2) edge node[right] {$\tilde{\gamma}^0$} (m-4-2);
\path[->,thick] (m-3-3) edge (m-4-3) edge node[right] {$\tilde{\gamma}^1$} (m-4-3);
\path[->,thick] (m-3-4) edge (m-4-4) edge node[right] {$\tilde{\gamma}^2$} (m-4-4);
\path[->,thick] (m-3-5) edge (m-4-5) edge node[right] {$\tilde{\gamma}^3$} (m-4-5);
\path[->,thick] (m-3-6) edge (m-4-6) edge node[right] {$\tilde{\gamma}^4$} (m-4-6);

\end{tikzpicture}
\end{center}
\caption{The $E_1$-sheet of the Koszul spectral sequence for codimension $3$.}
\label{figure6} 
\end{figure}
\begin{figure}[tb]
\begin{center}
\begin{tikzpicture}[every node/.style={anchor=center}]

\matrix (m) [matrix of nodes, row sep= 1.5em, column sep=1.5em, column 1/.style={nodes={minimum width = 3em},column sep = 0em}]
    { 
       $\mathcal{L}^\prime$ & $E_{(2)}^{3,0}$ & $E_{(2)}^{3,1}$ & $E_{(2)}^{3,2}$ & $E_{(2)}^{3,3}$ & $E_{(2)}^{3,4}$ \\
		   $\mathcal{V}_2$      & $E_{(2)}^{2,0}$ & $E_{(2)}^{2,1}$ & $E_{(2)}^{2,2}$ & $E_{(2)}^{2,3}$ & $E_{(2)}^{2,4}$ \\
			 $\mathcal{V}_1$      & $E_{(2)}^{1,0}$ & $E_{(2)}^{1,1}$ & $E_{(2)}^{1,2}$ & $E_{(2)}^{1,3}$ & $E_{(2)}^{1,4}$ \\
			 $\mathcal{L}$ \vphantom{${E}_{(4)}^{0,4}$} & $E_{(2)}^{0,0}$ & $E_{(2)}^{0,1}$ & $E_{(2)}^{0,2}$ & $E_{(2)}^{0,3}$ & $E_{(2)}^{0,4}$ \\
		             \hphantom{$\mathcal{V}_2$}       & $H^0$ & $H^1$ & $H^2$ & $H^3$ & $H^4$ \\	};
\draw[color=black, thick] ([xshift=-0.5em]m-1-1.north east)--([xshift=-0.5em]m-5-1.south east);
\draw[color=black, thick] ([yshift=-0.75em]m-4-1.south west)--([yshift=-0.75em]m-4-6.south east);
\draw[color=black, thick] ([yshift=-0.95em]m-4-1.south west)--([yshift=-0.95em]m-4-6.south east);

\draw[->,thick] (m-1-3)--(m-3-2);
\draw[->,thick] (m-1-4)--(m-3-3);
\draw[->,thick] (m-1-5)--(m-3-4);
\draw[->,thick] (m-1-6)--(m-3-5);
\draw[->,thick] (m-2-3)--(m-4-2);
\draw[->,thick] (m-2-4)--(m-4-3);
\draw[->,thick] (m-2-5)--(m-4-4);
\draw[->,thick] (m-2-6)--(m-4-5);

\draw[->,thick] (m-1-1)-- (m-2-1);
\draw[->,thick] (m-2-1)-- (m-3-1);
\draw[->,thick] (m-3-1)-- (m-4-1);

\end{tikzpicture}
\end{center}
\caption{The $E_2$-sheet of the Koszul spectral sequence for codimension $3$.}
\label{figure7} 
\end{figure}
\begin{figure}[tb]
\begin{center}

\begin{tikzpicture}[every node/.style={anchor=center}]

\matrix (m) [matrix of nodes, row sep= 1.5em, column sep=1.5em, column 1/.style={nodes={minimum width = 3em},column sep = 0em}]
    { 
       $\mathcal{L}^\prime$ & $E_{(3)}^{3,0}$ & $E_{(3)}^{3,1}$ & $E_{(3)}^{3,2}$ & $E_{(3)}^{3,3}$ & $E_{(3)}^{3,4}$ \\
		   $\mathcal{V}_2$ & $E_{(3)}^{2,0}$ & $E_{(3)}^{2,1}$ & $E_{(3)}^{2,2}$ & $E_{(3)}^{2,3}$ & $E_{(3)}^{2,4}$ \\
			 $\mathcal{V}_1$      & $E_{(3)}^{1,0}$ & $E_{(3)}^{1,1}$ & $E_{(3)}^{1,2}$ & $E_{(3)}^{1,3}$ & $E_{(3)}^{1,4}$ \\
			 $\mathcal{L}$ \vphantom{${E}_{(4)}^{0,4}$}     & $E_{(3)}^{0,0}$ & $E_{(3)}^{0,1}$ & $E_{(3)}^{0,2}$ & $E_{(3)}^{0,3}$ & $E_{(3)}^{0,4}$ \\
		   \hphantom{$\mathcal{V}_2$}       & $H^0$ & $H^1$ & $H^2$ & $H^3$ & $H^4$ \\
		};

\draw[color=black, thick] ([xshift=-0.5em]m-1-1.north east)--([xshift=-0.5em]m-5-1.south east);
\draw[color=black, thick] ([yshift=-0.75em]m-4-1.south west)--([yshift=-0.75em]m-4-6.south east);
\draw[color=black, thick] ([yshift=-0.95em]m-4-1.south west)--([yshift=-0.95em]m-4-6.south east);

\draw[->,thick,dashed] (m-1-4)--(m-4-2);
\draw[->,thick,dashed] (m-1-5)--(m-4-3);
\draw[->,thick,dashed] (m-1-6)--(m-4-4);

\draw[->,thick] (m-1-1)-- (m-2-1);
\draw[->,thick] (m-2-1)-- (m-3-1);
\draw[->,thick] (m-3-1)-- (m-4-1);

\end{tikzpicture}

\end{center}
\caption{The $E_3$-sheet of the Koszul spectral sequence for codimension $3$.}
\label{figure8}
\end{figure}
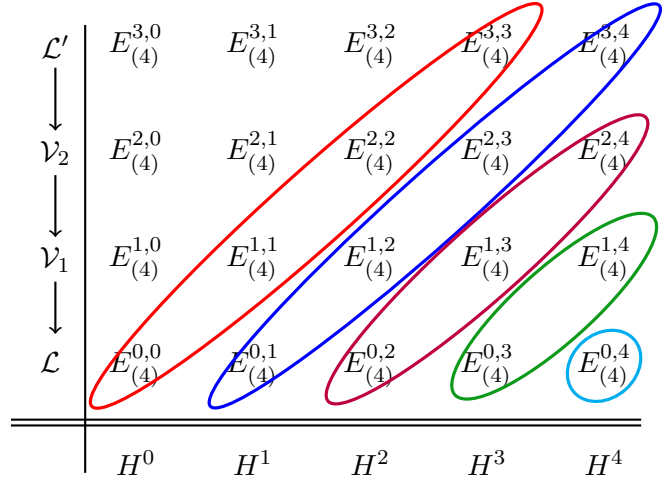
\begin{figure}[tb]
\begin{center}

\begin{tikzpicture}[every node/.style={anchor=center}]

\matrix (m) [matrix of nodes, row sep= 1.5em, column sep=1.5em, column 1/.style={nodes={minimum width = 3em},column sep = 0em}]
    { 
       $\mathcal{L}^\prime$ & $E_{(4)}^{3,0}$ & $E_{(4)}^{3,1}$ & $E_{(4)}^{3,2}$ & $E_{(4)}^{3,3}$ & $E_{(4)}^{3,4}$ \\
		   $\mathcal{V}_2$ & $E_{(4)}^{2,0}$ & $E_{(4)}^{2,1}$ & $E_{(4)}^{2,2}$ & $E_{(4)}^{2,3}$ & $E_{(4)}^{2,4}$ \\
			 $\mathcal{V}_1$ & $E_{(4)}^{1,0}$ & $E_{(4)}^{1,1}$ & $E_{(4)}^{1,2}$ & $E_{(4)}^{1,3}$ & $E_{(4)}^{1,4}$ \\
			 $\mathcal{L}$ \vphantom{${E}_{(4)}^{0,4}$} & $E_{(4)}^{0,0}$ & $E_{(4)}^{0,1}$ & $E_{(4)}^{0,2}$ & $E_{(4)}^{0,3}$ & $E_{(4)}^{0,4}$ \\
		   \hphantom{$\mathcal{V}_2$}       & $H^0$ & $H^1$ & $H^2$ & $H^3$ & $H^4$ \\
		};

\draw[color=black, thick] ([xshift=-0.5em]m-1-1.north east)--([xshift=-0.5em]m-5-1.south east);
\draw[color=black, thick] ([yshift=-0.75em]m-4-1.south west)--([yshift=-0.75em]m-4-6.south east);
\draw[color=black, thick] ([yshift=-0.95em]m-4-1.south west)--([yshift=-0.95em]m-4-6.south east);

\draw[->,thick] (m-1-1)-- (m-2-1);
\draw[->,thick] (m-2-1)-- (m-3-1);
\draw[->,thick] (m-3-1)-- (m-4-1);

\draw[rotate around={-48:(-0.13,0.58)},red,very thick] (-0.13,0.58) ellipse (14pt and 113pt);
\draw[rotate around={-48:(1.43,0.58)},blue,very thick] (1.43,0.58) ellipse (13pt and 113pt);
\draw[rotate around={-48:(m-3-5)},purple,very thick] (m-3-5) ellipse (14pt and 80pt);
\draw[rotate around={-48:(3.0,-0.75)},darkgreen,very thick] (3.0,-0.75) ellipse (14pt and 50pt);
\draw[rotate around={-48:(m-4-6)},cyan,very thick] (m-4-6) ellipse (12pt and 15pt);


\end{tikzpicture}

\end{center}
\caption{The $E_4$-sheet of the Koszul spectral sequence for codimension $3$. Note that $H^0 \left( C, \left. \cal L \right|_C \right)$ is obtained by adding the red spaces, $H^1 \left( C, \left. \cal L \right|_C \right)$ is given by the sum of the dark blue spaces and so on.}
\label{figure9}
\end{figure}Note that this spectral sequence makes use of the affine open cover $\mathcal{U}$ of $X_\Sigma$ given by
\bea
\mathcal{U} = \left\{U_\sigma \; , \; \sigma \in \Sigma_{\text{max}} \right\}.
\eea
which allows to compute sheaf cohomology from \v{C}ech cohomology \cite{cox2011toric}.

The vertical maps in \autoref{figure5} are naturally induced from the sheaf homomorphisms that appear in the Koszul sequence (\ref{equ:Koszul3CoDimension}). Note that the vertical maps preserve closure. The maps in the $E_1$-sheet are thus the corresponding maps on the \v{C}ech cohomologies. For example we therefore have
\bea
\tilde{\alpha}^0 \colon \check{H}^0 \left( \mathcal{U}, \mathcal{L}^\prime \right) \to \check{H}^1 \left( \mathcal{U}, \mathcal{V}_2 \right) \; , \; \left[ x \right] \mapsto \left[ \alpha^0 \left( x \right) \right]. 
\eea
The construction of the so-called Knight's moves in the $E_2$-sheet, which is displayed in \autoref{figure7}, is according to the following principle.
\begin{enumerate}
       \item Pick $x \in E^{3,1}_{(2)}$.
			 \item Represent $x$ by $\tilde{x} \in \check{C}^1 \left( \mathcal{U}, \mathcal{L}^\prime \right)$ with $\delta \tilde{x} = 0$ and $\alpha^1 \left( \tilde{x} 
			      \right) = \delta \left( \tilde{y} \right)$ for a suitable $\tilde{y} \in \check{C}^0 \left( \mathcal{U}, \mathcal{V}_2 \right)$.
			 \item Set $\tilde{z} := \beta^0 \left( \tilde{y} \right)$ and realise that $\delta \left( \tilde{z} \right) = \gamma^0 \left( \tilde{z} \right) = 0$.
			 \item Consequently $\tilde{z}$ gives rise to $z \in E^{1,0}_{(2)}$.
			 \item Define the Knight's move as the map $d_{(2)}^{3,1} \colon E_{(2)}^{3,1} \to E_{(2)}^{1,0} \; , \; x \mapsto z$. This map is independent of the choice of $
			      \tilde{y}$.
\end{enumerate}
Similarly one constructs the maps in the $E_3$-sheet.

From the spaces on the $E_4$-sheet one can finally compute the desired pullback cohomologies via \cite{distler1988aspects}
\bea
H^i \left( C, \left. \mathcal{L} \right|_C \right) = \bigoplus_{j = 0}^{\infty}{E_{(4)}^{j,i+j}}.
\eea
For more details on spectral sequences we refer the interested reader to Appendix C of \cite{cox2011toric}.

Whilst the construction of the maps in the $E_2$- and $E_3$-sheet along the presented lines is cumbersome, this is in principle the way to construct these maps. Note in particular that in order to follow this principle, one needs knowledge about the $E_0$-sheet. This information cannot be obtained from \emph{cohomCalg} as this algorithm immediately computes the cohomology classes in the $E_1$-sheet. In order to extract this information one has to follow the classical chamber counting approach \cite{cox2011toric} which was also followed in \cite{Cvetic:2010rq}.

However, sometimes the maps in the sheets $E_2$ and $E_3$ can be constructed by simpler means. For example it was used in \cite{Anderson:2008ex} that the cohomology classes of holomorphic line bundles on $\mathbb{CP}^n$ are labeled by representations of $U \left( 1 \right) \times U \left( n \right)$ \cite{hubsch1994calabi} and that this extends to direct products of $\mathbb{CP}^n$s via the K\"unneth formula. The resulting (anti)-symmetrisation properties of the cohomology classes can then be used to construct the maps in the sheets $E_r$ with $r \geq 2$ more easily.

We have mentioned already that based on \emph{cohomCalg}, the computation of a basis of cohomology for the ambient space cohomologies is possible. Also the maps in the $E_1$-sheet are therefore accessible. Our approach is to use the existing C++ implementation of \emph{cohomCalg} whose option \emph{integrated} produces an output styled for use in \emph{Mathematica}. We have then written a \emph{Mathematica} notebook that computes the $E_1$-sheet and all maps therein. Details of the maps on the $E_1$-sheet are given in appendix \ref{app-E1}. If the spectral sequence converges on the $E_2$-sheet, this is sufficient to compute the cohomologies of the pullback bundle.

Note that our notebook represents the maps in the $E_1$-sheet as matrices whose non-zero entries are determined by the complex coefficients of the polynomials $\tilde{s}_i$. This is also demonstrated in appendix \ref{app-E1}. Therefore the functionality of the notebook is not limited to generic cases. Rather in leaving these coefficients unconstrained, the notebook provides all the information necessary to study the dependence of the pullback cohomologies on the complex structure of a hypersurface.
 
 \subsection{An explicit example \label{sec:AConcreteExample}}

As a concrete example for our base space $B_3$, we consider a hypersurface within the toric ambient space $X_{\Sigma} = \mathbb{CP}^2 \times \mathbb{CP}^1 \times \mathbb{CP}^1$ with ${K}_\Sigma = -3 H_1 - 2 H_2 - 2 H_3$. Here $H_i$, $i=1,2,3$ represent the respective hyperplane classes on the three factors.  In the sequel we will abbreviate a class $a H_1 + b H_2 + c H_3$ as $(a,b,c)$. 
The base $B_3$ is taken to be the vanishing locus of the bundle with class  
$D_{B_3} = \left( 1,2,1 \right)$
such that ${K}_{B_3} = (-2,0,-1)|_{B_3}$.
The $SU(5)$ divisor is in the class  $D_{\text{GUT}} = \left( 1,0,1 \right)|_{B_3}$.
%
%
Since the GUT divisor is not given by a toric divisor, we rewrite the base in terms of a complete intersection such that in the new description it is realised torically \cite{Collinucci:2010gz}. The variety is embedded in the ambient space with weight matrix
\begin{align}
\begin{array}{|c|c|c|c|c|c|c|c|c|}
\hline
x_1 & x_2 & x_3 & y_1 & y_2 & z_1 & z_2 &  \lambda_2& \lambda_1\\
\hline \hline
1 & 1 & 1 & 0 & 0 & 0 & 0 & 0 & 1 \\
\hline
0 & 0 & 0 & 1 & 1 & 0 & 0& 0 & 0 \\
\hline
0 & 0 & 0 & 0 & 0 & 1 & 1 & 0 & 1\\
\hline
0 & 0 & 0 & 0 & 0 & 0 & 0 & 1 & 1 \\\hline
\end{array}\label{eq:ci_base-weightmatrix}\,
\end{align}
and Stanley-Reisner ideal 
\bea
\textmd{SR-i}:\,\{y_1\,y_2,\, \lambda_1\,\lambda_2,\, z_1\,z_2, x_1\,x_2\,x_3\}\label{eq:ci_base-SR-I}\,.
\eea
The base is now the complete intersection of the `original' section
\begin{align}\label{eq:ci-equationI}
P_{(1,2,1,0)}(x_i,y_i,z_i)=0
\end{align} 
and a new equation\footnote{It is important to note that $\lambda_2=0$ does not solve \eqref{eq:ci-equationII}. Therefore, \eqref{eq:ci-equationII} defines a section in \eqref{eq:ci_base-weightmatrix}, i.e.\ in the $\mathbb P^1$ bundle over $X_\Sigma$.}
\begin{align}\label{eq:ci-equationII}
\lambda_2\,P^\textmd{GUT}_{(1,0,1,0)}(x_i,z_i)=\lambda_1\,.
\end{align}

With this description we can  torically construct the elliptically fibred Calabi-Yau 4-fold including the $SU(5)$ gauge theory along the GUT divisor $\{\lambda_1=0\}$. Since we are interested in the $U(1)$-restricted case \cite{Grimm:2010ez}, we take  polygon 11 of \cite{Bouchard:2003bu} as the toric ambient space for the fibre. To obtain the non-abelian gauge structure along $S_\textmd{GUT}$, we construct an $SU(5)$-top \cite{Bouchard:2003bu,Borchmann:2013hta} such that $\pi^{-1}(\lambda_1)=e_0\,e_1\,e_2\,e_3\,e_4$. The ambient sevenfold reads then as follows: 
\begin{align}
&\begin{array}{|c|c|c|c|c|c|c|c|c|c|c|c|c|c|c|c|c|}
\hline
y & x & z & s & x_1 & x_2 & x_3 & z_1 & z_2 & y_1 & y_2 &  \lambda_2& e_0 & e_4 & e_3 & e_2 & e_1\\
\hline \hline
0 & 1 & -2 & 1 & -4 & 0 & 0 & -2 & 0 & 0 & 0 & -2 & 0 & 1 & 2 & 2 & 1\\
1 & -1 & -1 & 0 & -2 & 0 & 0 & -1 & 0 & 0 & 0 & -1 & 0 & 1 & 1 & 0 & 0\\
0 & 0 & 0 & 0 & 1 & 0 & -1 & 0 & 0 & 0 & 0 & 0 & 0 & 0 & 0 & 0 & 0\\
0 & 0 & 0 & 0 & 0 & 1 & -1 & 0 & 0 & 0 & 0 & 0 & 0 & 0 & 0 & 0 & 0\\
0 & 0 & 0 & 0 & 0 & 0 & 0 & 0 & 0 & 1 & -1 & 0 & 0 & 0 & 0 & 0 & 0\\
0 & 0 & 0 & 0 & 0 & 0 & 1 & 0 & 1 & 0 & 0 & 1 & -1 & -1 & -1 & -1 & -1\\
0 & 0 & 0 & 0 & 0 & 0 & 1 & 1 & 0 & 0 & 0 & 1 & -1 & -1 & -1 & -1 & -1\\
\hline
\end{array}\\
& \textmd{SR-i}:\,\{y\,x,\, y\,e_2,\, y\,e_1,\, z\,s,\, z\,e_4,\, z\,e_3,\, z\,e_2,\, z\,e_1,\, y_1\,y_2,\, \lambda_2\,e_0,\,
\lambda_2\,e_4,\, \lambda_2\,e_3,\, \lambda_2\,e_2, \nonumber \\
&\qquad\qquad \lambda_2\,e_1,\, z_1\,z_2,\, s\,e_0,\, e_0\,e_2,\, s\,e_4,\, s\,e_2,\, s\,e_1,\, x\,e_4,\,
e_4\,e_2,\, e_4\,e_1,\, x_1\,x_2\,x_3,\\
&\qquad\qquad y\,e_0\,e_3,\, x\,e_0\,e_3,\, x\,e_3\,e_1\} \nonumber
\end{align}
Together with \eqref{eq:ci-equationI}, \eqref{eq:ci-equationI} and \eqref{tate_proper_transform} this defines the elliptically fibred Calabi-Yau fourfold with the sought-after gauge-symmetries as a complete intersection in a 7-dimensional toric ambient variety. The Euler number of this Calabi-Yau is $\chi(\hat Y_4)=960$.

We now wish to compute the spectrum of $U \left( 1 \right)_X$ charged zero modes that are localised on the curves $C_{\bf 10}$, $C_{\bf \bar 5_m}$, $C_{\bf 5_H}$ and $C_{\bf 1}$ described generally in (\ref{mattercurves1}). In the present geometry, they are given as complete intersections on $X_\Sigma$ of sections with classes
\bea
 C_{\bf  10} &=& \left( 1,2,1 \right) \cap \left( 1,0,1 \right) \cap \left( 2,0,1 \right), \qquad 
			C_{ \bf \bar {5}_m} = \left( 1,2,1 \right) \cap \left( 1,0,1 \right) \cap \left( 4,0,1 \right), \\
C_{\bf 5_H}&=& \left( 1,2,1 \right) \cap \left( 1,0,1 \right) \cap \left( 7,0,2 \right), \qquad 
			C_{\bf 1} = \left( 1,2,1 \right) \cap \left( 5,0,1 \right) \cap \left( 4, 0, 1 \right).
\eea

We consider a gauge configuration of the form (\ref{alphaXa}) with $f = \tilde f|_{B_3}$ for $\tilde f \in \CH^1(X_\Sigma)$.
Since $X_\Sigma$ is a toric variety, rational equivalence and homological equivalence happen to coincide, and $\tilde f$ is uniquely characterized by its cohomology class $\gamma(\tilde f) \in H^{1,1}(X_\Sigma)$.
Whilst we leave it for future work to scan over all admissable choices of classes $\gamma(\tilde f) \in H^{1,1}(X_\Sigma)$,
 we here pick a specific  class $\tilde f$ which is particularly suited to illustrate the general philosophy. Our choice is 
\[ \gamma(\tilde f) = \frac{1}{2} \left( -7 ,0 ,9 \right), \]
which is indeed in agreement with the quantization condition (\ref{quantex1}). 
Note that this choice is motivated such as to produce three families of chiral matter; the 4-form flux $G_4 = \gamma(\tilde f)$, however, overshoots the D3-brane tadpole cancellation condition. This has no impact whatsoever on our computation so that our setup suffices as a toy model to illustrate the computation of the massless spectrum in a fully-fledged 4-fold geometry.

According to (\ref{cohom-bundles}) the number of chiral- and anti-chiral modes charged under the $U \left( 1 \right)_X$ symmetry and located on the respective curves are encoded in the cohomologies of the following line bundles:
\begin{align}
{\bf 10} &\leftrightarrow H^i(C_{\bf 10}, {\cal L}_1|_{C_{\bf 10}}), & \mathcal{L}_1 &= \mathcal{O}_{X_\Sigma} \left( 4,0, -4 \right), \\
{\bf \bar 5_m} &\leftrightarrow H^i(C_{\bf \bar 5_m},  {\cal L}_2 |_{C_{\bf \bar 5_m}}), & \mathcal{L}_2 &= \mathcal{O}_{X_\Sigma} \left( -9,0, 14 \right), \\
{\bf 5_H}& \leftrightarrow H^i(C_{\bf  5_H}, {\cal L}_3|_{C_{\bf  5_H}}) , & \mathcal{L}_3 &= \mathcal{O}_{X_\Sigma} \left( 4,0,10 \right), \\
{\bf 1} &\leftrightarrow H^i(C_{\bf 1},  {\cal L}_4|_{C_{\bf  1}}), & \mathcal{L}_4 &= \mathcal{O}_{X_\Sigma} \left( -14, 0, 23 \right).
 \end{align}

\begin{figure}[tb]
\subfloat[The $E_1$-sheet.]{
\begin{tikzpicture}[every node/.style={anchor=base,text height=.8em, text depth=.2em,minimum size=7mm},scale=0.8]

\matrix (m) [matrix of nodes, row sep= 1.5em, column sep=1em, column 2/.style={anchor= base west}, ampersand replacement=\&]
    {  
       $\mathcal{L}^\prime$ \& 0   \& 0   \& 6   \& 0   \& 0 \\
		   $\mathcal{V}_2$      \& 0   \& 15  \& 45  \& 0   \& 0 \\
			 $\mathcal{V}_1$      \& 0   \& 64  \& 40  \& 0   \& 0 \\
			 $\mathcal{L}_1$      \& 0   \& 45  \& 0   \& 0   \& 0 \\
		                        \& $H^0$ \& $H^1$ \& $H^2$ \& $H^3$ \& $H^4$ \\
		};

\draw[color=black, thick] ([xshift=-0.5em]m-1-2.north west)--([xshift=-0.5em]m-5-2.south west);
\draw[color=black, thick] ([yshift=-0.75em]m-4-1.south west)--([yshift=-0.75em]m-4-6.south east);
\draw[color=black, thick] ([yshift=-0.95em]m-4-1.south west)--([yshift=-0.95em]m-4-6.south east);

\draw[->,thick] (m-1-1)-- (m-2-1);
\draw[->,thick] (m-2-1)-- (m-3-1);
\draw[->,thick] (m-3-1)-- (m-4-1);

\draw[->,thick] (m-1-2)-- (m-2-2);
\draw[->,thick] (m-2-2)-- (m-3-2);
\draw[->,thick] (m-3-2)-- (m-4-2);

\draw[->,thick] (m-1-3)-- (m-2-3);
\draw[->,thick] (m-2-3)-- (m-3-3);
\draw[->,thick] (m-3-3)-- (m-4-3);

\draw[->,thick] (m-1-4)-- (m-2-4);
\draw[->,thick] (m-2-4)-- (m-3-4);
\draw[->,thick] (m-3-4)-- (m-4-4);

\draw[->,thick] (m-1-5)-- (m-2-5);
\draw[->,thick] (m-2-5)-- (m-3-5);
\draw[->,thick] (m-3-5)-- (m-4-5);

\draw[->,thick] (m-1-6)-- (m-2-6);
\draw[->,thick] (m-2-6)-- (m-3-6);
\draw[->,thick] (m-3-6)-- (m-4-6);

\end{tikzpicture}
}
\qquad
\subfloat[The $E_2$-sheet.]{

\begin{tikzpicture}[every node/.style={anchor=base,text height=.8em, text depth=.2em,minimum size=7mm},scale=0.8]

\matrix (m) [matrix of nodes, row sep= 1.5em, column sep=1em, column 2/.style={anchor= base west}, ampersand replacement=\&]
    { 
       $\mathcal{L}^\prime$ \& 0   \& 0   \& 0   \& 0   \& 0 \\
		   $\mathcal{V}_2$      \& 0   \& 0   \& 0   \& 0   \& 0 \\
			 $\mathcal{V}_1$      \& 0   \& 4   \& 1   \& 0   \& 0 \\
			 $\mathcal{L}_1$      \& 0   \& 0   \& 0   \& 0   \& 0 \\
		                        \& $H^0$ \& $H^1$ \& $H^2$ \& $H^3$ \& $H^4$ \\
		};

\draw[color=black, thick] ([xshift=-0.5em]m-1-2.north west)--([xshift=-0.5em]m-5-2.south west);
\draw[color=black, thick] ([yshift=-0.75em]m-4-1.south west)--([yshift=-0.75em]m-4-6.south east);
\draw[color=black, thick] ([yshift=-0.95em]m-4-1.south west)--([yshift=-0.95em]m-4-6.south east);

\draw[->,thick] (m-1-1)-- (m-2-1);
\draw[->,thick] (m-2-1)-- (m-3-1);
\draw[->,thick] (m-3-1)-- (m-4-1);

\draw[->,thick] (m-1-3)-- (m-3-2);
\draw[->,thick] (m-2-3)-- (m-4-2);

\draw[->,thick] (m-1-4)-- (m-3-3);
\draw[->,thick] (m-2-4)-- (m-4-3);

\draw[->,thick] (m-1-5)-- (m-3-4);
\draw[->,thick] (m-2-5)-- (m-4-4);

\draw[->,thick] (m-1-6)-- (m-3-5);
\draw[->,thick] (m-2-6)-- (m-4-5);

\end{tikzpicture}

}
\caption{Computation of the cohomologies $h^i \left( {C_{\bf 10}} , \mathcal{L}_1 \right)$.}
\label{figure1}
\end{figure}
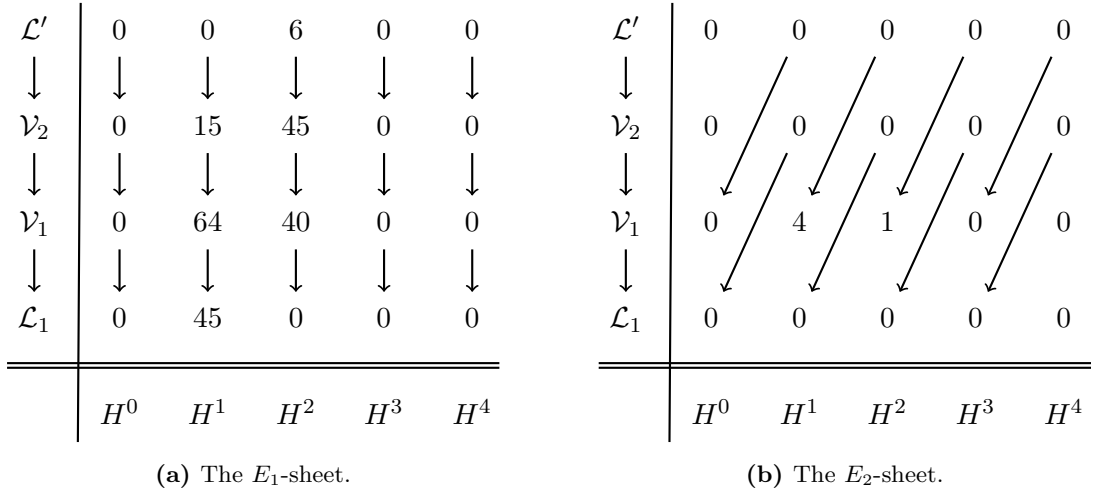

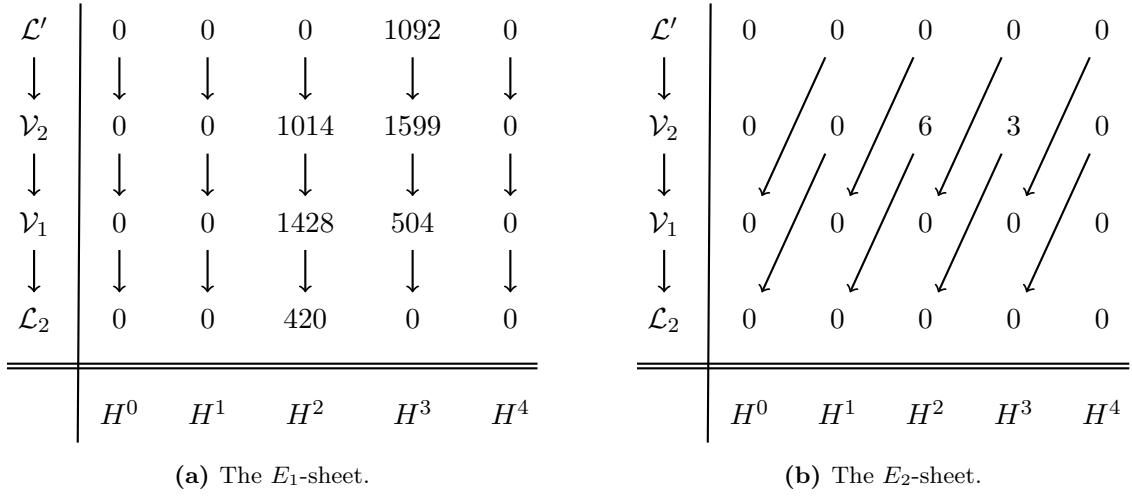
\begin{figure}[tb]
\subfloat[The $E_1$-sheet.]{

\begin{tikzpicture}[every node/.style={anchor=base,text height=.8em, text depth=.2em,minimum size=7mm},scale=0.8]

\matrix (m) [matrix of nodes, row sep= 1.5em, column sep=1em, column 2/.style={anchor= base west}, ampersand replacement=\&]
    { 
       $\mathcal{L}^\prime$ \& 0   \& 0  \& 0     \& 1092   \& 0 \\
		   $\mathcal{V}_2$      \& 0   \& 0  \& 1014  \& 1599   \& 0 \\
			 $\mathcal{V}_1$      \& 0   \& 0  \& 1428  \& 504    \& 0 \\
			 $\mathcal{L}_2$      \& 0   \& 0  \& 420   \& 0      \& 0 \\
		                        \& $H^0$ \& $H^1$ \& $H^2$ \& $H^3$ \& $H^4$ \\
		};

\draw[color=black, thick] ([xshift=-0.5em]m-1-2.north west)--([xshift=-0.5em]m-5-2.south west);
\draw[color=black, thick] ([yshift=-0.75em]m-4-1.south west)--([yshift=-0.75em]m-4-6.south east);
\draw[color=black, thick] ([yshift=-0.95em]m-4-1.south west)--([yshift=-0.95em]m-4-6.south east);

\draw[->,thick] (m-1-1)-- (m-2-1);
\draw[->,thick] (m-2-1)-- (m-3-1);
\draw[->,thick] (m-3-1)-- (m-4-1);

\draw[->,thick] (m-1-2)-- (m-2-2);
\draw[->,thick] (m-2-2)-- (m-3-2);
\draw[->,thick] (m-3-2)-- (m-4-2);

\draw[->,thick] (m-1-3)-- (m-2-3);
\draw[->,thick] (m-2-3)-- (m-3-3);
\draw[->,thick] (m-3-3)-- (m-4-3);

\draw[->,thick] (m-1-4)-- (m-2-4);
\draw[->,thick] (m-2-4)-- (m-3-4);
\draw[->,thick] (m-3-4)-- (m-4-4);

\draw[->,thick] (m-1-5)-- (m-2-5);
\draw[->,thick] (m-2-5)-- (m-3-5);
\draw[->,thick] (m-3-5)-- (m-4-5);

\draw[->,thick] (m-1-6)-- (m-2-6);
\draw[->,thick] (m-2-6)-- (m-3-6);
\draw[->,thick] (m-3-6)-- (m-4-6);

\end{tikzpicture}

}
\qquad
\subfloat[The $E_2$-sheet.]{

\begin{tikzpicture}[every node/.style={anchor=base,text height=.8em, text depth=.2em,minimum size=7mm},scale=0.8]

\matrix (m) [matrix of nodes, row sep= 1.5em, column sep=1em, column 2/.style={anchor= base west}, ampersand replacement=\&]
    { 
       $\mathcal{L}^\prime$ \& 0   \& 0   \& 0   \& 0   \& 0 \\
		   $\mathcal{V}_2$      \& 0   \& 0   \& 6   \& 3   \& 0 \\
			 $\mathcal{V}_1$      \& 0   \& 0   \& 0   \& 0   \& 0 \\
			 $\mathcal{L}_2$      \& 0   \& 0   \& 0   \& 0   \& 0 \\
		                        \& $H^0$ \& $H^1$ \& $H^2$ \& $H^3$ \& $H^4$ \\
		};

\draw[color=black, thick] ([xshift=-0.5em]m-1-2.north west)--([xshift=-0.5em]m-5-2.south west);
\draw[color=black, thick] ([yshift=-0.75em]m-4-1.south west)--([yshift=-0.75em]m-4-6.south east);
\draw[color=black, thick] ([yshift=-0.95em]m-4-1.south west)--([yshift=-0.95em]m-4-6.south east);

\draw[->,thick] (m-1-1)-- (m-2-1);
\draw[->,thick] (m-2-1)-- (m-3-1);
\draw[->,thick] (m-3-1)-- (m-4-1);

\draw[->,thick] (m-1-3)-- (m-3-2);
\draw[->,thick] (m-2-3)-- (m-4-2);

\draw[->,thick] (m-1-4)-- (m-3-3);
\draw[->,thick] (m-2-4)-- (m-4-3);

\draw[->,thick] (m-1-5)-- (m-3-4);
\draw[->,thick] (m-2-5)-- (m-4-4);

\draw[->,thick] (m-1-6)-- (m-3-5);
\draw[->,thick] (m-2-6)-- (m-4-5);

\end{tikzpicture}

}
\caption{Computation of the cohomologies $h^i \left(C_{\bf \bar 5_m} , \mathcal{L}_2 \right)$.}
\label{figure2}
\end{figure}

Note that a proper definition of the model requires explicitly specifying the holomorphic section
$Q \in H^0(X_\Sigma, B_3)$ defining the base $B_3$ via (\ref{Qpoly}) as well as the Tate polynomials $a_{i,j}$ and the GUT brane normal section $w$, which enter the definition of the matter curves via (\ref{mattercurves1}).
If we view these sections as polynomials in the homogenous coordinates of $X_\Sigma$, the coefficients of the various monomials give an (in general redundant) description of the complex structure moduli of $B_3$ and the 7-brane moduli.
Therefore these coefficients have an influence on the cohomology groups of ${L}_R |_{C_R}$ appearing in (\ref{cohom-bundles}). Our choice is to pick pseudo-random numbers in the parameter range $\left( 0, 1 \right)$ for all these coefficients. Note in particular that none of the coefficients vanishes. This setup is often referred to as 'maximally generic'. \\
For such a situation, we have automatised the computation of the $E_1$-sheet of the Koszul spectral sequence in a \emph{Mathematica} notebook. The computed $E_1$-sheets are displayed in \autoref{figure1}, \autoref{figure2}, \autoref{figure3} and \autoref{figure4}.
 
\begin{figure}[tb]
\subfloat[The $E_1$-sheet.]{

\begin{tikzpicture}[every node/.style={anchor=base,text height=.8em, text depth=.2em,minimum size=7mm},scale=0.8]

\matrix (m) [matrix of nodes, row sep= 1.5em, column sep=1em, column 2/.style={anchor= base west}, ampersand replacement=\&]
    { 
       $\mathcal{L}^\prime$ \& 0   \& 0  \& 0    \& 462   \& 0 \\
		   $\mathcal{V}_2$      \& 0   \& 0  \& 440  \& 530   \& 0 \\
			 $\mathcal{V}_1$      \& 0   \& 0  \& 465  \& 60    \& 0 \\
			 $\mathcal{L}_3$      \& 0   \& 0  \& 33   \& 0     \& 0 \\
		                        \& $H^0$ \& $H^1$ \& $H^2$ \& $H^3$ \& $H^4$ \\
		};

\draw[color=black, thick] ([xshift=-0.5em]m-1-2.north west)--([xshift=-0.5em]m-5-2.south west);
\draw[color=black, thick] ([yshift=-0.75em]m-4-1.south west)--([yshift=-0.75em]m-4-6.south east);
\draw[color=black, thick] ([yshift=-0.95em]m-4-1.south west)--([yshift=-0.95em]m-4-6.south east);

\draw[->,thick] (m-1-1)-- (m-2-1);
\draw[->,thick] (m-2-1)-- (m-3-1);
\draw[->,thick] (m-3-1)-- (m-4-1);

\draw[->,thick] (m-1-2)-- (m-2-2);
\draw[->,thick] (m-2-2)-- (m-3-2);
\draw[->,thick] (m-3-2)-- (m-4-2);

\draw[->,thick] (m-1-3)-- (m-2-3);
\draw[->,thick] (m-2-3)-- (m-3-3);
\draw[->,thick] (m-3-3)-- (m-4-3);

\draw[->,thick] (m-1-4)-- (m-2-4);
\draw[->,thick] (m-2-4)-- (m-3-4);
\draw[->,thick] (m-3-4)-- (m-4-4);

\draw[->,thick] (m-1-5)-- (m-2-5);
\draw[->,thick] (m-2-5)-- (m-3-5);
\draw[->,thick] (m-3-5)-- (m-4-5);

\draw[->,thick] (m-1-6)-- (m-2-6);
\draw[->,thick] (m-2-6)-- (m-3-6);
\draw[->,thick] (m-3-6)-- (m-4-6);

\end{tikzpicture}

}
\qquad
\subfloat[The $E_2$-sheet.]{

\begin{tikzpicture}[every node/.style={anchor=base,text height=.8em, text depth=.2em,minimum size=7mm},scale=0.8]

\matrix (m) [matrix of nodes, row sep= 1.5em, column sep=1em, column 2/.style={anchor= base west}, ampersand replacement=\&]
    { 
       $\mathcal{L}^\prime$ \& 0   \& 0   \& 0   \& 0   \& 0 \\
		   $\mathcal{V}_2$      \& 0   \& 0   \& 9   \& 8   \& 0 \\
			 $\mathcal{V}_1$      \& 0   \& 0   \& 1   \& 0   \& 0 \\
			 $\mathcal{L}_3$      \& 0   \& 0   \& 0   \& 0   \& 0 \\
		                        \& $H^0$ \& $H^1$ \& $H^2$ \& $H^3$ \& $H^4$ \\
		};

\draw[color=black, thick] ([xshift=-0.5em]m-1-2.north west)--([xshift=-0.5em]m-5-2.south west);
\draw[color=black, thick] ([yshift=-0.75em]m-4-1.south west)--([yshift=-0.75em]m-4-6.south east);
\draw[color=black, thick] ([yshift=-0.95em]m-4-1.south west)--([yshift=-0.95em]m-4-6.south east);

\draw[->,thick] (m-1-1)-- (m-2-1);
\draw[->,thick] (m-2-1)-- (m-3-1);
\draw[->,thick] (m-3-1)-- (m-4-1);

\draw[->,thick] (m-1-3)-- (m-3-2);
\draw[->,thick] (m-2-3)-- (m-4-2);

\draw[->,thick] (m-1-4)-- (m-3-3);
\draw[->,thick] (m-2-4)-- (m-4-3);

\draw[->,thick] (m-1-5)-- (m-3-4);
\draw[->,thick] (m-2-5)-- (m-4-4);

\draw[->,thick] (m-1-6)-- (m-3-5);
\draw[->,thick] (m-2-6)-- (m-4-5);

\end{tikzpicture}

}
\caption{Computation of the cohomologies $h^i \left(C_{\bf  5_H} , \mathcal{L}_3 \right)$.}
\label{figure3}
\end{figure}

\begin{figure}[tb]
\subfloat[The $E_1$-sheet.]{

\begin{tikzpicture}[every node/.style={anchor=base,text height=.8em, text depth=.2em,minimum size=7mm},scale=0.8]

\matrix (m) [matrix of nodes, row sep= 1.5em, column sep=1em, column 2/.style={anchor= base west}, ampersand replacement=\&]
    { 
       $\mathcal{L}^\prime$ \& 0   \& 0  \& 0     \& 5313   \& 0 \\
		   $\mathcal{V}_2$      \& 0   \& 0  \& 5082  \& 7128   \& 0 \\
			 $\mathcal{V}_1$      \& 0   \& 0  \& 6647  \& 2093   \& 0 \\
			 $\mathcal{L}_4$      \& 0   \& 0  \& 1872  \& 0      \& 0 \\
		                        \& $H^0$ \& $H^1$ \& $H^2$ \& $H^3$ \& $H^4$ \\
		};

\draw[color=black, thick] ([xshift=-0.5em]m-1-2.north west)--([xshift=-0.5em]m-5-2.south west);
\draw[color=black, thick] ([yshift=-0.75em]m-4-1.south west)--([yshift=-0.75em]m-4-6.south east);
\draw[color=black, thick] ([yshift=-0.95em]m-4-1.south west)--([yshift=-0.95em]m-4-6.south east);

\draw[->,thick] (m-1-1)-- (m-2-1);
\draw[->,thick] (m-2-1)-- (m-3-1);
\draw[->,thick] (m-3-1)-- (m-4-1);

\draw[->,thick] (m-1-2)-- (m-2-2);
\draw[->,thick] (m-2-2)-- (m-3-2);
\draw[->,thick] (m-3-2)-- (m-4-2);

\draw[->,thick] (m-1-3)-- (m-2-3);
\draw[->,thick] (m-2-3)-- (m-3-3);
\draw[->,thick] (m-3-3)-- (m-4-3);

\draw[->,thick] (m-1-4)-- (m-2-4);
\draw[->,thick] (m-2-4)-- (m-3-4);
\draw[->,thick] (m-3-4)-- (m-4-4);

\draw[->,thick] (m-1-5)-- (m-2-5);
\draw[->,thick] (m-2-5)-- (m-3-5);
\draw[->,thick] (m-3-5)-- (m-4-5);

\draw[->,thick] (m-1-6)-- (m-2-6);
\draw[->,thick] (m-2-6)-- (m-3-6);
\draw[->,thick] (m-3-6)-- (m-4-6);

\end{tikzpicture}

}
\qquad
\subfloat[The $E_2$-sheet.]{

\begin{tikzpicture}[every node/.style={anchor=base,text height=.8em, text depth=.2em,minimum size=7mm},scale=0.8]

\matrix (m) [matrix of nodes, row sep= 1.5em, column sep=1em, column 2/.style={anchor= base west}, ampersand replacement=\&]
    { 
       $\mathcal{L}^\prime$ \& 0   \& 0   \& 0   \& 278 \& 0 \\
		   $\mathcal{V}_2$      \& 0   \& 0   \& 307 \& 0   \& 0 \\
			 $\mathcal{V}_1$      \& 0   \& 0   \& 0   \& 0   \& 0 \\
			 $\mathcal{L}_4$      \& 0   \& 0   \& 0   \& 0   \& 0 \\
		                        \& $H^0$ \& $H^1$ \& $H^2$ \& $H^3$ \& $H^4$ \\
		};

\draw[color=black, thick] ([xshift=-0.5em]m-1-2.north west)--([xshift=-0.5em]m-5-2.south west);
\draw[color=black, thick] ([yshift=-0.75em]m-4-1.south west)--([yshift=-0.75em]m-4-6.south east);
\draw[color=black, thick] ([yshift=-0.95em]m-4-1.south west)--([yshift=-0.95em]m-4-6.south east);

\draw[->,thick] (m-1-1)-- (m-2-1);
\draw[->,thick] (m-2-1)-- (m-3-1);
\draw[->,thick] (m-3-1)-- (m-4-1);

\draw[->,thick] (m-1-3)-- (m-3-2);
\draw[->,thick] (m-2-3)-- (m-4-2);

\draw[->,thick] (m-1-4)-- (m-3-3);
\draw[->,thick] (m-2-4)-- (m-4-3);

\draw[->,thick] (m-1-5)-- (m-3-4);
\draw[->,thick] (m-2-5)-- (m-4-4);

\draw[->,thick] (m-1-6)-- (m-3-5);
\draw[->,thick] (m-2-6)-- (m-4-5);

\end{tikzpicture}

}
\caption{Computation of the cohomologies $h^i \left(C_{\bf 1} , \mathcal{L}_4 \right)$.}
\label{figure4}
\end{figure}
In particular this shows that all four spectral sequences converge on the $E_2$-sheet. We display the so-obtained spectrum of chiral multiplets in table (\ref{table1}).
\begin{table}[tb]
\begin{center}
\begin{tabular}{|c||c|c|c|c|}
\toprule
curve & \multicolumn{2}{|c|}{$H^0 \left( C, \left. \cal L \right|_C \right)$} & \multicolumn{2}{|c|}{$H^1 \left( C, \left. \cal L \right|_C \right)$} \\
& $h^0 \left( C, \left. \cal L \right|_C \right)$ & representation & $h^1 \left( C, \left. \cal L \right|_C \right)$ & representation \\
\hline \hline
$C_{\bf 10}$ & 4 & $\bf{10}_{-1}$ & 1 & $\overline{\bf{10}}_{\bf +1}$ \\
$C_{\bf \bar 5_m}$ & 6 & $\overline{\bf{5}}_{\bf 3}$ & 3 & $\bf{5}_{\bf -3}$ \\
$C_{\bf 5_H}$ & 9 & $\bf{5}_2$ & 9 & $\overline{\bf{5}}_{\bf -2}$ \\
$C_{\bf{1}}$ & 585 & $\bf{1}_5$ & 0 & $\overline{\bf{1}}_{\bf -5}$ \\
\bottomrule
\end{tabular}
\end{center}
\caption{The chiral $\mathcal{N} = 1$ multiplets in the example presented in subsection \ref{sec:AConcreteExample}.}
\label{table1}
\end{table}
We find 4 chiral $\mathcal{N} = 1$ multiplets in the representation $\bf{10}_{-1}$ localised on $C_{\bf{10}}$. In addition there is one chiral $\mathcal{N} = 1$ multiplet in the $\overline{\bf{10}}_{1}$ representation on $C_{\bf 10}$. Similarly there are 6 multiplets in representation $\overline{\bf{5}}_3$ and 3 in the representation $\bf{5}_{-3}$ localised on $C_{\bf \bar 5_m}$. This gives us 3 chiral families. On $C_{\bf 5_H}$ we have 9 Higgs like doublets.

\section{Conclusions and Outlook} \label{sec_Concl.}

In this work we have described a method to extract  the number of localised charged massless ${\cal N}=1$ chiral multiplets in an F-theory compactification on an elliptic 4-fold $\hat Y_4$.
Our starting point has been to specify the 3-form data by means of a rational equivalence class of algebraic complex dimension-two cycles $\alpha \in \CH^2(\hat Y_4)$.
The refined cycle map assigns to $\alpha$ an element of the Deligne cohomology group $H^4_D(\hat Y_4, \mathbb Z(2))$, which is known \cite{Curio:1998bva,Donagi:1998vw} to correctly account for the 3-form data in F/M-theory. Working at the level of Chow groups has the practical advantage that it allows us to very concretely specify the 3-form data.
In particular we have described how the intersection product within the Chow ring of $\hat Y_4$ can be used to extract from this data a line bundle $L_R$ on the matter curve $C_R$ which is the natural object counting the number of massless chiral multiplets via $H^i(C_R, L_R \otimes \sqrt{K_{C_R}})$. Our approach is general and does not rely on any local approximation. 

Extracting the line bundle $L_R$ on $C_R$ is most immediate if $\hat Y_4$ is embedded into a toric ambient space $X_\Sigma$ and the 3-form data $\alpha$ arises essentially by pullback from $\CH^2(X_\Sigma)$. In this case,  the equivalence of the Chow and the Dolbeault cohomology rings on the toric variety $X_\Sigma$ implies that we can work within cohomology. We have exemplified this procedure for the 3-form data associated with a $U(1)$ gauge flux and explicitly computed the number of massless charged matter states in an $SU(5) \times U(1)$ toy model. 

The obvious next step would be to apply the proposed procedure to a 3-form configuration not of the pullback type focused on in the example of the present paper.
This would require carrying out the intersection computation to extract the line bundle $L_R$ directly within the Chow ring of $\hat Y_4$. An example of such an F-theory 4-form gauge flux has been introduced in \cite{Braun:2011zm}.  It would be interesting to use our formalism to extend the homological computations of the flux performed in \cite{Braun:2011zm} to the level of the underlying 3-form data.
Relatedly, the examples considered in the present paper have the simplifying property that they do not induce a Gukow-Vafa-Witten superpotential in the low-energy effective action. In particular, they do not obstruct any of the complex structure moduli of the fourfold. The rich relation between the number of charged localised zero modes and the critical points of this superpotential has been explored in \cite{oai:arXiv.org:1203.6662} in the context of a representation of the 3-form data via Cheeger-Simons differential forms. A task left for future work is to address configurations with superpotential inducing fluxes in our framework.

Throughout this article we have worked with a smooth resolution $\hat Y_4$ of the singular elliptic fibration. As an advantage, this allows us to study 3-form data within the context of the Deligne cohomology of a \emph{smooth} 4-fold. On the other hand, since the resolution corresponds to moving in the Coulomb branch of the associated 3-dimensional effective action of M-theory compactified on  $\hat Y_4$, the resolved geometry necessarily misses inherently non-abelian aspects of the gauge dynamics. For example the configurations considered here do not include any non-trivial gluing data \cite{Cecotti:2010bp,Donagi:2011jy,Donagi:2011dv}.\footnote{See however \cite{Braun:2013cb} for the effect of gluing-related monodromies on the resolution geometry.}  Interestingly, \cite{Anderson:2013rka} provides evidence for a definition of the Deligne cohomology of  a \emph{singular} elliptic fibration via the Higgs bundle which at least locally captures the non-abelian degrees of freedom on the 7-branes \cite{oai:arXiv.org:0802.2969,oai:arXiv.org:0802.3391,oai:arXiv.org:0904.1218}. 
An alternative treatment could be to work not in the resolved, but in the deformed geometry as studied recently in \cite{Grassi:2013kha}.

Despite these slightly more formal considerations to be explored further, we would like to stress that the orignal motivation for this work was phenomenological: The computation of the number of charged massless vectorlike pairs (in addition to the chiral index) is of obvious importance in model building.
The biggest advantage of our approach is therefore its direct applicability to a sufficiently rich set of 3-form data which are employed the explicit construction of realistic string vacua, e.g. of GUT type. 
Indeed a great deal of work has been invested in the computation of the massless spectrum of realistic heterotic compactifications with cohomological methods  - for an incomplete list see e.g.
 \cite{Donagi:2004ia,Donagi:2004ub,Bouchard:2005ag,Braun:2005zv,Blumenhagen:2006wj,Anderson:2011ns,Anderson:2012yf} and references therein.
In this spirit we look forward to putting Chow groups and Deligne cohomology to work for phenomenological F-theory model building in the near future.

\subsection*{Acknowledgements}

We would like to thank Lara Anderson, Hans Jockers, Luca Martucci, Dave Morrison and Eran Palti for useful discussions and correspondence.
This work was supported in part by Transregio TR 33 'The Dark Universe' and by the Partnership Mathematics and Physics at Heidelberg University.

\appendix

\section{Details on the $E_1$-sheet} \label{app-E1}
In this appendix to section \ref{sub_cohom} we explain in more detail the construction of the $d_1$-maps appearing on the $E_1$-sheet. To state the steps to be performed let us consider the map $\tilde{\alpha}^2 \colon H^2 \left( X_\Sigma, \mathcal{L}^\prime \right) \to H^2 \left( X_\Sigma, \mathcal{V}_2 \right)$, which we construct by performing the following steps.
\begin{enumerate}
       \item Determine the rationom basis of $H^2 \left( X_\Sigma, \mathcal{L}^\prime \right)$ and $H^2 \left( X_\Sigma, \mathcal{V}_2 \right)$ based on \emph{cohomCalg}.
			 \item Multiply the general element $x$ of $H^2 \left( X_\Sigma, \mathcal{L}^\prime \right)$ by the matrix
			      \bea
						    \alpha = \left( \begin{array}{c} \tilde{s}_1 \\ - \tilde{s}_2 \\ \tilde{s}_3 \end{array} \right).
						\eea
			 \item The part of $\alpha \cdot x$ not expressable as a linear combination over $\mathbb{C}$ of the basis of $H^2 \left( X_\Sigma, \mathcal{V}_2 \right)$ is 
			      cohomologically zero. Therefore this part can be dropped.
			 \item Finally express the remaining polynomial vector as a linear combination over $\mathbb{C}$ of the basis of $H^2 \left( X_\Sigma, \mathcal{V}_2 \right)$ and 
			      thereby identify the mapping matrix.
\end{enumerate}
Let us exemplify this strategy by considering the hypersurface $C$ in $\mathbb{CP}^2$ given by
\bea 
C := \left\{ \left[ x_1 , x_2 , x_3 \right] \in \mathbb{CP}^2 \; ,\; \tilde{s} \left( x_1, x_2, x_3 \right) := C_1 x_1 + C_2 x_2 + C_3 x_3 = 0 \right\}. 
\eea
We now wish to compute the pullback cohomologies of the holomorphic line bundle $\mathcal{O}_{\mathbb{CP}^2} \left( -3 \right)$ onto this hypersurface $C$. To this end we make use of the sheaf exact Koszul sequence
\bea 
0 \to \underbrace{\mathcal{O}_{\mathbb{CP}^2} \left( - 4 \right)}_{= \mathcal{L}^\prime} \stackrel{\otimes \tilde{s}}{\rightarrow} \underbrace{\mathcal{O}_{\mathbb{CP}^2} \left( -3 \right) }_{= \mathcal{L}} \stackrel{\text{r}}{\rightarrow} \left. \mathcal{O}_{\mathbb{CP}^2} \left( -3 \right) \right|_C \to 0. \label{equ:Koszul2} 
\eea
and display the $E_1$-sheet of the associated Koszul spectral sequence in \autoref{figure20}. From this picture it is immediately clear that the spectral sequence does converge on the $E_2$-sheet and that we need only knowledge about the map
\bea 
\tilde{\alpha}^2 \colon P_1 \to P_2
\eea
in order to compute the pullback cohomologies.
\begin{figure}[tb]
\begin{center}

\begin{tikzpicture}[every node/.style={anchor=center}]

\matrix (m) [matrix of nodes, row sep= 1.5em, column sep=1.5em, column 1/.style={nodes={minimum width = 3em},column sep = 0em}]
    { 
       $\mathcal{L}^\prime$ & $0$ & $0$ & $P_1$ \\
 			 $\mathcal{L}$ \vphantom{$P_2$} & $0$ & $0$ & $P_2$ \\
		  \hphantom{$\mathcal{L}^\prime$} & $H^0$ & $H^1$ & $H^2$ \\
		};

\draw[color=black, thick] ([xshift=-0.5em]m-1-1.north east)--([xshift=-0.5em]m-3-1.south east);
\draw[color=black, thick] ([yshift=-0.75em]m-2-1.south west)--([yshift=-0.75em]m-2-4.south east);
\draw[color=black, thick] ([yshift=-0.95em]m-2-1.south west)--([yshift=-0.95em]m-2-4.south east);

\draw[->,thick] (m-1-1)-- (m-2-1);

\path[->,thick] (m-1-2) edge (m-2-2) edge node[right] {$\tilde{\alpha}^0$} (m-2-2);
\path[->,thick] (m-1-3) edge (m-2-3) edge node[right] {$\tilde{\alpha}^1$} (m-2-3);
\path[->,thick] (m-1-4) edge (m-2-4) edge node[right] {$\tilde{\alpha}^2$} (m-2-4);

\end{tikzpicture}
\end{center}
\caption{The $E_1$-sheet of the Koszul spectral sequence for the hypersurface example.}
\label{figure20}
\end{figure}
The spaces $P_1$ and $P_2$ are computed based on \emph{cohomCalg} to be
\bea 
P_1 &=& \left\{ A_1 \left[ \frac{1}{x_1 x_2 x_3^2} \right] + A_2 \left[ \frac{1}{x_1 x_2^2 x_3} \right] + A_3 \left[ \frac{1}{x_1^2 x_2 x_3} \right] \; , \; A_i \in \mathbb{C} \right\} \cong \mathbb{C}^3, \\
P_2 &=& \left\{ B \left[ \frac{1}{x_1 x_2 x_3} \right] \; , \; B \in \mathbb{C} \right\} \cong \mathbb{C}, \label{equ:RationomSpaceP2}
\eea
where $\left[ \cdot \right]$ represents an equivalence class with respect to the \v{C}ech complex. On the level of \v{C}ech complexes we have the map
\bea
\alpha^2 \colon \check{C}^2 \left( \mathcal{U}, \mathcal{L}^\prime \right) \to \check{C}^2 \left( \mathcal{U} , \mathcal{L} \right) \; , \; x \mapsto \tilde{s} \cdot x .
\eea
From the commutativity of the $E_0$-sheet it follows
\bea 
\alpha^3 \circ \delta \left( x \right) = \delta \circ \alpha^2 \left( x \right). 
\eea
Hence if $x$ is closed, so is $\alpha^2 \left( x \right)$. This enables us to consider the natural map on the \v{C}ech cohomologies given by
\bea \tilde{\alpha}^2 \colon \check{H}^2 \left( \mathcal{U}, \mathcal{L}^\prime \right) \to \check{H}^2 \left( \mathcal{U}, \mathcal{L} \right) \; , \; \left[ x \right] \mapsto \left[ \alpha^2 \left( x \right) \right] = \left[ \tilde{s} \cdot x \right] .
\eea
In particular we find for $x := \left[ \frac{A_1}{x_1 x_2 x_3^2} + \frac{A_2}{x_1 x_2^2 x_3} + \frac{A_3}{x_1^2 x_2 x_3} \right]$ that $\tilde{\alpha}^2 \left( x \right) = \left[ M_1 \right] + \left[ M_2 \right]$ where
\bea
M_1 &=& \frac{A_1 C_3}{x_1 x_2 x_3} + \frac{A_2 C_2}{x_1 x_2 x_3} + \frac{A_3 C_1}{x_1 x_2 x_3}, \\
M_2 &=& \frac{A_2 C_3}{x_1 x_2^2} + \frac{A_3 C_3}{x_1^2 x_2} + \frac{A_1 C_2}{x_1 x_3^2} + \frac{A_1 C_1}{x_2 x_3^2} + \frac{A_3 C_2}{x_1^2 x_3} + \frac{A_2 C_1}{x_2^2 x_3}. \label{equ:AdditionalRationomSapceM2}
\eea
The crucial observation is that all rationoms in $M_2$ lie in the image of the \v{C}ech coboundary
\bea
\delta \colon \check{C}^1 \left( \mathcal{U}, \mathcal{L} \right) {\rightarrow} \check{C}^2 \left( \mathcal{U} , \mathcal{L} \right). 
\eea
This can be verified explicitly by computing the \v{C}ech cochains as described in \cite{cox2011toric} (see also \cite{Cvetic:2010rq}). Therefore we ignore the rationoms in $M_2$ and find
\bea 
\tilde{\alpha}^2 \left( x \right) = A_1 C_3 \left[ \frac{1}{x_1 x_2 x_3} \right] + A_2 C_2 \left[ \frac{1}{x_1 x_2 x_3} \right] + A_3 C_1 \left[ \frac{1}{x_1 x_2 x_3} \right].
\eea
From this we can represent the map $\tilde{\alpha}^2$ by the matrix $M_{\tilde{\alpha}^2} = \left( C_3 , C_2 , C_1 \right)$. In conclusion we thus find
\bea
H^0 \left( C, \left. \mathcal{L} \right|_C \right) = 0, \quad H^1 \left( C, \left. \mathcal{L} \right|_C \right) = \text{ker} \left( \tilde{\alpha}^2 \right), \quad H^2 \left( C, \left. \mathcal{L} \right|_C \right) = \text{coker} \left( \tilde{\alpha}^2 \right).
\eea
Note that for equ. (\ref{equ:Koszul2}) to be sheaf exact, the section $\tilde{s}$ must be non-trivial so that $\tilde{s}$ indeed cuts out a hypersurface. Consequently at least one of the three parameters $C_1$, $C_2$, $C_3$ must be non-zero. In the case at hand this suffices to conclude $\text{im} \left( \tilde{\alpha}^2 \right) \cong \mathbb{C}$ and therefore
\bea
h^0 \left( C, \left. \mathcal{L} \right|_C \right) = 0, \quad h^1 \left( C, \left. \mathcal{L} \right|_C \right) = 2, \quad h^2 \left( C, \left. \mathcal{L} \right|_C \right) = 0.
\eea
The vanishing of $h^2 \left( C, \left. \mathcal{L} \right|_C \right)$ is expected from the finiteness theorem, whilst the dimensions of $h^0$ and $h^1$ are easily confirmed with the Koszul extension of \emph{cohomCalg} \cite{cohomCalg:Implementation}.

This method is easily generalised e.g. to the ambient space $X_{\Sigma} = \mathbb {CP}^2 \times \mathbb {CP}^1 \times \mathbb {CP}^1$ considered in the example presented in section \ref{sec:AConcreteExample}.

\section{Deligne cohomology from the Deligne-Beilinson
  complex \label{app-DeligneCohom}}

Recall that we introduced Deligne cohomology in section  \ref{ssec:GaugeDataDeligne}
as a generalization of the exact sequence
$0 \to J^1(X) \to H^1(X,\mathcal{O}_X^\ast) \to H^{1,1}_{\mathbb{Z}}(X) \to 0$ 
to
\begin{equation}
\label{eq:deligne-short-exact}
  0 \to J^p(X) \to H^{2p}_D(X,\mathbb{Z}(p)) \xrightarrow{\hat c_p}
  H^{p,p}_\mathbb{Z}(X) \to 0,
\end{equation}
that is as an extension of the group of Hodge cycles $H^{p,p}_\mathbb{Z}(X)$ by the intermediate Jacobian $J^p(X)$.
In this appendix we will review how to arrive at a definition of Deligne Cohomology by thinking about how that short sequence might arise.

Recall that on a K\"ahler manifold $X$, there is a Hodge decomposition
\begin{equation}
H^k(X,\mathbb{C}) = \bigoplus_{p + q = k} H^{p,q}(X,\mathbb{C})
\end{equation}
which allows us to define a filtration
\begin{equation}
F^p H^k(X) = \bigoplus_{p' \geq p} H^{p',k-p'} =
H^{k ,0} \oplus H^{k-1,1} \oplus \ldots \oplus H^{p,k-p}.
\end{equation}
The $p$-th intermediate Jacobian of $X$ is defined to be
\begin{equation}
\label{eq:6c}
  J^p(X) =  \frac{H^{2p-1}(X,\mathbb{C})}{F^pH^{2p-1}(X,\mathbb{C})
    + H^{2p - 1}(X,\mathbb{Z})}.
\end{equation}
For $p = 1$ this is
\begin{equation}
  J^1(X) =  \frac{H^{1}(X,\mathbb{C})}{H^{1,0}(X,\mathbb{C})
    + H^{1}(X,\mathbb{Z})} = H^{0,1}(X,\mathbb{C})/H^{1}(X,\mathbb{Z}),
\end{equation}
so it generalizes the classical notion of the Jacobian.


Now note that the integral Hodge cycles can be described as a kernel
$$H^{p,p}_\mathbb{Z}(X) = {\Ker}(H^{2p}(X,\mathbb{Z}) \to
H^{2p}(X,\mathbb{C})/F^pH^{2p}(X,\mathbb{C}))$$
and likewise the $p$-th intermediate Jacobian is a cokernel
$$J^p(X) = \coker(H^{2p-1}(X,\mathbb{Z}) \to
H^{2p-1}(X,\mathbb{C})/F^pH^{2p-1}(X,\mathbb{C})),$$
so the exact sequence (\ref{eq:deligne-short-exact}) could be deduced from the long exact sequence:
\begin{equation}
  \label{eq:3}
  \cdots \to H^k_D(X,\mathbb{Z}(p)) \to H^k(X,\mathbb{Z}) \to
  H^k(X,\mathbb{C}) / F^pH^k(X,\mathbb{C}) \to
  H^{k+1}_D(X,\mathbb{Z}(p)) \to \cdots.
\end{equation}

This suggests  reexpressing  $H^k(X,\mathbb{C}) /
F^pH^k(X,\mathbb{C})$. In order to do this we will need the notion of
\emph{hypercohomology of a complex}.

The \emph{holomorphic deRham complex} is the complex of sheaves of holomorphic
differential forms $\Omega^\bullet_X$ on $X$,
\begin{equation}
0 \to \mathcal{O}_X \to \Omega^1_X \to \cdots \to \Omega^n_X \to 0.
\end{equation}
Its \emph{hypercohomology} $\mathbb{H}^k(\Omega^\bullet_X) =
H^k(X,\mathbb{C})$ is just ordinary cohomology with complex
coefficients by the Dolbeault lemma\footnote{In the same way that the hypercohomology of the
  deRahm complex $0 \to \mathcal{C}^\infty_X \to \mathcal{A}^1_X \to
  \ldots$ is just ordinary cohomology with real coefficients, by the
  Poincar\'e lemma.}.
Moreover it is possible to show that the hypercohomology
$\mathbb{H}^k(X,\Omega_X^{\geq p})$ of the
truncated holomorphic deRahm complex $\Omega_X^{\geq p} = \Omega_X^p
\to \cdots \to \Omega_X^n$ is isomorphic to $F^pH^k(X,\mathbb{C})$. So
by the short exact sequence of complexes
\begin{equation}
  \label{eq:11}
  0 \to \Omega^{\geq p}_X \to \Omega^\bullet_X \to \Omega^{\leq p-1}_X
  \to 0 
\end{equation}
the hypercohomology of the truncated deRahm complex $\Omega^{\leq p -1}(X)$ is 
$$\mathbb{H}^k(\Omega^{\leq
  p-1}(X)) = H^k(X,\mathbb{C})/F^pH^k(X,\mathbb{C}).$$
Deligne had the insight to consider the following complex, now called
the \emph{Deligne-Beilinson complex},
\begin{equation}
  \label{eq:Deligne-Complex}
  \mathbb{Z}(p)_D = 0 \to \mathbb{Z} \xrightarrow{(2\pi i)^p} \mathcal{O}_X \to
  \Omega^1_X \to \cdots \to \Omega^{p-1}_X.
\end{equation}
By construction, there is a morphism from the truncated holomorphic deRahm complex
$\Omega^{\leq p-1}_X[1]$ with degree shifted by one to it, as well as a
morphism to $\mathbb{Z}$, viewed as a complex with one entry. In other
words there is a short exact sequence of complexes
\begin{equation}
  \label{eq:14}
  0 \to \Omega^{\leq p-1}_X[1] \to \mathbb{Z}_D(p) \to \mathbb{Z} \to 0.
\end{equation}
The crucial insight is that the induced long exact sequence in hypercohomology is precisely
the exact sequence (\ref{eq:deligne-short-exact}). That is if we define Deligne cohomology to
be the hypercohomology of the \emph{Deligne complex}
$\mathbb{Z}^\bullet_D(p)$,
\begin{equation}
H^k_D(X,\mathbb{Z}(p)) = \mathbb{H}^k(Z^\bullet_D(p)),
\end{equation}
then the induced long exact sequence reads
\begin{equation}
 \cdots \to H^k_D(X,\mathbb{Z}(p)) \to H^k(X,\mathbb{Z}) \to
  H^k(X,\mathbb{C}) / F^pH^k(X,\mathbb{C}) \to
  H^{k+1}_D(X,\mathbb{Z}(p)) \to \cdots.
\end{equation}

There is another way to arrive at the definition of Deligne
cohomology. Consider the two embeddings $\alpha_1 \colon \mathbb{Z}
\to \Omega^\bullet_X$ and $\alpha_2 \colon \Omega^{\geq p}_X  \to
\Omega^\bullet_X$ in the holomorphic deRahm complex, where $\alpha_1$
is $(2 \pi i)^p$ times the natural inclusion of $\mathbb{Z}$ in
$\mathcal{O}_X$ and $\alpha_2$ is, up to sign, the natural
inclusion. Form the cone of the morphism of complexes of sheaves over
$X$:

\begin{equation}
  \label{eq:Cone-Deligne}
  \mathbb{Z} \oplus \Omega^{\geq p}_X \xrightarrow{\alpha_1 - \alpha_2} \Omega^\bullet_X.
\end{equation}

It can be shown that this cone is quasiisomorphic to the Deligne
complex (\ref{eq:Deligne-Complex}) defined above \cite{EsnaultDeligne} and therefore their hypercohomologies coincide.

\newpage
\bibliography{papers}  

\begin{thebibliography}{10}
\expandafter\ifx\csname url\endcsname\relax
  \def\url#1{{\tt #1}}\fi
\expandafter\ifx\csname urlprefix\endcsname\relax\def\urlprefix{URL }\fi
\providecommand{\eprint}[2][]{\url{#2}}

\bibitem{Vafa:1996xn}
C.~Vafa, {\it {Evidence for F-Theory}\/}, {\it Nucl. Phys.\/} {\bf B469} (1996)
  403--418, \href{http://www.arxiv.org/abs/hep-th/9602022}{{\tt
  [hep-th/9602022]}}.

\bibitem{oai:arXiv.org:hep-th/9602114}
D.~R. Morrison and C.~Vafa, {\it {Compactifications of F theory on Calabi-Yau
  threefolds. 1}\/}, {\it Nucl.Phys.\/} {\bf B473} (1996) 74--92,
  \href{http://www.arxiv.org/abs/hep-th/9602114}{{\tt [hep-th/9602114]}}.

\bibitem{oai:arXiv.org:hep-th/9603161}
D.~R. Morrison and C.~Vafa, {\it {Compactifications of F theory on Calabi-Yau
  threefolds. 2.}\/}, {\it Nucl.Phys.\/} {\bf B476} (1996) 437--469,
  \href{http://www.arxiv.org/abs/hep-th/9603161}{{\tt [hep-th/9603161]}}.

\bibitem{oai:arXiv.org:0802.2969}
R.~Donagi and M.~Wijnholt, {\it {Model Building with F-Theory}\/}, {\it
  Adv.Theor.Math.Phys.\/} {\bf 15} (2011) 1237--1318,
  \href{http://www.arxiv.org/abs/0802.2969}{{\tt [0802.2969]}}.

\bibitem{oai:arXiv.org:0802.3391}
C.~Beasley, J.~J. Heckman and C.~Vafa, {\it {GUTs and Exceptional Branes in
  F-theory - I}\/}, {\it JHEP\/} {\bf 0901} (2009) 058,
  \href{http://www.arxiv.org/abs/0802.3391}{{\tt [0802.3391]}}.

\bibitem{Beasley:2008kw}
C.~Beasley, J.~J. Heckman and C.~Vafa, {\it {GUTs and Exceptional Branes in
  F-theory - II: Experimental Predictions}\/}, {\it JHEP\/} {\bf 01} (2009)
  059, \href{http://www.arxiv.org/abs/0806.0102}{{\tt [0806.0102]}}.

\bibitem{Donagi:2008kj}
R.~Donagi and M.~Wijnholt, {\it {Breaking GUT Groups in F-Theory}\/}, {\it
  Adv.Theor.Math.Phys.\/} {\bf 15} (2011) 1523--1604,
  \href{http://www.arxiv.org/abs/0808.2223}{{\tt [0808.2223]}}.

\bibitem{Katz:1996xe}
S.~H. Katz and C.~Vafa, {\it {Matter from geometry}\/}, {\it Nucl.Phys.\/} {\bf
  B497} (1997) 146--154, \href{http://www.arxiv.org/abs/hep-th/9606086}{{\tt
  [hep-th/9606086]}}.

\bibitem{oai:arXiv.org:0904.1218}
R.~Donagi and M.~Wijnholt, {\it {Higgs Bundles and UV Completion in
  F-Theory}\/} \href{http://www.arxiv.org/abs/0904.1218}{{\tt [0904.1218]}}.

\bibitem{oai:arXiv.org:1108.1794}
J.~Marsano and S.~Schafer-Nameki, {\it {Yukawas, G-flux, and Spectral Covers
  from Resolved Calabi-Yau's}\/}, {\it JHEP\/} {\bf 1111} (2011) 098,
  \href{http://www.arxiv.org/abs/1108.1794}{{\tt [1108.1794]}}.

\bibitem{Braun:2011zm}
A.~P. Braun, A.~Collinucci and R.~Valandro, {\it {G-flux in F-theory and
  algebraic cycles}\/}, {\it Nucl. Phys.\/} {\bf B856} (2012) 129--179,
  \href{http://www.arxiv.org/abs/1107.5337}{{\tt [1107.5337]}}.

\bibitem{Krause:2011xj}
S.~Krause, C.~Mayrhofer and T.~Weigand, {\it {$G_4$ flux, chiral matter and
  singularity resolution in F-theory compactifications}\/}, {\it Nucl.Phys.\/}
  {\bf B858} (2012) 1--47, \href{http://www.arxiv.org/abs/1109.3454}{{\tt
  [1109.3454]}}.

\bibitem{oai:arXiv.org:1202.3138}
S.~Krause, C.~Mayrhofer and T.~Weigand, {\it {Gauge Fluxes in F-theory and Type
  IIB Orientifolds}\/}, {\it JHEP\/} {\bf 1208} (2012) 119,
  \href{http://www.arxiv.org/abs/1202.3138}{{\tt [1202.3138]}}.

\bibitem{Clingher:2012rg}
A.~Clingher, R.~Donagi and M.~Wijnholt, {\it {The Sen Limit}\/}
  \href{http://www.arxiv.org/abs/1212.4505}{{\tt [1212.4505]}}.

\bibitem{oai:arXiv.org:1111.1232}
T.~W. Grimm and H.~Hayashi, {\it {F-theory fluxes, Chirality and Chern-Simons
  theories}\/}, {\it JHEP\/} {\bf 1203} (2012) 027,
  \href{http://www.arxiv.org/abs/1111.1232}{{\tt [1111.1232]}}.

\bibitem{Diaconescu:2003bm}
E.~Diaconescu, G.~W. Moore and D.~S. Freed, {\it {The M theory three form and
  E(8) gauge theory}\/} \href{http://www.arxiv.org/abs/hep-th/0312069}{{\tt
  [hep-th/0312069]}}.

\bibitem{Freed:2004yc}
D.~S. Freed and G.~W. Moore, {\it {Setting the quantum integrand of
  M-theory}\/}, {\it Commun.Math.Phys.\/} {\bf 263} (2006) 89--132,
  \href{http://www.arxiv.org/abs/hep-th/0409135}{{\tt [hep-th/0409135]}}.

\bibitem{oai:arXiv.org:hep-th/0409158}
G.~W. Moore, {\it {Anomalies, Gauss laws, and Page charges in M-theory}\/},
  {\it Comptes Rendus Physique\/} {\bf 6} (2005) 251--259,
  \href{http://www.arxiv.org/abs/hep-th/0409158}{{\tt [hep-th/0409158]}}.

\bibitem{Cheeger-Simons}
J.~Cheeger and J.~Simons, {\it {Differential characters and geometric
  invariants}\/}, {\it Lecture Notes in Math.\/} {\bf 1167} (1985) 50--80.

\bibitem{Curio:1998bva}
G.~Curio and R.~Y. Donagi, {\it {Moduli in N=1 heterotic / F theory
  duality}\/}, {\it Nucl.Phys.\/} {\bf B518} (1998) 603--631,
  \href{http://www.arxiv.org/abs/hep-th/9801057}{{\tt [hep-th/9801057]}}.

\bibitem{Donagi:1998vw}
R.~Donagi, {\it {Heterotic / F theory duality: ICMP lecture}\/}
  \href{http://www.arxiv.org/abs/hep-th/9802093}{{\tt [hep-th/9802093]}}.

\bibitem{EsnaultDeligne}
H.~Esnault and E.~Viehweg, {\it Beilinson's conjectures on special values of
  L-functions\/}, chapter Deligne-Beilinson cohomology, Waltham: Academic
  Press, 1988.

\bibitem{oai:arXiv.org:1203.6662}
K.~Intriligator, H.~Jockers, P.~Mayr, D.~R. Morrison and M.~R. Plesser, {\it
  {Conifold Transitions in M-theory on Calabi-Yau Fourfolds with Background
  Fluxes}\/} \href{http://www.arxiv.org/abs/1203.6662}{{\tt [1203.6662]}}.

\bibitem{Anderson:2013rka}
L.~B. Anderson, J.~J. Heckman and S.~Katz, {\it {T-Branes and Geometry}\/}
  \href{http://www.arxiv.org/abs/1310.1931}{{\tt [1310.1931]}}.

\bibitem{GreenMurreVoisin}
M.~Green, J.~Murre and C.~Voisin, {\it Algebraic Cycles and Hodge Theory\/},
  Springer, 1994.

\bibitem{EisenbudInt}
D.~Eisenbud and J.~Harris, {\it 3264 \& All that, Intersection Theory and
  Algebraic Geometry\/}, 2013.

\bibitem{FultonInt}
W.~Fulton, {\it {Intersection Theory}\/}, {\it Princeton University Press
  1993\/} .

\bibitem{Katz:2002gh}
S.~H. Katz and E.~Sharpe, {\it {D-branes, open string vertex operators, and Ext
  groups}\/}, {\it Adv.Theor.Math.Phys.\/} {\bf 6} (2003) 979--1030,
  \href{http://www.arxiv.org/abs/hep-th/0208104}{{\tt [hep-th/0208104]}}.

\bibitem{Donagi:2011jy}
R.~Donagi and M.~Wijnholt, {\it {Gluing Branes, I}\/}, {\it JHEP\/} {\bf 1305}
  (2013) 068, \href{http://www.arxiv.org/abs/1104.2610}{{\tt [1104.2610]}}.

\bibitem{Donagi:2011dv}
R.~Donagi and M.~Wijnholt, {\it {Gluing Branes II: Flavour Physics and String
  Duality}\/}, {\it JHEP\/} {\bf 1305} (2013) 092,
  \href{http://www.arxiv.org/abs/1112.4854}{{\tt [1112.4854]}}.

\bibitem{Marsano:2012yc}
J.~Marsano, H.~Clemens, T.~Pantev, S.~Raby and H.-H. Tseng, {\it {A Global
  SU(5) F-theory model with Wilson line breaking}\/}, {\it JHEP\/} {\bf 1301}
  (2013) 150, \href{http://www.arxiv.org/abs/1206.6132}{{\tt [1206.6132]}}.

\bibitem{Grimm:2010ez}
T.~W. Grimm and T.~Weigand, {\it {On Abelian Gauge Symmetries and Proton Decay
  in Global F-theory GUTs}\/}, {\it Phys.Rev.\/} {\bf D82} (2010) 086009,
  \href{http://www.arxiv.org/abs/1006.0226}{{\tt [1006.0226]}}.

\bibitem{Grimm:2011fx}
T.~W. Grimm and H.~Hayashi, {\it {F-theory fluxes, Chirality and Chern-Simons
  theories}\/}, {\it JHEP\/} {\bf 1203} (2012) 027,
  \href{http://www.arxiv.org/abs/1111.1232}{{\tt [1111.1232]}}.

\bibitem{oai:arXiv.org:1208.2695}
D.~R. Morrison and D.~S. Park, {\it {F-Theory and the Mordell-Weil Group of
  Elliptically-Fibered Calabi-Yau Threefolds}\/}, {\it JHEP\/} {\bf 1210}
  (2012) 128, \href{http://www.arxiv.org/abs/1208.2695}{{\tt [1208.2695]}}.

\bibitem{oai:arXiv.org:1210.6034}
M.~Cvetic, T.~W. Grimm and D.~Klevers, {\it {Anomaly Cancellation And Abelian
  Gauge Symmetries In F-theory}\/}, {\it JHEP\/} {\bf 1302} (2013) 101,
  \href{http://www.arxiv.org/abs/1210.6034}{{\tt [1210.6034]}}.

\bibitem{Mayrhofer:2012zy}
C.~Mayrhofer, E.~Palti and T.~Weigand, {\it {U(1) symmetries in F-theory GUTs
  with multiple sections}\/}, {\it JHEP\/} {\bf 1303} (2013) 098,
  \href{http://www.arxiv.org/abs/1211.6742}{{\tt [1211.6742]}}.

\bibitem{Braun:2013yti}
V.~Braun, T.~W. Grimm and J.~Keitel, {\it {New Global F-theory GUTs with U(1)
  symmetries}\/}, {\it JHEP\/} {\bf 1309} (2013) 154,
  \href{http://www.arxiv.org/abs/1302.1854}{{\tt [1302.1854]}}.

\bibitem{Borchmann:2013jwa}
J.~Borchmann, C.~Mayrhofer, E.~Palti and T.~Weigand, {\it {Elliptic fibrations
  for SU(5) x U(1) x U(1) F-theory vacua}\/}
  \href{http://www.arxiv.org/abs/1303.5054}{{\tt [1303.5054]}}.

\bibitem{Cvetic:2013nia}
M.~Cvetic, D.~Klevers and H.~Piragua, {\it {F-Theory Compactifications with
  Multiple U(1)-Factors: Constructing Elliptic Fibrations with Rational
  Sections}\/}, {\it JHEP\/} {\bf 1306} (2013) 067,
  \href{http://www.arxiv.org/abs/1303.6970}{{\tt [1303.6970]}}.

\bibitem{Braun:2013nqa}
V.~Braun, T.~W. Grimm and J.~Keitel, {\it {Geometric Engineering in Toric
  F-Theory and GUTs with U(1) Gauge Factors}\/}, {\it JHEP\/} {\bf 1312} (2013)
  069, \href{http://www.arxiv.org/abs/1306.0577}{{\tt [1306.0577]}}.

\bibitem{Cvetic:2013uta}
M.~Cveti\v{c}, A.~Grassi, D.~Klevers and H.~Piragua, {\it {Chiral
  Four-Dimensional F-Theory Compactifications With SU(5) and Multiple
  U(1)-Factors}\/} \href{http://www.arxiv.org/abs/1306.3987}{{\tt
  [1306.3987]}}.

\bibitem{Borchmann:2013hta}
J.~Borchmann, C.~Mayrhofer, E.~Palti and T.~Weigand, {\it {SU(5) Tops with
  Multiple U(1)s in F-theory}\/} \href{http://www.arxiv.org/abs/1307.2902}{{\tt
  [1307.2902]}}.

\bibitem{Cvetic:2013qsa}
M.~Cvetic, D.~Klevers, H.~Piragua and P.~Song, {\it {Elliptic Fibrations with
  Rank Three Mordell-Weil Group: F-theory with U(1) x U(1) x U(1) Gauge
  Symmetry}\/} \href{http://www.arxiv.org/abs/1310.0463}{{\tt [1310.0463]}}.

\bibitem{Blumenhagen:2010pv}
R.~Blumenhagen, B.~Jurke, T.~Rahn and H.~Roschy, {\it {Cohomology of Line
  Bundles: A Computational Algorithm}\/}, {\it J. Math. Phys.\/} {\bf 51}
  (2010) 103525, \href{http://www.arxiv.org/abs/1003.5217}{{\tt [1003.5217]}}.

\bibitem{cohomCalg:Implementation}
{\it {cohomCalg package}\/}, Download link, 2010,
  \urlprefix\url{http://wwwth.mppmu.mpg.de/members/blumenha/cohomcalg/},
  high-performance line bundle cohomology computation based on
  \cite{Blumenhagen:2010pv}.

\bibitem{2011JMP....52c3506J}
S.-Y. {Jow}, {\it {Cohomology of toric line bundles via simplicial Alexander
  duality}\/}, {\it Journal of Mathematical Physics\/} {\bf 52}, no.~3 (2011)
  033506, \href{http://www.arxiv.org/abs/1006.0780}{{\tt [1006.0780]}}.

\bibitem{Rahn:2010fm}
T.~Rahn and H.~Roschy, {\it {Cohomology of Line Bundles: Proof of the
  Algorithm}\/}, {\it J.Math.Phys.\/} {\bf 51} (2010) 103520,
  \href{http://www.arxiv.org/abs/1006.2392}{{\tt [1006.2392]}}.

\bibitem{Blumenhagen:2010ed}
R.~Blumenhagen, B.~Jurke, T.~Rahn and H.~Roschy, {\it {Cohomology of Line
  Bundles: Applications}\/}, {\it J.Math.Phys.\/} {\bf 53} (2012) 012302,
  \href{http://www.arxiv.org/abs/1010.3717}{{\tt [1010.3717]}}.

\bibitem{Aspinwall:2004jr}
P.~S. Aspinwall, {\it {D-branes on Calabi-Yau manifolds}\/} pp. 1--152,
  \href{http://www.arxiv.org/abs/hep-th/0403166}{{\tt [hep-th/0403166]}}.

\bibitem{Distler:1987ee}
J.~Distler and B.~R. Greene, {\it {Aspects of (2,0) String
  Compactifications}\/}, {\it Nucl.Phys.\/} {\bf B304} (1988) 1.

\bibitem{Freed:1999vc}
D.~S. Freed and E.~Witten, {\it {Anomalies in string theory with D-branes}\/}
  \href{http://www.arxiv.org/abs/hep-th/9907189}{{\tt [hep-th/9907189]}}.

\bibitem{Blumenhagen:2009yv}
R.~Blumenhagen, T.~W. Grimm, B.~Jurke and T.~Weigand, {\it {Global F-theory
  GUTs}\/}, {\it Nucl. Phys.\/} {\bf B829} (2010) 325--369,
  \href{http://www.arxiv.org/abs/0908.1784}{{\tt [0908.1784]}}.

\bibitem{Grimm:2009yu}
T.~W. Grimm, S.~Krause and T.~Weigand, {\it {F-Theory GUT Vacua on Compact
  Calabi-Yau Fourfolds}\/}, {\it JHEP\/} {\bf 07} (2010) 037,
  \href{http://www.arxiv.org/abs/0912.3524}{{\tt [0912.3524]}}.

\bibitem{Chen:2010ts}
C.-M. Chen, J.~Knapp, M.~Kreuzer and C.~Mayrhofer, {\it {Global SO(10) F-theory
  GUTs}\/}, {\it JHEP\/} {\bf 1010} (2010) 057,
  \href{http://www.arxiv.org/abs/1005.5735}{{\tt [1005.5735]}}.

\bibitem{oai:arXiv.org:1011.6388}
A.~Collinucci and R.~Savelli, {\it {On Flux Quantization in F-Theory}\/}, {\it
  JHEP\/} {\bf 1202} (2012) 015, \href{http://www.arxiv.org/abs/1011.6388}{{\tt
  [1011.6388]}}.

\bibitem{Knapp:2011wk}
J.~Knapp, M.~Kreuzer, C.~Mayrhofer and N.-O. Walliser, {\it {Toric Construction
  of Global F-Theory GUTs}\/}, {\it JHEP\/} {\bf 1103} (2011) 138,
  \href{http://www.arxiv.org/abs/1101.4908}{{\tt [1101.4908]}}.

\bibitem{Esole:2011sm}
M.~Esole and S.-T. Yau, {\it {Small resolutions of SU(5)-models in F-theory}\/}
  \href{http://www.arxiv.org/abs/1107.0733}{{\tt [1107.0733]}}.

\bibitem{Lawrie:2012gg}
C.~Lawrie and S.~SchŠfer-Nameki, {\it {The Tate Form on Steroids: Resolution
  and Higher Codimension Fibers}\/}, {\it JHEP\/} {\bf 1304} (2013) 061,
  \href{http://www.arxiv.org/abs/1212.2949}{{\tt [1212.2949]}}.

\bibitem{Hayashi:2013lra}
H.~Hayashi, C.~Lawrie and S.~Schafer-Nameki, {\it {Phases, Flops and F-theory:
  SU(5) Gauge Theories}\/}, {\it JHEP\/} {\bf 1310} (2013) 046,
  \href{http://www.arxiv.org/abs/1304.1678}{{\tt [1304.1678]}}.

\bibitem{Hayashi:2014kca}
H.~Hayashi, C.~Lawrie, D.~R. Morrison and S.~Schafer-Nameki, {\it {Box Graphs
  and Singular Fibers}\/} \href{http://www.arxiv.org/abs/1402.2653}{{\tt
  [1402.2653]}}.

\bibitem{oai:arXiv.org:hep-th/9605053}
K.~Becker and M.~Becker, {\it {M theory on eight manifolds}\/}, {\it
  Nucl.Phys.\/} {\bf B477} (1996) 155--167,
  \href{http://www.arxiv.org/abs/hep-th/9605053}{{\tt [hep-th/9605053]}}.

\bibitem{oai:arXiv.org:hep-th/9606122}
S.~Sethi, C.~Vafa and E.~Witten, {\it {Constraints on low dimensional string
  compactifications}\/}, {\it Nucl.Phys.\/} {\bf B480} (1996) 213--224,
  \href{http://www.arxiv.org/abs/hep-th/9606122}{{\tt [hep-th/9606122]}}.

\bibitem{oai:arXiv.org:hep-th/9908088}
K.~Dasgupta, G.~Rajesh and S.~Sethi, {\it {M theory, orientifolds and G -
  flux}\/}, {\it JHEP\/} {\bf 9908} (1999) 023,
  \href{http://www.arxiv.org/abs/hep-th/9908088}{{\tt [hep-th/9908088]}}.

\bibitem{oai:arXiv.org:hep-th/9609122}
E.~Witten, {\it {On flux quantization in M theory and the effective action}\/},
  {\it J.Geom.Phys.\/} {\bf 22} (1997) 1--13,
  \href{http://www.arxiv.org/abs/hep-th/9609122}{{\tt [hep-th/9609122]}}.

\bibitem{oai:arXiv.org:1203.4542}
A.~Collinucci and R.~Savelli, {\it {On Flux Quantization in F-Theory II:
  Unitary and Symplectic Gauge Groups}\/}, {\it JHEP\/} {\bf 1208} (2012) 094,
  \href{http://www.arxiv.org/abs/1203.4542}{{\tt [1203.4542]}}.

\bibitem{Intriligator:2012ue}
K.~Intriligator, H.~Jockers, P.~Mayr, D.~R. Morrison and M.~R. Plesser, {\it
  {Conifold Transitions in M-theory on Calabi-Yau Fourfolds with Background
  Fluxes}\/} \href{http://www.arxiv.org/abs/1203.6662}{{\tt [1203.6662]}}.

\bibitem{VoisinHodgeTheoryI}
C.~Voisin, {\it Hodge Theory and Complex Algebraic Geometry I\/}, {\it
  Cambridge University Press 2002\/} .

\bibitem{BredonAT}
G.~E. Bredon, {\it Topology and Geometry\/}, Springer, 1997.

\bibitem{Donagi:2009ra}
R.~Donagi and M.~Wijnholt, {\it {Higgs Bundles and UV Completion in
  F-Theory}\/} \href{http://www.arxiv.org/abs/0904.1218}{{\tt [0904.1218]}}.

\bibitem{FultonTor}
W.~Fulton, {\it {Introduction to Toric Varieties}\/}, {\it Springer 1997\/} .

\bibitem{Klemm:1996hh}
A.~Klemm, P.~Mayr and C.~Vafa, {\it {BPS states of exceptional noncritical
  strings}\/} \href{http://www.arxiv.org/abs/hep-th/9607139}{{\tt
  [hep-th/9607139]}}.

\bibitem{Aldazabal:1996du}
G.~Aldazabal, A.~Font, L.~E. Ibanez and A.~Uranga, {\it {New branches of string
  compactifications and their F theory duals}\/}, {\it Nucl.Phys.\/} {\bf B492}
  (1997) 119--151, \href{http://www.arxiv.org/abs/hep-th/9607121}{{\tt
  [hep-th/9607121]}}.

\bibitem{Candelas:1997pv}
P.~Candelas, E.~Perevalov and G.~Rajesh, {\it {Comments on A, B, C chains of
  heterotic and type II vacua}\/}, {\it Nucl.Phys.\/} {\bf B502} (1997)
  594--612, \href{http://www.arxiv.org/abs/hep-th/9703148}{{\tt
  [hep-th/9703148]}}.

\bibitem{Berglund:1998va}
P.~Berglund, A.~Klemm, P.~Mayr and S.~Theisen, {\it {On type IIB vacua with
  varying coupling constant}\/}, {\it Nucl.Phys.\/} {\bf B558} (1999) 178--204,
  \href{http://www.arxiv.org/abs/hep-th/9805189}{{\tt [hep-th/9805189]}}.

\bibitem{Shioda}
T.~Shioda, {\it {On the Mordell-Weil lattices}\/}, {\it Comment. Math. Univ.
  St. Pauli\/} {\bf 39} (1990) 211--240.

\bibitem{RahnPhd}
T.~Rahn, {\it {Heterotic Target Space Dualities with Line Bundle
  Cohomology}\/}, Ph.D. thesis, Ludwig-Maximilians-Universität München, 2012,
  \urlprefix\url{http://edoc.ub.uni-muenchen.de/14344/2/Rahn_Thorsten.pdf}.

\bibitem{Anderson:2013qca}
L.~B. Anderson, J.~Gray, A.~Lukas and B.~Ovrut, {\it {Vacuum Varieties,
  Holomorphic Bundles and Complex Structure Stabilization in Heterotic
  Theories}\/}, {\it JHEP\/} {\bf 1307} (2013) 017,
  \href{http://www.arxiv.org/abs/1304.2704}{{\tt [1304.2704]}}.

\bibitem{eisenbud1995commutative}
D.~Eisenbud, {\it Commutative Algebra: With a View Toward Algebraic
  Geometry\/}, Graduate Texts in Mathematics, Springer, 1995, ISBN
  9780387942698, \urlprefix\url{http://books.google.de/books?id=Fm\_yPgZBucMC}.

\bibitem{cox2011toric}
D.~Cox, J.~Little and H.~Schenck, {\it Toric Varieties\/}, Graduate studies in
  mathematics, American Mathematical Society, 2011, ISBN 9780821884263,
  \urlprefix\url{http://books.google.de/books?id=eXLGwYD4pmAC}.

\bibitem{distler1988aspects}
J.~Distler and B.~Greene, {\it Aspects of (2, 0) string compactifications\/},
  {\it Nuclear Physics B\/} {\bf 304} (1988) 1--62.

\bibitem{Cvetic:2010rq}
M.~Cveti\v{c}, I.~Garcia-Etxebarria and J.~Halverson, {\it {Global F-theory
  Models: Instantons and Gauge Dynamics}\/}, {\it JHEP\/} {\bf 1101} (2011)
  073, \href{http://www.arxiv.org/abs/1003.5337}{{\tt [1003.5337]}}.

\bibitem{Anderson:2008ex}
L.~B. Anderson, {\it {Heterotic and M-theory Compactifications for String
  Phenomenology}\/} \href{http://www.arxiv.org/abs/0808.3621}{{\tt
  [0808.3621]}}.

\bibitem{hubsch1994calabi}
T.~H{\"u}bsch, {\it Calabi-Yau Manifolds: A Bestiary for Physicists\/}, World
  Scientific, 1994, ISBN 9789810219277,
  \urlprefix\url{http://books.google.de/books?id=Z5zSbFktn1EC}.

\bibitem{Collinucci:2010gz}
A.~Collinucci and R.~Savelli, {\it {On Flux Quantization in F-Theory}\/}, {\it
  JHEP\/} {\bf 1202} (2012) 015, \href{http://www.arxiv.org/abs/1011.6388}{{\tt
  [1011.6388]}}.

\bibitem{Bouchard:2003bu}
V.~Bouchard and H.~Skarke, {\it {Affine Kac-Moody algebras, CHL strings and the
  classification of tops}\/}, {\it Adv. Theor. Math. Phys.\/} {\bf 7} (2003)
  205--232, \href{http://www.arxiv.org/abs/hep-th/0303218}{{\tt
  [hep-th/0303218]}}.

\bibitem{Cecotti:2010bp}
S.~Cecotti, C.~Cordova, J.~J. Heckman and C.~Vafa, {\it {T-Branes and
  Monodromy}\/}, {\it JHEP\/} {\bf 1107} (2011) 030,
  \href{http://www.arxiv.org/abs/1010.5780}{{\tt [1010.5780]}}.

\bibitem{Braun:2013cb}
A.~P. Braun and T.~Watari, {\it {On Singular Fibres in F-Theory}\/}, {\it
  JHEP\/} {\bf 1307} (2013) 031, \href{http://www.arxiv.org/abs/1301.5814}{{\tt
  [1301.5814]}}.

\bibitem{Grassi:2013kha}
A.~Grassi, J.~Halverson and J.~L. Shaneson, {\it {Matter From Geometry Without
  Resolution}\/} \href{http://www.arxiv.org/abs/1306.1832}{{\tt [1306.1832]}}.

\bibitem{Donagi:2004ia}
R.~Donagi, Y.-H. He, B.~A. Ovrut and R.~Reinbacher, {\it {The Particle spectrum
  of heterotic compactifications}\/}, {\it JHEP\/} {\bf 0412} (2004) 054,
  \href{http://www.arxiv.org/abs/hep-th/0405014}{{\tt [hep-th/0405014]}}.

\bibitem{Donagi:2004ub}
R.~Donagi, Y.-H. He, B.~A. Ovrut and R.~Reinbacher, {\it {The Spectra of
  heterotic standard model vacua}\/}, {\it JHEP\/} {\bf 0506} (2005) 070,
  \href{http://www.arxiv.org/abs/hep-th/0411156}{{\tt [hep-th/0411156]}}.

\bibitem{Bouchard:2005ag}
V.~Bouchard and R.~Donagi, {\it {An SU(5) heterotic standard model}\/}, {\it
  Phys.Lett.\/} {\bf B633} (2006) 783--791,
  \href{http://www.arxiv.org/abs/hep-th/0512149}{{\tt [hep-th/0512149]}}.

\bibitem{Braun:2005zv}
V.~Braun, Y.-H. He, B.~A. Ovrut and T.~Pantev, {\it {Vector bundle extensions,
  sheaf cohomology, and the heterotic standard model}\/}, {\it
  Adv.Theor.Math.Phys.\/} {\bf 10} (2006) 525--589,
  \href{http://www.arxiv.org/abs/hep-th/0505041}{{\tt [hep-th/0505041]}}.

\bibitem{Blumenhagen:2006wj}
R.~Blumenhagen, S.~Moster, R.~Reinbacher and T.~Weigand, {\it {Massless Spectra
  of Three Generation U(N) Heterotic String Vacua}\/}, {\it JHEP\/} {\bf 0705}
  (2007) 041, \href{http://www.arxiv.org/abs/hep-th/0612039}{{\tt
  [hep-th/0612039]}}.

\bibitem{Anderson:2011ns}
L.~B. Anderson, J.~Gray, A.~Lukas and E.~Palti, {\it {Two Hundred Heterotic
  Standard Models on Smooth Calabi-Yau Threefolds}\/}, {\it Phys.Rev.\/} {\bf
  D84} (2011) 106005, \href{http://www.arxiv.org/abs/1106.4804}{{\tt
  [1106.4804]}}.

\bibitem{Anderson:2012yf}
L.~B. Anderson, J.~Gray, A.~Lukas and E.~Palti, {\it {Heterotic Line Bundle
  Standard Models}\/}, {\it JHEP\/} {\bf 1206} (2012) 113,
  \href{http://www.arxiv.org/abs/1202.1757}{{\tt [1202.1757]}}.

\end{thebibliography}
\bibliographystyle{custom1}

\end{document}